\documentclass[structabstract]{raa_twocloumn}
\usepackage{graphicx,times}
\usepackage{threeparttable}
\usepackage{natbib,amssymb}
\usepackage{float}
\usepackage[figuresright]{rotating}
\usepackage{amsmath}

\begin{document}
   \title{The local properties of supernova explosion and their host galaxies}
   \volnopage{ {\bf 2019} Vol.\ {\bf XX} No. {\bf XXX}, 000--000}
   \setcounter{page}{1}
   \author{Li Zhou \inst{1,2,3}, Yan-Chun Liang \inst{2}, Jun-Qiang Ge \inst{2}, Xu Shao \inst{2}, Xiao-Yan Chen \inst{2}, Li-Cai Deng \inst{2}}
   \institute{Shandong Provincial Key Laboratory of Optical Astronomy and
  Solar-Terrestrial Environment, School of Space Science and Physics,
  Shandong University at Weihai, Weihai 264209, China;\\
          \and
  National Astronomical Observatories, Chinese Academy of Sciences,
          20A Datun Road, Chaoyang District, Beijing 100012, China;{\it ycliang@nao.cas.cn}\\
          \and
          Graduate University of the Chinese Academy of Sciences, Huairou District, 101408, Beijing,
  China}

   \date{Received; accepted}

\abstract{
We aim to understand the properties at the locations of supernova (SN) explosion in their host galaxies and compare with the global properties of the host galaxies. We use the integral field spectrograph (IFS) of Mapping Nearby Galaxies (MaNGA) at Apache Point Observatory (APO) to get the 2D maps of the parameter properties for eleven SN host galaxies. The sample galaxies are analyzed one by one in details on their properties of velocity field, star
formation rate, oxygen abundance and stellar mass etc. This sample of SN host galaxies have redshifts around $z$ $\sim$ 0.03, which is higher than those of the previous related works. The higher redshift distribution allows us to obtain the properties of more distant SN host galaxies. 
Metallicity (gas-phase oxygen abundance) estimated from integrated spectra could represent the local metallicity at SN explosion sites with small bias. 
All the host galaxies in our sample are metal-rich galaxies (12+log(O/H) $>$ 8.5) except for NGC 6387, which means supernovae (SNe) may be more inclined to explode in rich-metallicity galaxies. There is a positive relation
between global gas-phase oxygen abundance and the stellar mass of host galaxies.
We also try to compare the differences of the host galaxies between SN Ia and SN II.
In our sample, both SNe Ia and SNe II could explode in normal galaxies, while SNe II also could explode in an interactive or merger system, which has star formation in the galaxy. 
 }
\authorrunning{Zhou et al.}
\titlerunning{The local properties of supernovae explosion and their host galaxies}
\maketitle

\section{Introduction}
\label{sec:intro}

Supernovae explosions are one of the most important processes, which mark the end of a stellar's life.
According to the presence or absence of various features in the spectra,
supernovae (SNe) are classified into different types \citep{fi97}.
In the spectra of SNe I, hydrogen is absent. While in the SNe II spectra, hydrogen is present. Among Type I SNe, Si line is present in the spectra of SNe Ia, while the Si line is absent in the spectra of SNe Ib/c. The presence or absence of He line distinguishes SNe Ib from SNe Ic \citep{ha02,tu03}. According to the light curves, SNe II could be classified into SNe IIP and IIL \citep{ba79}. For the light curves of SNe IIP, there is a fast rise to peak \citep{2016ApJ...820...33R} and followed by a long plateau (~90 days). For SNe IIL, the light curves present a fast rise to peak and then followed by a linear decline ($>$ 1.4 mag/100 days). There are strong narrow or intermediate-width hydrogen emission lines superimposed on an otherwise smooth blue continuum in the spectra of SNe IIn \citep{sc90}.

It is generally known that in the end of a star's life, which has the initial mass larger than about 8 M$_\odot$, the star will explode due to the gravitational collapse and leave the remnant as a neutron star or black hole \citep{be79,ar89}. This is the core-collapse supernova (CCSN, including SN II and Ib/c). \citet{bi80} showed that if the initial stellar mass of a star is less than 8M${_\odot}$, the star will explode to a degenerate carbon-oxygen (CO) white dwarf, which may be ignited when its mass increases to 1.4 M$_\odot$ due to accreting materials from its non-degenerate companion star, and it can be completely disrupted in a bright thermonuclear explosion that produces a SN Ia \citep{hf60}. SNe Ia are known as the standard candle to estimate the distance of cosmology and they have led to the discovery of the accelerating expansion of the Universe and dark energy \citep{ri98,pe99}.

The relations between different types of SNe and their host galaxies have been studied, such as on the luminosity, color and the explosion environments of SNe, both on SNe Ia (\citealt{ha10}) and core-collapse supernovae (CCSNe, \citealt{ke12}). \citet{pr08} studied the metallicity of host galaxies of different types of SNe and found that SNe Ib/c tend to explode in galaxies with higher metallicity than SNe II. \citet{sh14} has compared the properties of the host galaxies of SN Ia, SN II and SN Ibc based on the SDSS DR7 data, with 213 host galaxies as working sample and 689 host galaxies as comparison ones. The sample galaxies of these works are from fiber spectra.

Considering the different stellar populations of HII regions and clumps within one galaxy, the integral field spectroscopy (IFS) of SN host galaxies is necessary to get robust conclusions on the local environment of SNe and their progenitors. The IFS allows the combination of spatial and specific spectral information of the local SNe explosion sites and other sites in the host galaxies, and simultaneously to make it possible to investigate the properties of SN progenitors. Thanks to the improvements and integral-field unit (IFU) spectroscopic survey on modern telescopes, some works have made this possible and focused their
studies on the local environment or progenitors of SN.

With the wide-field IFU spectrograph PMAS/PPAK at the 3.5 m telescope of Calar Alto Observatory,  \citet{st12} compared the properties of 7 nearby Type Ia supernovae explosion sites with those of their host galaxies.
\citet{ku13a} presented their investigation of 11 SNe Ib/Ic explosion sites on the IFS observations obtained using the SuperNova  Integral Field Spectrograph (SNIFS) mounted at the University of Hawaii 2.2 m telescope (UH88) and the Gemini Multi-Object Spectrograph (GMOS) with the 8.1 m Gemini North telescope at Mauna Kea. \cite{ku13b} performed a similar study for 13 SNe II-P and II-L explosion sites in nearby galaxies.
\citet{ga16a} made analysis of the local explosion environments of 11 SNe exploded in 6 nearby galaxies (z $\leq$ 0.016), which were observed by the Multi-Unit Spectroscopy Explorer (MUSE) of Very Large Telescope (VLT).

To obtain more statistic analysis on hosts of different types of SNe, several researches have been enlarged the sample of the SN host galaxies.
Based on the Calar Alto Legacy
Integral Field Area (CALIFA), the PPAK IFS Nearby Galaxies Survey (PINGS) and some other observations, \citet{ga14} analyzed IFU spectroscopy of 81 galaxies that hosted 95 SNe with different types.
\citet{ga16b} took further extended SN host galaxies and analyzed the metallicity of 115 SN host galaxies.  With the PMAS/PPak Integral-field Supernova hosts COmpilation (PISCO), \citet{ga18} presented the properties of a sample of SN hosts with IFS, which included 232 SN host galaxies that hosted 272 SNe.
\citet{ku18} explored and analyzed 83 nearby core-collapse SN explosion sites with IFS. \citet{ly18} performed spectroscopic environmental
measurements for a sample of 37 SNe Iax (a peculiar SN class and differ from normal SNe Ia) and their host galaxies using both
IFS and long-slit data from VLT/MUSE and Nordic Optical
Telescope/Andalucia Faint Object Spectrograph and Camera
(NOT/ALFOSC).

The Mapping Nearby Galaxies at APO (MaNGA, \citealt{bu15} )
must make important contribution on understanding the properties of SN host galaxies, especially at the SN explosion sites, both on the individual target with detailed analyses and on the statistical properties based on a larger sample of host galaxies with exploration of different types of SNe. \citet{ch17} and \citet{iz18} have studied one most nearby superluminous supernova (SLSN) 2017egm using MaNGA IFU data and provide detailed analysis of the nearby environment of this supernova. It is necessary to analyze a larger sample size in detail.

In this paper, we will study the detailed properties of 11 SN host galaxies (4 Type Ia, 5 Type II and 2 unclassified types) by using their IFS observations from MaNGA. We will compare the properties in the region of SNe explosion sites with those of the global regions of the host galaxies, about the properties of star formation rates (SFRs), gas-phase oxygen abundances, stellar masses and stellar population ages etc. Most of our sample of galaxies have redshift around 0.03 following the selection of MaNGA sample galaxies. This redshift value is higher than the SN host galaxies in other work which normally have median value of redshift around 0.01. We will specially discuss this in Sect.~\ref{comparision}.
We will extend our sample size from MaNGA database and try to make a statistical conclusion on the properties of SN host galaxies in the following work.

The paper is
organized as follows: In Sect.~\ref{sample selection}, we present the method used to
select our sample. The data reduction of IFU observation is shown in Sect.~\ref{Sect:analysis}. The properties of environment for different types of SNe are arranged in Sect.~\ref{results}. Finally, we
will discuss our results in Sect.~\ref{discussion} and obtain conclusions in Sect.~\ref{conclusion}. Throughout this paper, we adopt
a cosmological model with $H_0$ = 70 $km s^{-1} Mpc ^{-1}$, $\Omega_M$ = 0.3, $\Omega_{\Lambda} $= 0.7.

\section{Sample selection}
\label{sample selection}
To select the SN host galaxies within the field of view of MaNGA, we match Asiago supernova catalogue (ASC) with 1390 IFU galaxies from the first MaNGA public data release, which is part of SDSS Data Release (DR13, \citealt{al17}). The details of cross-correlation will be described as follows.

\subsection{Asiago Supernova Catalogue}
\label{asc}
Asiago Supernova Catalogue (ASC) was initially developed by \citet{ba84}, which displayed some basic information about 568 supernovae and host galaxies explored from the year 1885 to 1983. The global number of 661 supernovae and host galaxies discovered before 1988 December 31 were listed in  \citet{ba89}. Ten years later, \citet{ba99} presented data of a larger number of 1447 supernovae and host galaxies. Asiago supernova group has been updating continually and has presented 6530 supernovae and their host galaxies discovered before 2016 January 1.

Compared with Sternberg Astronomical Institute (SAI) Supernova Catalogue, which has the latest date of modification of 2014-10-17, ASC has updated more recently, so it presents a lager number of SNe and has more exact information about SNe and their host galaxies. Meanwhile, ASC presents two decimal places in right ascension and declination of supernovae coordinate.

\subsection{MaNGA survey}
\label{manga}
The SDSS-IV/MaNGA survey is designed to mapping 10,000 nearby galaxies and intend to help clarify the process of present galaxies from birth, growth to their death finally  \citep{we15,bu15,la15,dr15,ba17}.
To realize the intention of clarifying the process, MaNGA will make good use of the SDSS-III BOSS spectrograph \citep{sm13}
to get 2D spectrograph maps.
Moreover, the spectra of MaNGA could characterize the internal
composition and the dynamical state of a sample of 10,000 galaxies,
whose stellar masses are greater than $\rm 10^9$ $\rm M_\odot$.

The sample of MaNGA is intended to have a Primary and Secondary sample,
which should cover more than 1.5 $R_e$ and 2.5 $R_e$, respectively.
The median redshift of Primary and Secondary sample is 0.03 and 0.045, respectively (see  \citealt{bu15,la15,wa17,ya16b} in more detail). \cite{la15} found that the best integrated field unit shape is a regular hexagonal form, and it realized 3 $\mu$m rms fiber placement using MaNGA hardware.
We can get the 2D maps of a lot of parameters from MaNGA, including $H\alpha$ velocity, gas-phase oxygen abundance, star formation rate (SFR), stellar mass etc., to trace the formation and evolution processes of galaxies.

\subsection{cross-correlations of catalogs and final sample}
\label{sample}
To study the local properties of SNe and their host galaxies, we cross-correlate the R.A. and DEC. of 6530 SNe from ASC displayed up to 2015 December 31
with the R.A. and DEC. of 1390 galaxies from MaNGA in SDSS DR13.
The largest diameter of IFU size is 32 $arcsec$, so here we adopt 15 $arcsec$ as the matching radius. There are 14 sample galaxies selected.
There are 3 out of 14 SN host galaxies excluded: 2 of 14 SNe can not been explored in the field of view of MaNGA, and the signal to noise ratio of 1 of 14 SN host galaxies
is too low to be analyzed. Finally, we select 11 galaxies in this matching radius that have been observed. The reason why the sample ratio in this work is smaller than that in \cite{ga16b} in CALIFA sample will be discussed in Sect.~\ref{comparision}.

\begin{figure}
\centering
\includegraphics[angle=0,width=7.8cm]{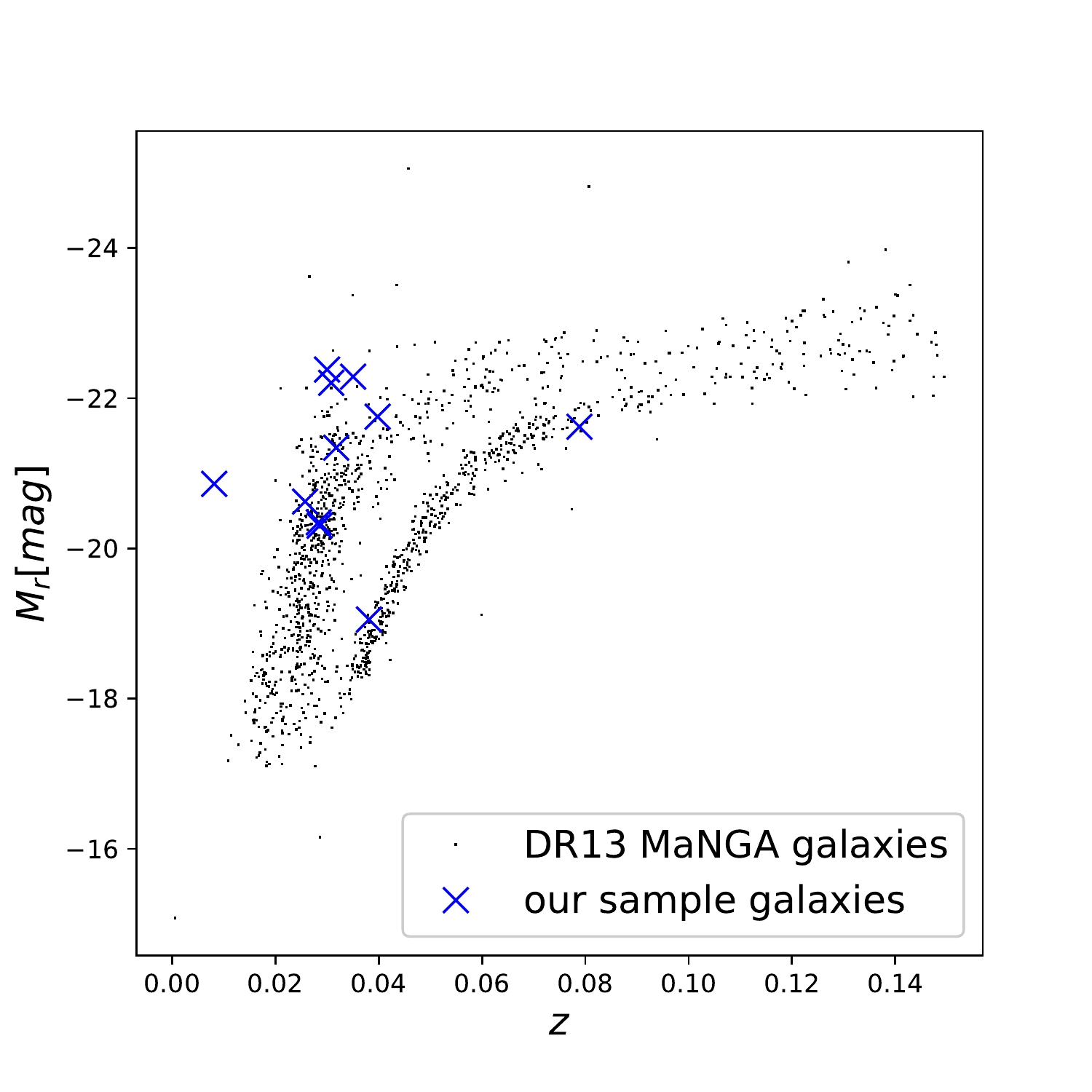}
\caption {The redshift and absolute magnitude in $r$ band distributions of all the 11 sample galaxies comparing with the DR13 MaNGA sample galaxies.}
\label{fig.z-Mr}
\end{figure}

Fig.~\ref{fig.z-Mr} shows the range of redshift and absolute magnitude in $r$ band ($M_r$, which are from MaNGA) of our sample galaxies and DR13 MaNGA galaxies marked by grey dots. There is a gap for DR13 MaNGA galaxies, which results from the sample selection of MaNGA (Primary and Secondary Sample, e.g. \citealt{wa15,be16}). From this figure, the $M_r$ of 11 galaxies range from -23 to -19 $mag$, and the redshifts are mostly around 0.03 but with one having 0.008 and another having 0.0789. Details of our 11 sample galaxies are presented in Table~\ref{table.allsamples}. From Table~\ref{table.allsamples}, we can see that there are 4 SNe Ia, 5 SNe II and 2 unclassified type of SNe in our sample. The host galaxies of SN 2004eb (Type II, NGC 6387) and 1999gw (unclassified type, UGC 04881NED02) are in merger system. Especially for SN 1999gw, the interaction region is inside the field of view of MaNGA.

\begin{table*}
\centering
\caption{Basic information from MaNGA and ASC of our 11 sample galaxies. RD: rotation disk; PR:perturbed rotation; CK: complex kinematics.}
\label{table.allsamples}
\begin{threeparttable}
\resizebox{\textwidth}{!}{
\begin{tabular}[b]{|l|l|l|r|r|l|r|r|r|r|r|r|r|r|r|}
\hline
  \multicolumn{1}{|c|}{Plateifu} &
  \multicolumn{1}{c|}{SN name} &
  \multicolumn{1}{c|}{SN type} &
  \multicolumn{1}{c|}{SN R.A.} &
  \multicolumn{1}{c|}{SN Dec.} &
  \multicolumn{1}{c|}{Host} &
  \multicolumn{1}{c|}{Type} &
  \multicolumn{1}{c|}{Redshift} &
  \multicolumn{1}{c|}{$b/a$} &
  \multicolumn{1}{c|}{PA} &
  \multicolumn{1}{c|}{Kinematics\tnote{e}}\\
\hline
  8261-12705 & 2007sw & Ia & 183.4037 & 46.4934 & UGC 7228 & Sbc\tnote{b} & 0.0257 & 0.41 &179.9 &  RD\\
  7975-6104 & 2006iq & Ia & 324.8906 & 10.4849 & PGC 1380172 & Sb\tnote{d} & 0.0789  & 0.87 &79.2 & RD\\
  8138-12704 & 2007R & Ia & 116.6564 & 44.7895 & PGC 21767 & S0/a\tnote{b} &  0.0308  & 0.69& 14.9 & RD\\
  8332-1902 & 2005cc & Ia pec & 209.2702 & 41.8449 & NGC 5383 & SBb\tnote{b} & 0.00814 &0.70 & 62.8 & RD\\
  8604-12701 & 2000cs & II pec & 245.8843 & 39.1248 & MCG +07-34-015 & Sb\tnote{d} & 0.0350  &  0.98&21.0 & RD \\
  7495-12702 & 2010ee & II & 205.0750 & 26.3533 & UGC 8652 & Sb\tnote{d} & 0.0284 & 0.29& 165.5 & RD\\
  8453-12702 & 2012al & IIn & 151.5485 & 47.2946 & PGC 213664 &Sb\tnote{d} & 0.0381 &0.55 & 12.7 & RD\\
  8588-6101 & 2011cc & IIn & 248.4560 & 39.2635 & IC 4612 & Sc\tnote{c}& 0.0318 &0.96 & 18.9 & RD\\
  7990-3703 & 2004eb & II & 262.1013 & 57.5460 & NGC 6387 & Sb\tnote{d} & 0.0286  &0.62&  92.1 & PR\\
  8250-12704 & 1999gw & U\tnote{a} & 138.9779 & 44.3319 & UGC 4881 & Interaction \tnote{c}&  0.0398  & 0.63& 108.9 &  CK\\
  8550-12705 & 1975K & U\tnote{a} & 249.1371 & 39.0301 & NGC 6195 & Sb\tnote{b}& 0.0300 &0.67 & 141.4 & RD \\
\hline\end{tabular}}
\begin{tablenotes}
 \footnotesize
 \item[a] unclassified. \item[b] Based on Asiago Supernova Catalogue. \item[c] Based on SDSS pseudo-images. \item[d]Based on \citet{2013OAP....26..187D}. \item[e]Based on \citet{ya08} and  \citet{2017sdg..book.....H}.

\end{tablenotes}
\end{threeparttable}
\end{table*}

\section{Data analysis}
\label{Sect:analysis}
\subsection{IFS analysis}
\label{ifs analysis}
We go through three steps from MaNGA data cube to galaxy emission line measurements and then use the emission line measurements to obtain the
parameters of galaxies, such as gas abundance, SFR etc.
\subsubsection{Voronoi 2d binning}
\label{binning}
Considering that the spectral S/N's are usually less than 1 at the edge of the MaNGA field of view, we therefore
spatially rebin those spaxels with low S/N together to improve the spectral S/N, which allow us to obtain
reliable spectral fitting results. For each spectrum, its S/N is defined as the median S/N of the whole
spectrum, which is dominated by the continuum, instead of those regions at the blue or red end, or those emission lines.

Here we apply the Voronoi 2D binning method \citep{cc03} to perform the spatial rebinning, and
set the rebinned spectra to have S/N=20, which make sure relative high spatial resolution for our science
and high enough S/N for reliable spectral analysis. Before the binning, those spectra with S/N$<3$ are removed
due to their little contribution to the spectral S/N improvement.

\subsubsection{Stellar population synthesis}
\label{sps}
In this work, we apply the {\tt STARLIGHT} code \citep{ci05} and the simple stellar population (SSP) library from BC03 model
with Chabrier initial mass function \citep{ch03} to perform the spectral fitting and stellar population analysis.
For the SSP library, the selected SSP templates include 24 different ages and
6 different metallicities as follows:
\begin{itemize}
\item The stellar ages: [ 0.001, 0.003, 0.005, 0.007, 0.009, 0.01,
0.014, 0.025, 0.04, 0.055, 0.1, 0.16, 0.29, 0.51, 0.9, 1.28, 1.43,
2.5, 4.25, 6.25, 7.5, 10.0, 13.0, 15.0] Gyr;
\item The metallicities ($Z/Z_{\odot}$): [0.005, 0.02, 0.2, 0.4, 1.0, 2.5].
\end{itemize}
The current selection on age and metallicity grids allows us to fit both the late and early
types of galaxies with different metal abundances.

We resample all the spectra with $\Delta \lambda = 1$\AA~ linearly before STARLIGHT fitting.
In the spectral fitting, we assume the Calzetti law \citep{ca00} for dust extinction
correction, and set the parameter $A_V$ as free. Taking into account the uncertainties
of flux calibration, we allow a negative $A_V$ in each spectral fitting.
Following the suggestions from \citet{ci05}, we de-redshift each
spectra to rest-frame wavelength. During the spectral fitting, we only take the spectra with the wavelength
ranging from 3700\AA~ to 7000\AA~, which includes those emission lines from [O~{\uppercase\expandafter{\romannumeral2}}]3727 to [S~{\uppercase\expandafter{\romannumeral2}}]6731 lines.

After the {\tt STARLIGHT} fitting, 99\% of the spectra in our sample have $A_V<1$. In this
case, the derived stellar population parameters are reliable and have no significant
biases \citep{ge2018,cf18}. Therefore, for the current sample, we can use the {\tt STARLIGHT} code
for both stellar population analysis and emission-line measurements.

\subsubsection{Emission-line measurements}
\label{emission line}
As was described in the SDSS spectral fitting paper \citep{ge12},
after the {\tt STARLIGHT} fitting, we resample the model spectra with 1\AA~ to
observed spectral resolution. We get the resampled model spectra with continuum and absorption lines subtracted
and finally obtain the pure-emission-line spectra.
Errors of the data points in the pure-emission-line
spectra are taken as the same as those in the MaNGA data cube.
In the case of weak emission lines, whether the continuum
of pure-emission-line spectra are zero will affect much to the emission-line
fitting. We use a linear fit to adjust the shape of continuum, then
treat each emission line as a Gaussian with three parameters:
line width, flux and offset (with respect to the rest-frame central
wavelength). To measure these weak emission lines more accurately,
we fixed the offset and width of all emission lines but with
flux free. And we also imposed two flux ratio constraints:
[O~{\uppercase\expandafter{\romannumeral3}}] $\lambda$5007/[O~{\uppercase\expandafter{\romannumeral3}}] $\lambda$4959=2.97 and
[N~{\uppercase\expandafter{\romannumeral2}}] $\lambda$6583/[N~{\uppercase\expandafter{\romannumeral3}}] $\lambda$6548=3.

\subsection{Emission line fluxes and extinction}
\label{flux}
With the above emission line fitting, we can derive emission-line fluxes of H$\alpha$, H$\beta$, [O~{\uppercase\expandafter{\romannumeral3}}] $\lambda$5007, [N~{\uppercase\expandafter{\romannumeral2}}] $\lambda$6583, [O~{\uppercase\expandafter{\romannumeral2}}] $\lambda$3727, [S~{\uppercase\expandafter{\romannumeral2}}] $\lambda\lambda$6716/6731.
To perform a robust analysis on the emission line related results (e.g. gas metallicity), we then exclude those spaxels that have low S/N ($<3$).
For instance, to estimate the gas-phase oxygen abundance using O3N2 method \citep{pp04}, only those spaxels that
have their emission lines H$\alpha$, H$\beta$, [O~{\sc iii}] $\lambda$5007 and
[N~{\sc ii}] $\lambda$6583
measured with S/N $>3$ simultaneously, are used to calculate the metallicity.

The emission lines fluxes are corrected for dust extinction in the direction of line of sight through galaxies, which could be estimated by H$\alpha$ and H$\beta$ emission lines \citep{of06}. In the case B recombination, the ratio of I(H$\alpha$) and I(H$\beta$) is 2.86, the electron density and electron temperature is $10^2cm^{-3}$ and 10,000 K, respectively \citep{of06}. Using the equation of Rv=Av/E(B-V) provided by \citet{fi99} and the value of Rv=3.1, we compute Av.
 We can estimate the gas velocity field from the strong H$\alpha$ emission lines using the method of \citet{kr06}.
Then the position angle (PA) and redshift of the host galaxies could be derived.

\subsection{Star formation rate}
\label{sfr}
We estimate the ongoing SFR from H$\alpha$ luminosity according to the relation given by  \citet{ke98}:
\begin{equation}
 SFR[M_{\odot}yr^{-1}] = 7.9 \times 10^{-42}L(H\alpha),
\end{equation}
where $L(H\alpha$)=4{$\pi$}{$d_L^2$}$F(H\alpha$), which is H$\alpha$ luminosity in units of $erg$  $s^{-1}$, $d_L$ is the luminosity distance to the galaxy in units of $cm$, $F(H\alpha)$ is the extinction-corrected flux of H$\alpha$ in $erg$ $s^{-1} cm^{-2}$. Given 2D maps of properties of galaxies, there are two methods to measure global ongoing SFR: summing the SFR of all the single spaxels in the field of view of MaNGA and estimating from the H$\alpha$ emission line flux of the global spectra in the field of view of MaNGA. The results of these two methods agree well for the spaxels with high S/N. Here we use the first one to estimate the global SFR of supernovae host galaxies.
Also, we estimate the
specific star formation rate $\rm sSFR$ = SFR/$M$, where $M$ is the relevant mass estimated from {\tt STARLIGHT} code. We generate 2D maps of $\rm sSFR$ to show the distributions.

\subsection{gas-phase oxygen abundance}
\label{oh}
There are several methods to estimate the oxygen abundance. The most straightforward approach is electron temperature method (Te method). It is often estimated from [O~{\uppercase\expandafter{\romannumeral3}}] $\lambda$5007, [O~{\uppercase\expandafter{\romannumeral3}}] $\lambda$4959/[O~{\uppercase\expandafter{\romannumeral3}}] $\lambda$4363(e.g. \citealt{st06,iz06,li06,li07,yi07}). But we should notice that the most important and temperature sensitive emission line [O~{\uppercase\expandafter{\romannumeral3}}] $\lambda$4363 is difficult to measure, especially in metal-rich environment \citep{iz06}. All of our sample galaxies don't have strong [O~{\uppercase\expandafter{\romannumeral3}] $\lambda$4363 line. Therefore, we have to use other strong line methods to estimate oxygen abundance, like O3N2, N2O2 and $\rm R_{23}$ methods.

$\rm R_{23}$ is one of the most commonly used methods (\citealt{pt05,pa79,mc91,za94,tr04,kd02,kk04,pi01}):
\begin{eqnarray}
\begin{aligned}
\rm{R_{23}}=&(I_{[O~{\uppercase\expandafter{\romannumeral2}}] \lambda 3727} +I_{[O~{\uppercase\expandafter{\romannumeral3}]} \lambda 4959}\\
&+I_{[O~{\uppercase\expandafter{\romannumeral3}] \lambda 5007}})/H\beta.
\end{aligned}
\end{eqnarray}
It has a special feature, sometimes also a severe defect that metallicity increases with increasing $\rm R_{23}$ in the metal-poor branch, but decreases in the metal-rich branch. So, the key problem of $\rm R_{23}$ is to choose the real relation to avoid double peaks.
Here we apply the relation in \citet{tr04} to estimate the gas-phase oxygen abundance:
\begin{eqnarray}
\begin{aligned}
&\rm{12+\log(O/H)} = 9.185 - 0.313 \times \log(R_{23}) - \\
&0.24 \times (\log(R_{23}))^2
- 0.321 \times (\log(R_{23}))^3. \\
\end{aligned}
\end{eqnarray}
This relation is only suitable for galaxies with metal-rich environment ($\rm 12+log(O/H) > 8.5$).

In this work, we also use O3N2 and N2O2 methods. \citet{al79} gave O3N2:
\begin{equation}
\rm{O3N2=\log(([O~{\uppercase\expandafter{\romannumeral3}}] \lambda 5007/H\beta)/([N~{\uppercase\expandafter{\romannumeral2}}] \lambda 6583/H\alpha))}.
\end{equation}
\citet{pp04} showed that there is an apparent and significant linear relation between O3N2 and log(O/H) when the value of O3N2 locates in the region between -1 and 1.9:
\begin{equation}
\rm{12+\log(O/H) = 8.73 - 0.32 \times O3N2}.
\end{equation}
\cite{li06} used a sample (
$\rm 8.4 \leq 12+\log(O/H) \leq  9.3$, $ \rm -1.2 \leq  \log([N~{\uppercase\expandafter{\romannumeral2}]}/[O~{\uppercase\expandafter{\romannumeral2}]}) \leq 0.7$) from SDSS to acquire the linear relation:
\begin{equation}
\rm{N2O2=\log(([N~{\uppercase\expandafter{\romannumeral2}}] \lambda 6583)/([O~{\uppercase\expandafter{\romannumeral2}}] \lambda 3727))}.
\end{equation}
\begin{equation}
\rm{12+\log(O/H) = 9.125 + 0.49 \times N2O2}.
\end{equation}

Compared with other methods, \citet{2017MNRAS.466.3217Z} indicated that metallicities estimated using N2O2 method have smallest bias and error for Diffuse Ionized Gas (DIG) and H II regions. According to \cite{2017MNRAS.466.3217Z}, the N2O2 method can only be affected by N/O abundance ratio and temperature, but it is not subject to ionization parameter and ionizing spectrum shape variation  (\citealt{do00,do13}). The key problem for this method is dust extinction should be corrected accurately.

\section{Results}
\label{results}
From emission line fluxes,
we obtain dust extinction, SFR, oxygen abundance, stellar mass and stellar population age. Fig.~\ref{fig.BPT} presents the Baldwin, Phillips \& Terlevich (BPT, \citealt{ba81}) diagram for our sample galaxies. The values of emission line fluxes ratio and parameters are presented from Table.~\ref{table.bptlines} to Table.~\ref{table.age}. The 2D maps of the properties for the host galaxies of 4 Type Ia, 5 Type II and 2 unclassified type of SNe are shown in Fig.~\ref{fig.SNIa-havd}, Fig.~\ref{fig.SNII-havd} and  Fig.~\ref{fig.unSN-havd}, respectively. The global values of parameters are estimated using all the useful spaxels of the host galaxies in the field of view of MaNGA. We take a circular region with 4 $arcsec$ diameter around the host galaxies centers and the SNe positions to estimate the values of parameters at the central regions and SN locations, respectively. Here we should note that for SN 1999gw, which explodes in the merger system, the global values of the parameters are estimated using all useful spaxels of the system in the field of view of MaNGA.

\subsection{BPT diagnostic diagram}
\label{bpt}
\begin{figure}
\centering
\includegraphics[angle=0,width=7.8cm]{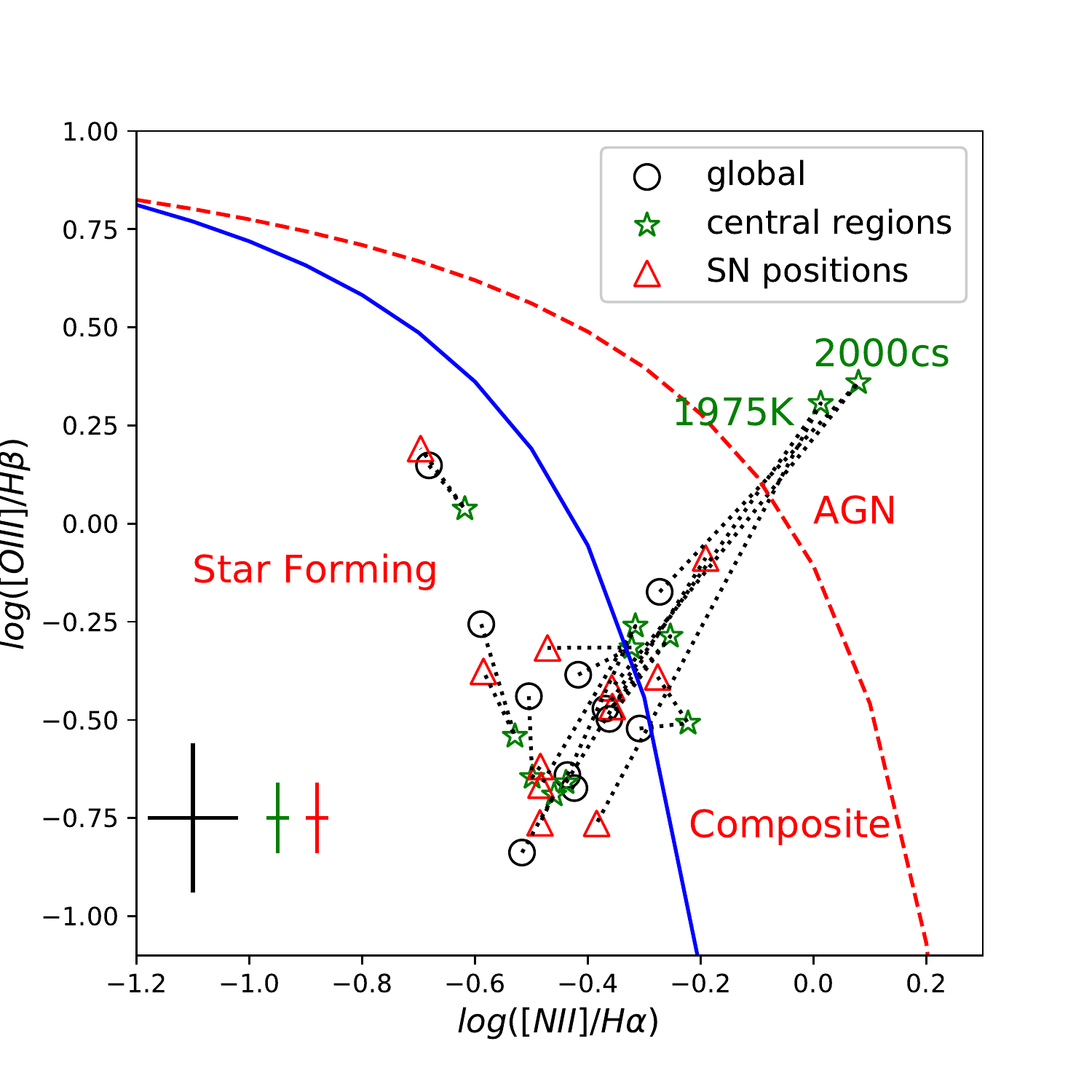}
\caption {The BPT diagram of the sample galaxies. We plot the emission-line flux ratio of
  log([O~{\sc iii}]/H$\beta$) versus the ratio log([N~{\sc ii}]/H$\alpha$) for all the galaxies in our sample.
The errorbars are the average standard deviation of the log([O~{\sc iii}]/H$\beta$) and log([N~{\sc ii}]/H$\alpha$) for the global SN host galaxies (black circles), central regions of SN host galaxies (green stars) and the explosion sites (red triangles), respectively.}
\label{fig.BPT}
\end{figure}

\begin{table*}
\caption{The values of $\rm {\log([N~{\uppercase\expandafter{\romannumeral2}}]/H\alpha)}$ and $\rm  log([O~{\uppercase\expandafter{\romannumeral3}}]/H\beta)$ estimated from global galaxy spectra, central and SN positions.}
\label{table.bptlines}
\centering
\begin{threeparttable}
\resizebox{\textwidth}{!}{
\begin{tabular}[b]{|l|l|l|r|r|r|r|r|r|}
\hline
  \multicolumn{1}{|c|}{SN} &
  \multicolumn{1}{c|}{SN type} &
  \multicolumn{1}{c|}{Host} &
  \multicolumn{3}{c|}{$\rm \log([N~{\uppercase\expandafter{\romannumeral2}}]/H\alpha)$}&
  \multicolumn{3}{c|}{$\rm \log([O~{\uppercase\expandafter{\romannumeral3}}]/H\beta)$} \\
  \multicolumn{1}{|c|}{} &
  \multicolumn{1}{|c|}{} &
  \multicolumn{1}{|c|}{} &
  \multicolumn{1}{|c|}{global} &
  \multicolumn{1}{|c|}{center} &
  \multicolumn{1}{c|}{local} &
  \multicolumn{1}{c|}{global} &
  \multicolumn{1}{c|}{center} &
  \multicolumn{1}{c|}{local}  \\
  \hline
  2007sw &  Ia &  UGC 7228 &  -0.50$\pm$0.07 & -0.50$\pm$0.02 & -0.48$\pm$0.02 & -0.44$\pm$0.18 & -0.65$\pm$0.09 &  -0.62$\pm$0.10\\
 2006iq &  Ia & PGC 1380172 &  -0.42$\pm$0.07 & -0.44$\pm$0.03 & -0.19$\pm$0.01 & -0.67$\pm$0.18 & -0.66$\pm$0.05 &  -0.09$\pm$0.11\\
 2007R &  Ia & PGC 21767 &  -0.44$\pm$0.08 & -0.32$\pm$0.05 & -0.48$\pm$0.02 & -0.64$\pm$0.19 & -0.26$\pm$0.11 & -0.67$\pm$0.04\\
 2005cc &   Ia pec &NGC 5383 & -0.52$\pm$0.05 & -0.46$\pm$0.02 & -0.48$\pm$0.01 & -0.84$\pm$0.11 &  -0.69$\pm$0.05 &  -0.76$\pm$0.06\\
 2000cs &  II pec & MCG +07-34-015 & -0.27$\pm$0.14 & 0.08$\pm$0.03& -0.36$\pm$0.02 & -0.17$\pm$0.26  &  0.36$\pm$0.07 &  -0.42$\pm$0.07\\
 2010ee &   II &UGC 8652 &  -0.42$\pm$0.08 &-0.32$\pm$0.02 & -0.47$\pm$0.03 & -0.38$\pm$0.26 & -0.32$\pm$0.04 &  -0.32$\pm$0.14\\
 2012al &   IIn &PGC 213664 &  -0.59$\pm$0.07 & -0.53$\pm$0.02 &-0.59$\pm$0.05 & -0.26$\pm$0.17 &  -0.54$\pm$0.10 &  -0.38$\pm$0.15 \\
 2011cc &   IIn &IC 4612 & -0.36$\pm$0.07 &  -0.25$\pm$0.05 & -0.36$\pm$0.05 & -0.50$\pm$0.21 &-0.29$\pm$0.11 &  -0.47$\pm$0.12\\
 2004eb & II &  NGC 6387 &  -0.68$\pm$0.08 & -0.20$\pm$0.01 & -0.70$\pm$0.04 & 0.15$\pm$0.09 &  0.04$\pm$0.01 &  0.19$\pm$0.09\\
   1999gw &U& UGC 4881 &  -0.31$\pm$0.14 & -0.22$\pm$0.05 & -0.28$\pm$0.05 & -0.52$\pm$0.23 & -0.51$\pm$0.12 & -0.39$\pm$0.09\\
   1975K &U& NGC 6195 &  -0.37$\pm$0.10 & 0.01$\pm$0.02 & -0.38$\pm$0.02 & -0.47$\pm$0.23 &  0.31$\pm$0.13 & -0.76$\pm$0.07\\
  \hline\end{tabular}
}
\end{threeparttable}
\end{table*}

When estimating the gas-phase oxygen abundance, AGNs should be excluded from calculating and analysis since the existence of AGNs somewhat bias the measurements.
To distinguish AGNs contamination from star-forming regions in the center of galaxies, we adopt the BPT diagram,
which can separate emission line galaxies from AGNs according to different excitation mechanism  (\citealt{ba81}). Fig.~\ref{fig.BPT} shows the BPT diagram of our sample. The horizontal axis is the value of log([N~$\rm {\uppercase\expandafter{\romannumeral2}}$] $\lambda$6583/H$\rm \alpha$) and the vertical axis represents the value of log([O~$\rm {\uppercase\expandafter{\romannumeral3}}$] $\lambda$5007/H$\beta$). The dashed line shows the boundary between AGNs and composite galaxies and it is taken from \citet{ke01}. The blue solid line separates composite galaxies from star-forming galaxies and it is taken from \citet{ka03a}.
The air-corn circles, green stars and red triangles  represent the positions of the global galaxy spectrum, the galaxy nuclei, and the supernovae locations, respectively. For every single sample galaxy, these three points are connected using dotted lines. The average standard deviation of log([N~$\rm {\uppercase\expandafter{\romannumeral2}}$] $\lambda$6583/H$\alpha$) and log([O~$\rm {\uppercase\expandafter{\romannumeral3}}$] $\lambda$5007/H$\beta$) for the global SN host galaxies, central regions of SN host galaxies and SNe explosion sites are about 0.08 and 0.19, 0.02 and 0.09, 0.02 and 0.09 dex, respectively. The errorbars are marked in the left corner of the BPT diagram.

The values of emission line fluxes ratio of the global, central regions and SNe explosion regions are presented in Table.~\ref{table.bptlines}. According to Fig.~\ref{fig.BPT} and Table.~\ref{table.bptlines}, the centers of the host galaxy of unclassified type of SN 1975K (NGC 6195) and the host galaxy of SN II 2000cs (MCG +07-34-015) are in the AGN regions. However, their global spectrum locate in the star-forming galaxies and composites regions.
We can infer that for these two galaxies, the effect of AGNs is too small to change the global galaxy spectra (\citealt{st12}).
The difference of the emission line ratio in Fig.~\ref{fig.BPT} between supernova location and global spectrum is a little larger for SNe 2006iq, 2007sw and 1975K.
The global spectrum of the host galaxy of SN 2006iq (PGC 1380172) locates in the star forming region, while the supernova 2006iq locates in the composite region.
For the other sample galaxies, the difference of the emission line ratio between local and global is small.

\subsection{H$\alpha$ velocity field}
\label{velocity}

\begin{figure*}[]
\centering
\begin{minipage}{1.0\textwidth}
\hspace{1.5cm}
\includegraphics[width=20mm]{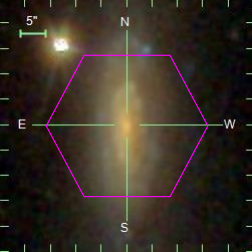}
\hspace{0.2cm}
\includegraphics[width=30mm]{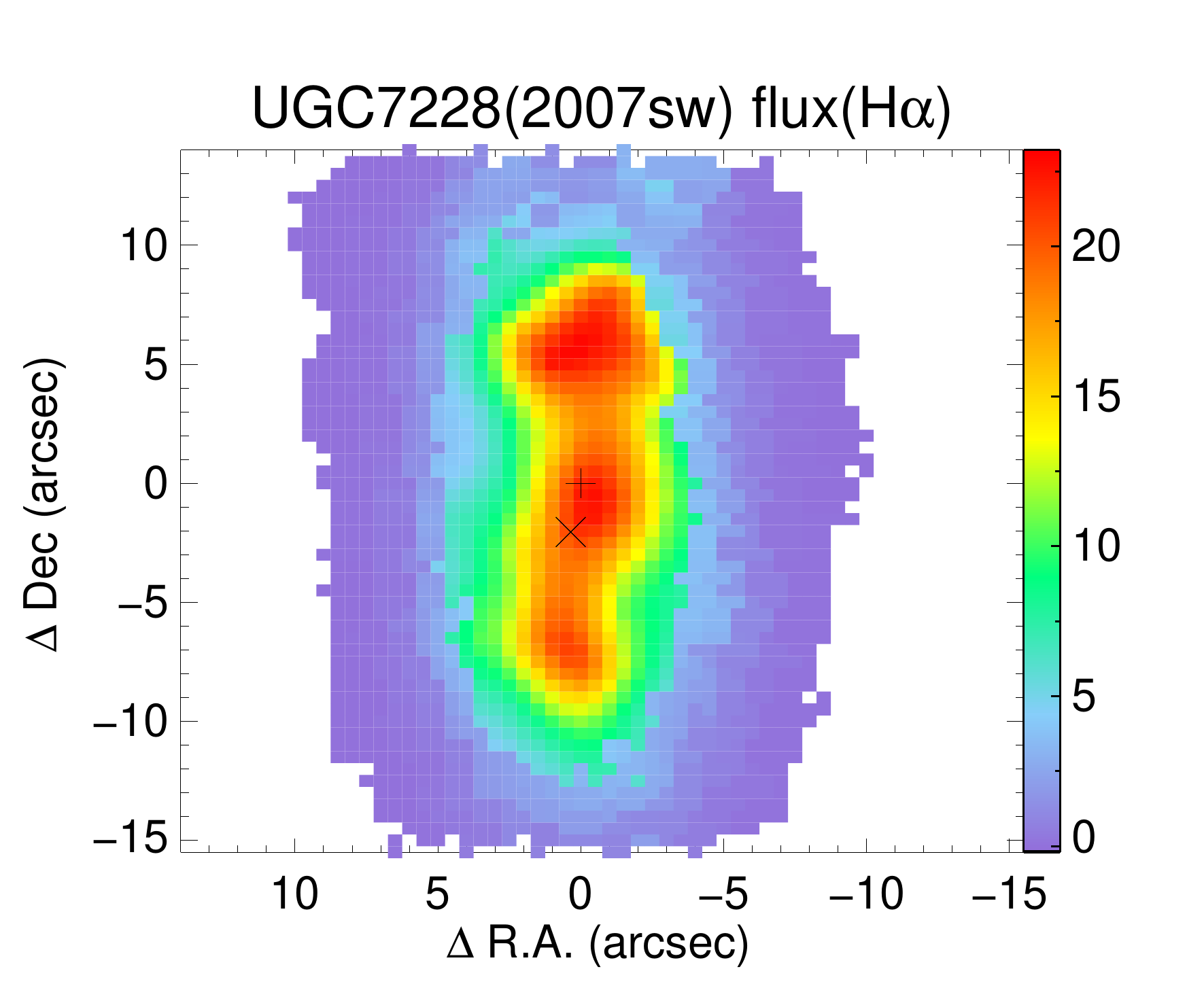}
\hspace{0.2cm}
\includegraphics[width=30mm]{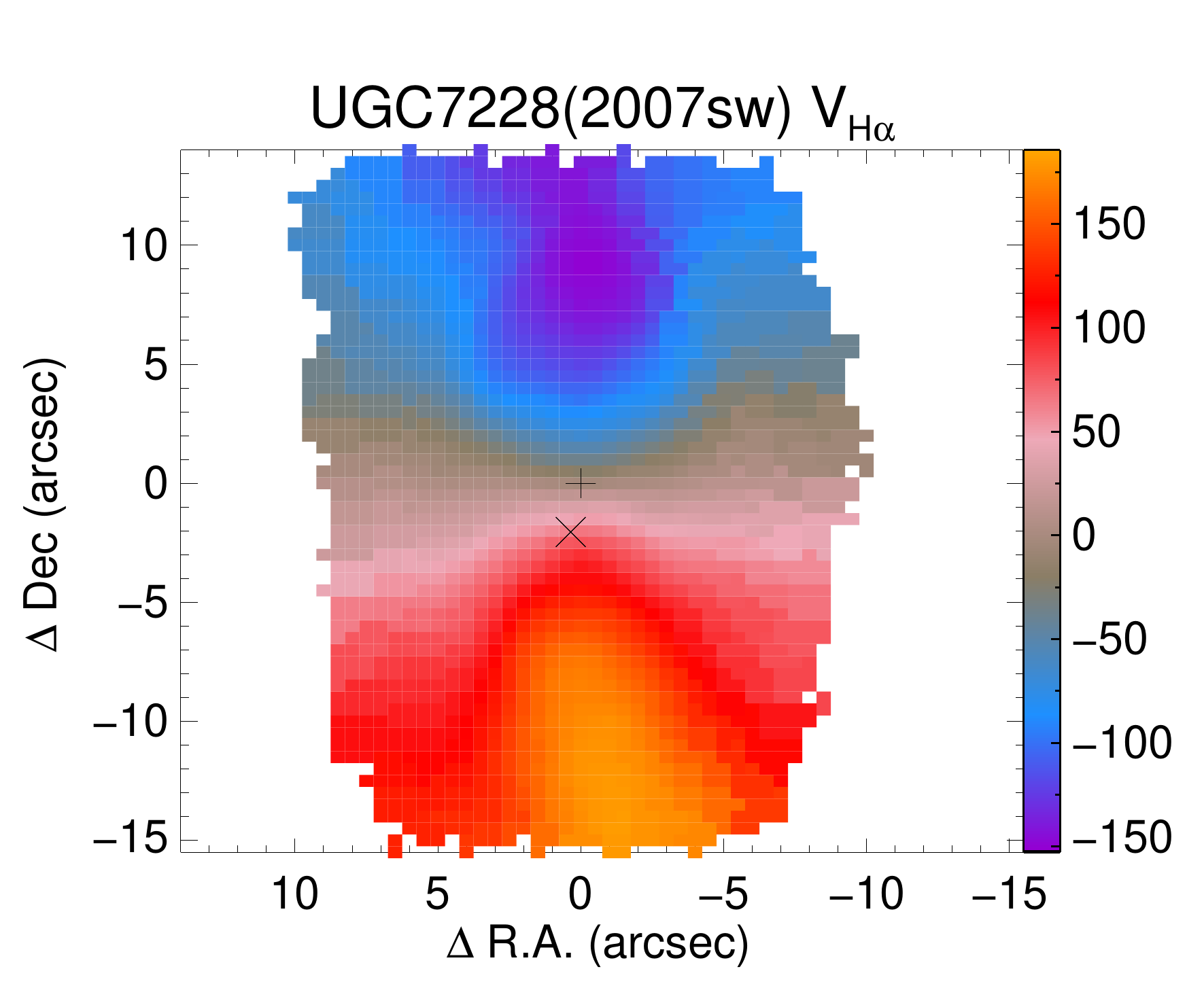}
\hspace{0.2cm}
\includegraphics[width=30mm]{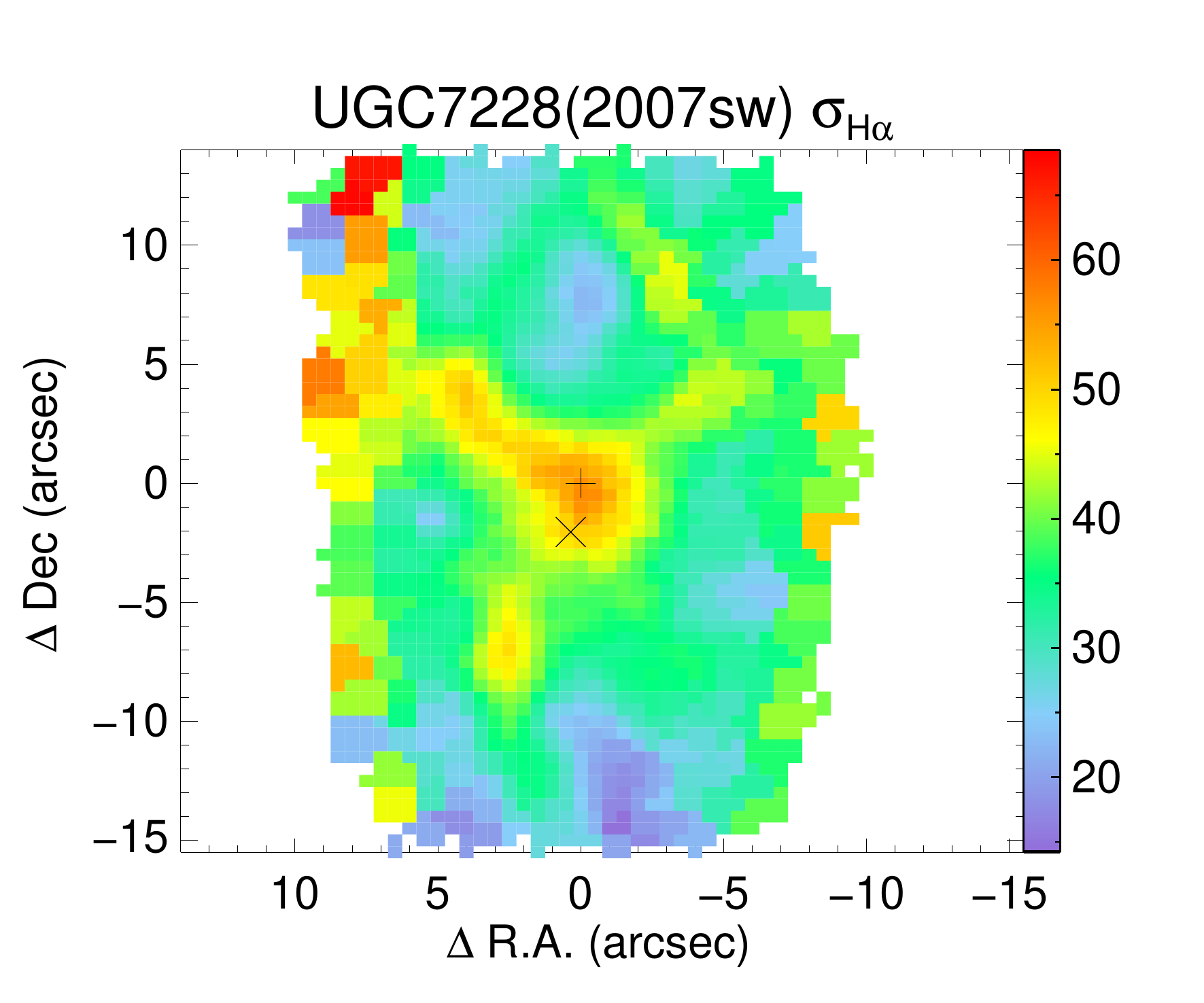}
\end{minipage}
\begin{minipage}{1.0\textwidth}
\hspace{0.9cm}
\includegraphics[width=30mm]{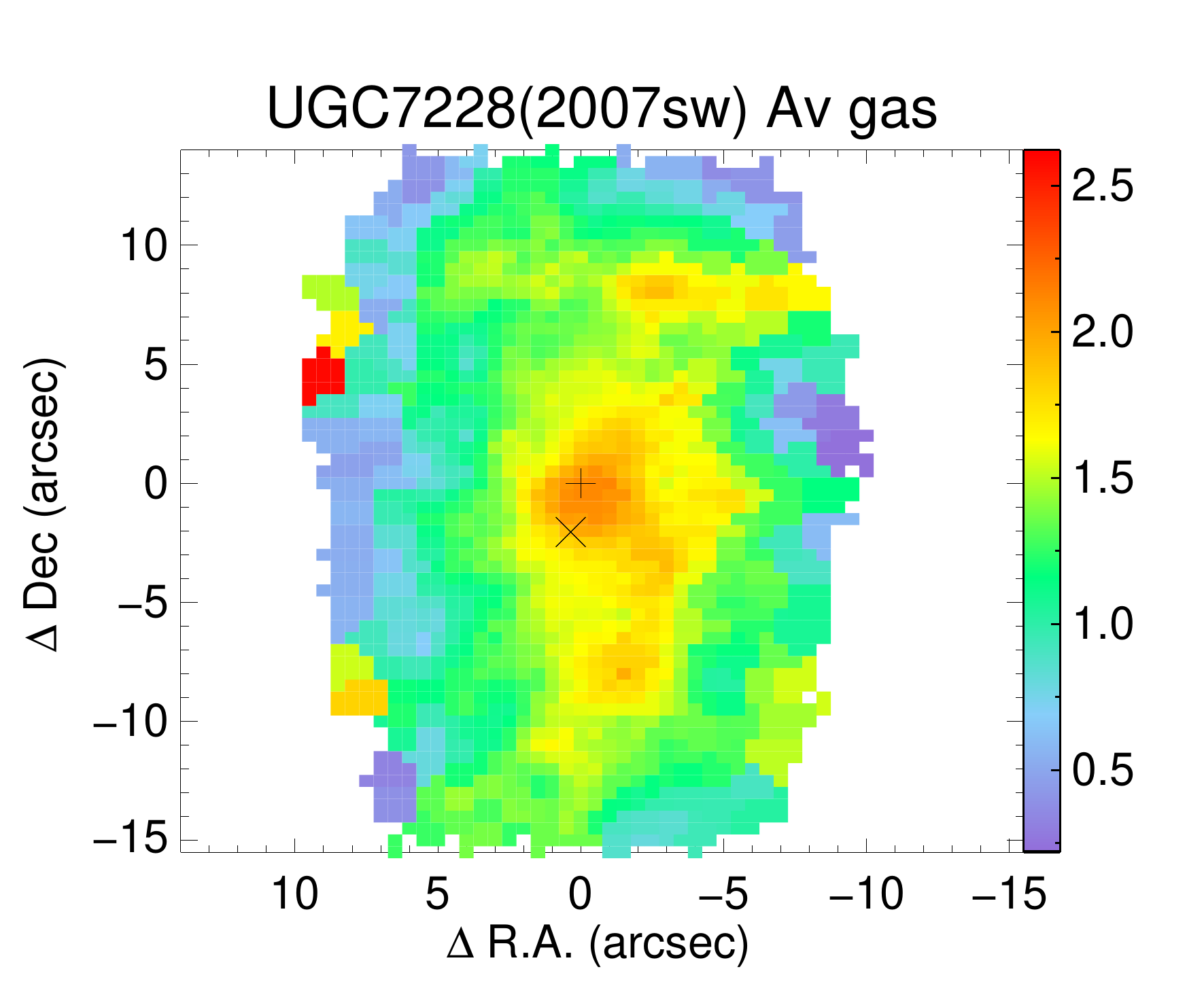}
\hspace{-0.2cm}
\includegraphics[width=30mm]{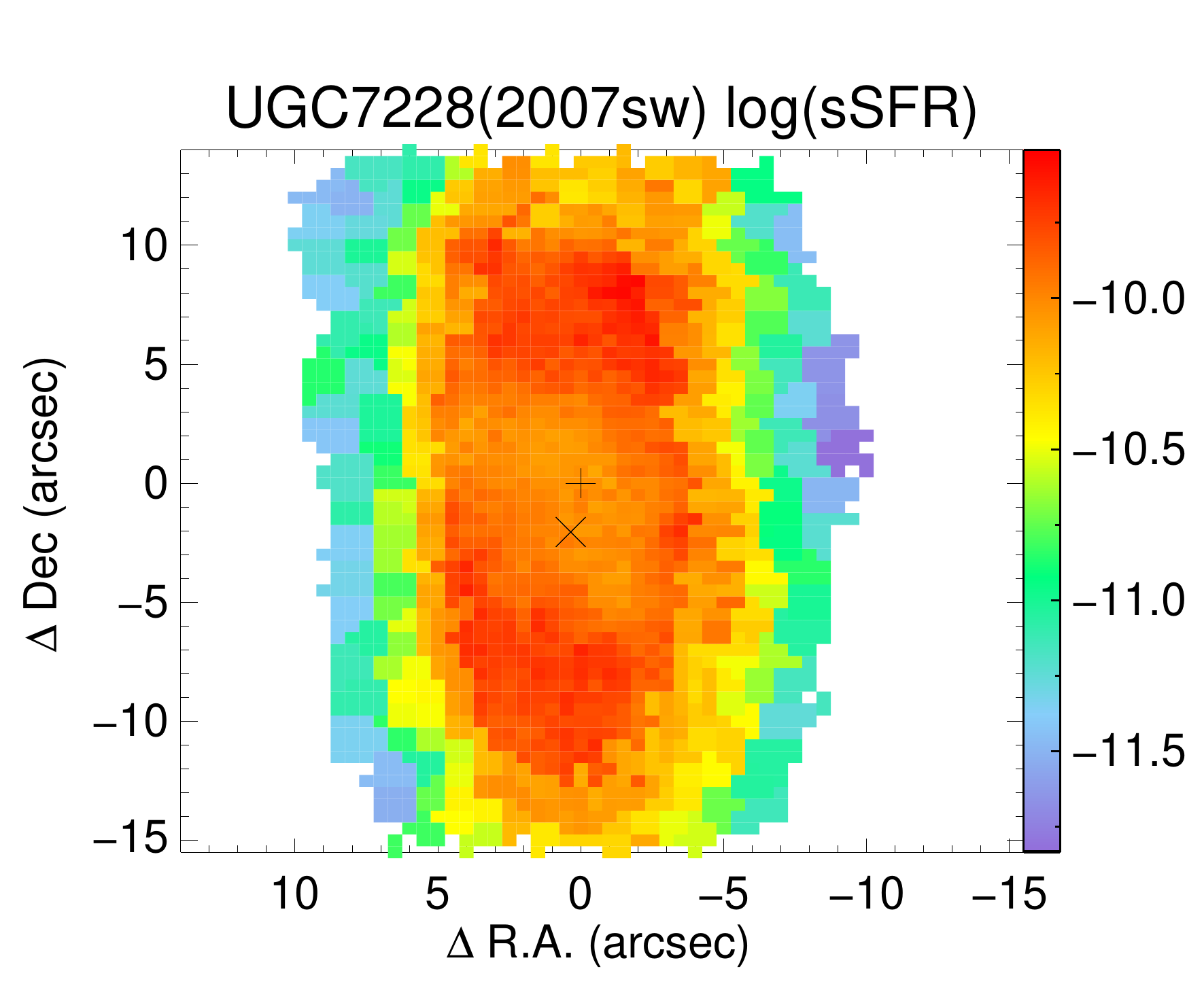}
\hspace{0.2cm}
\includegraphics[width=30mm]{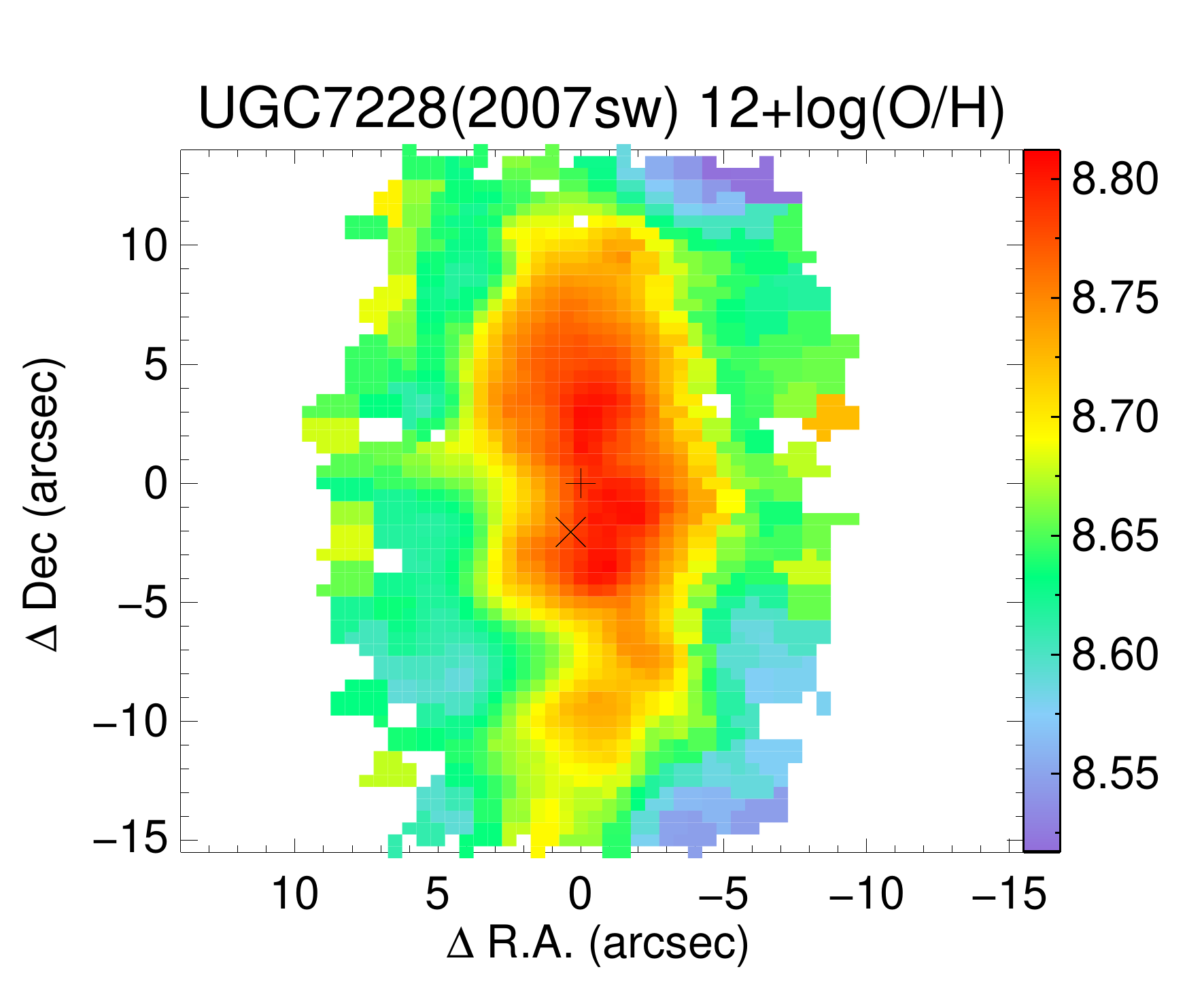}
\hspace{0.2cm}
\includegraphics[width=30mm]{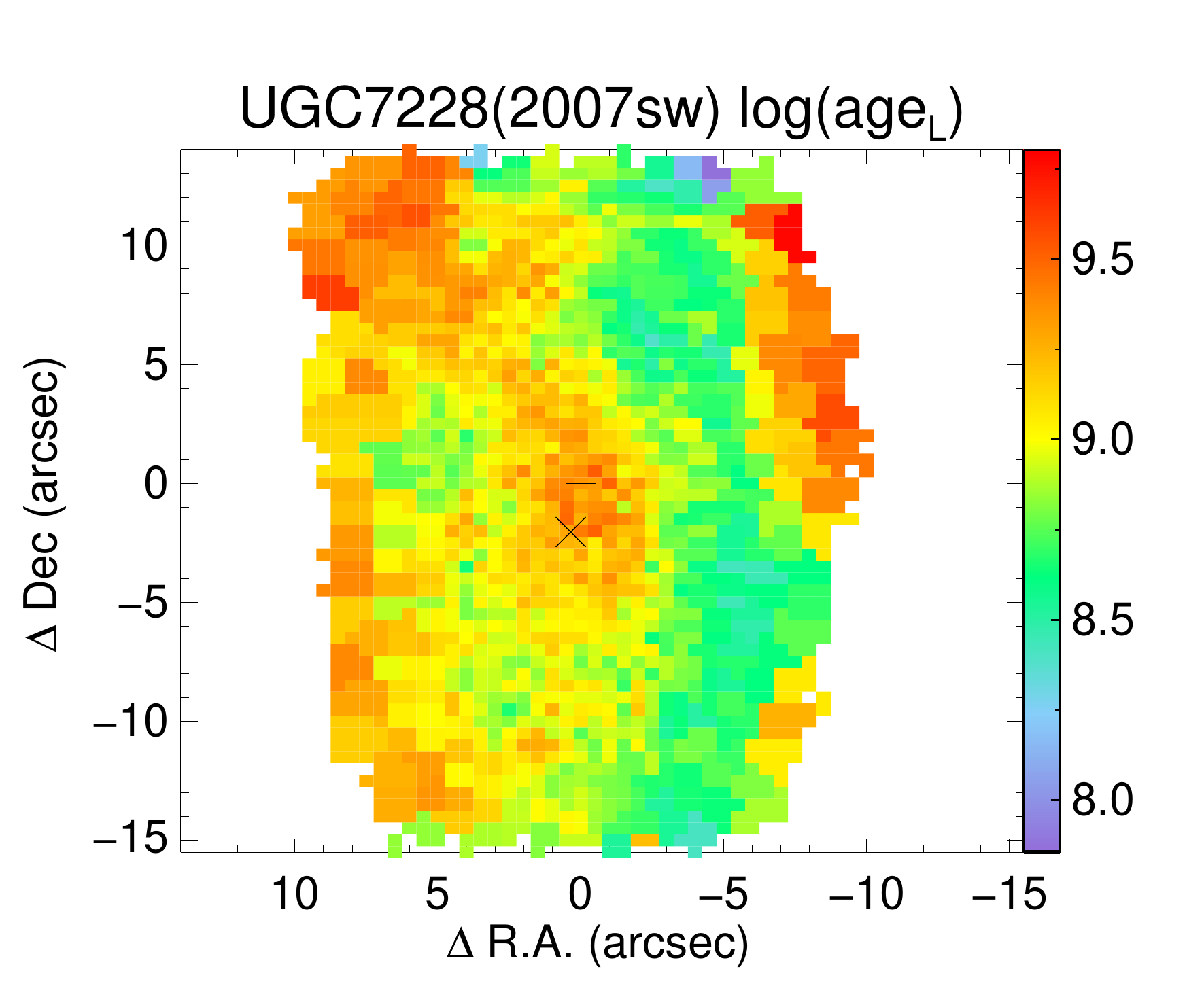}
\end{minipage}

\centering
\begin{minipage}{1.0\textwidth}
\hspace{1.5cm}
\includegraphics[width=20mm]{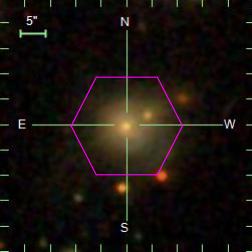}
\hspace{0.2cm}
\includegraphics[width=30mm]{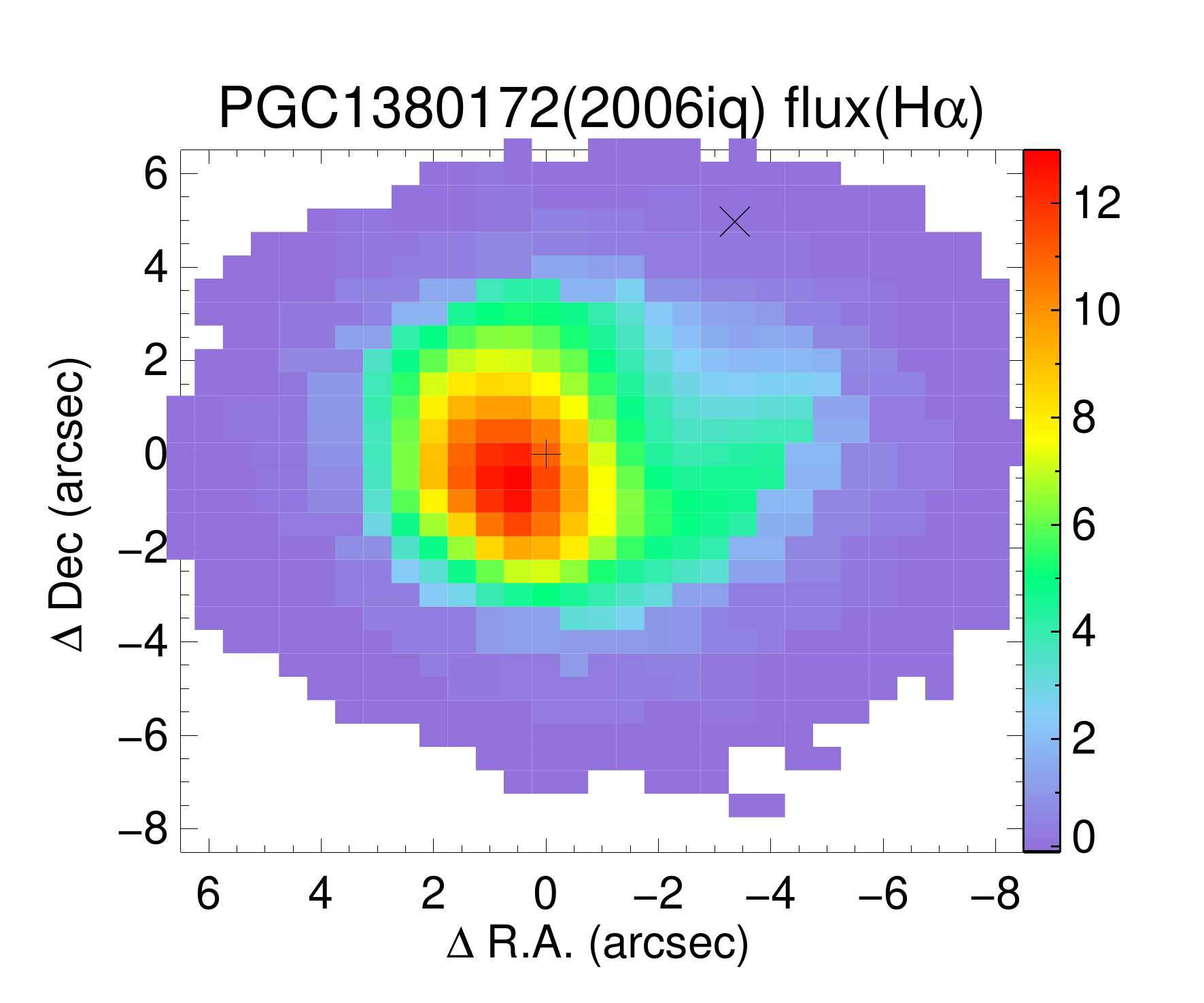}
\hspace{0.2cm}
\includegraphics[width=30mm]{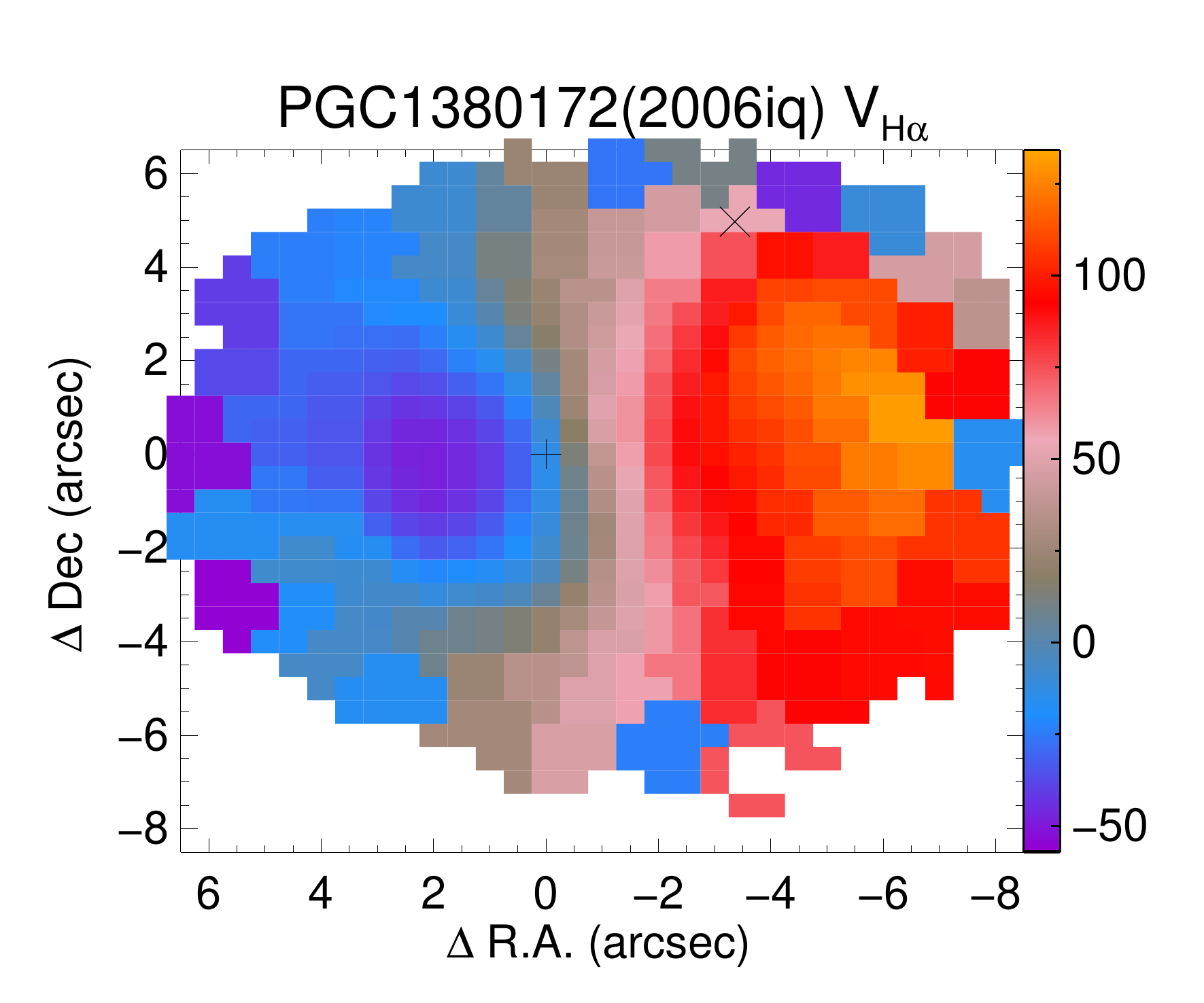}
\hspace{0.2cm}
\includegraphics[width=30mm]{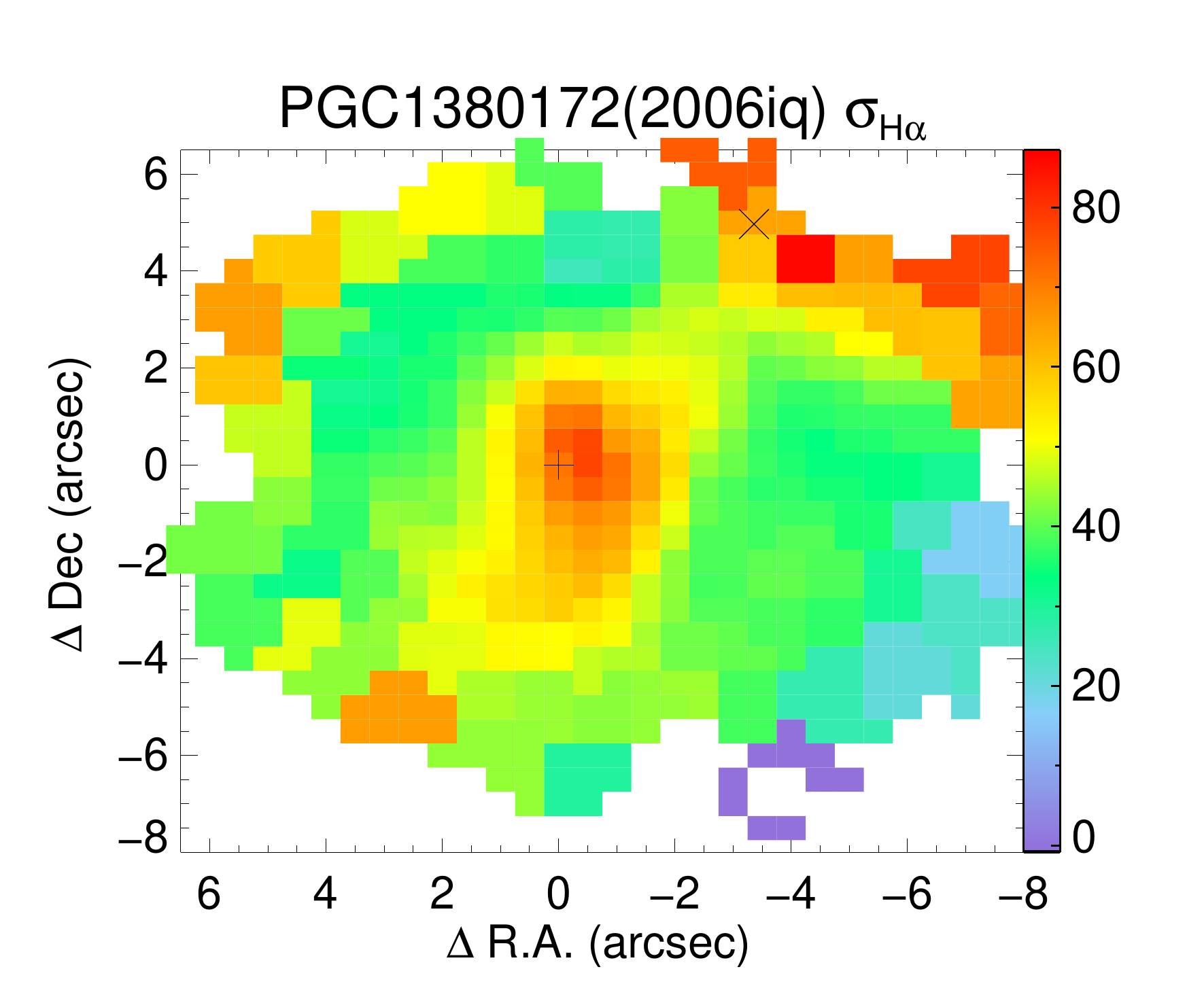}
\end{minipage}
\begin{minipage}{1.0\textwidth}
\hspace{0.9cm}
\includegraphics[width=30mm]{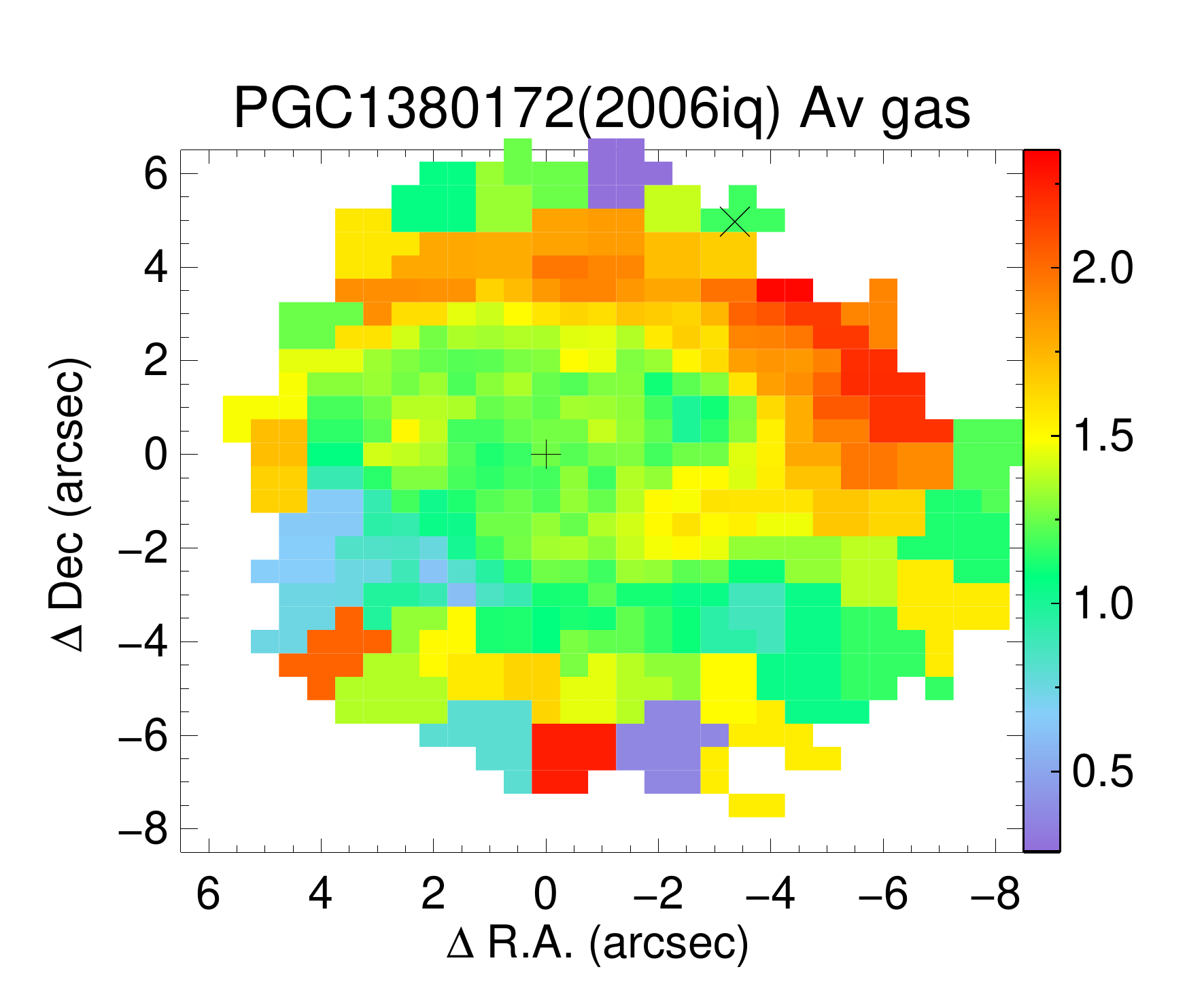}
\hspace{-0.2cm}
\includegraphics[width=30mm]{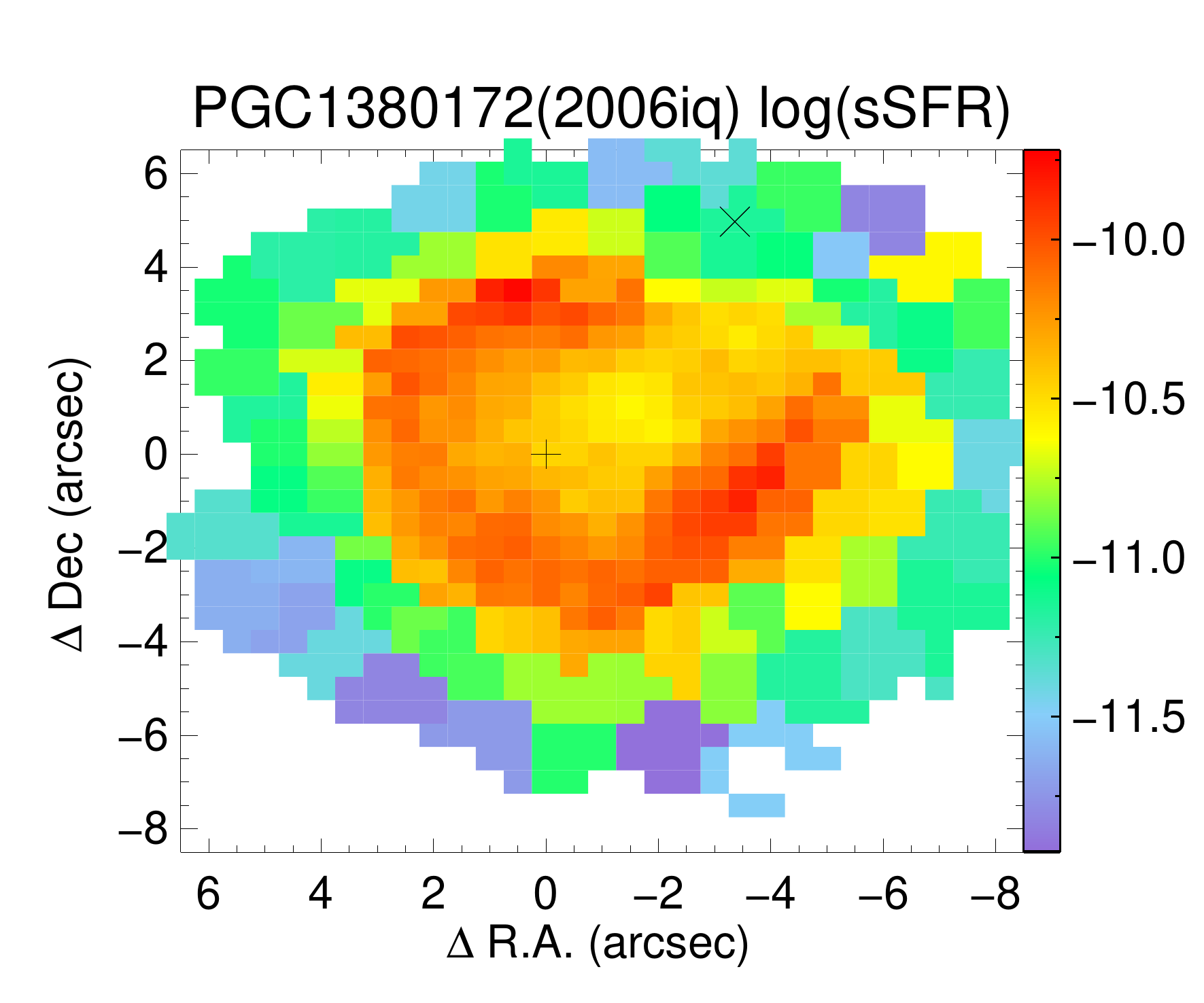}
\hspace{0.2cm}
\includegraphics[width=30mm]{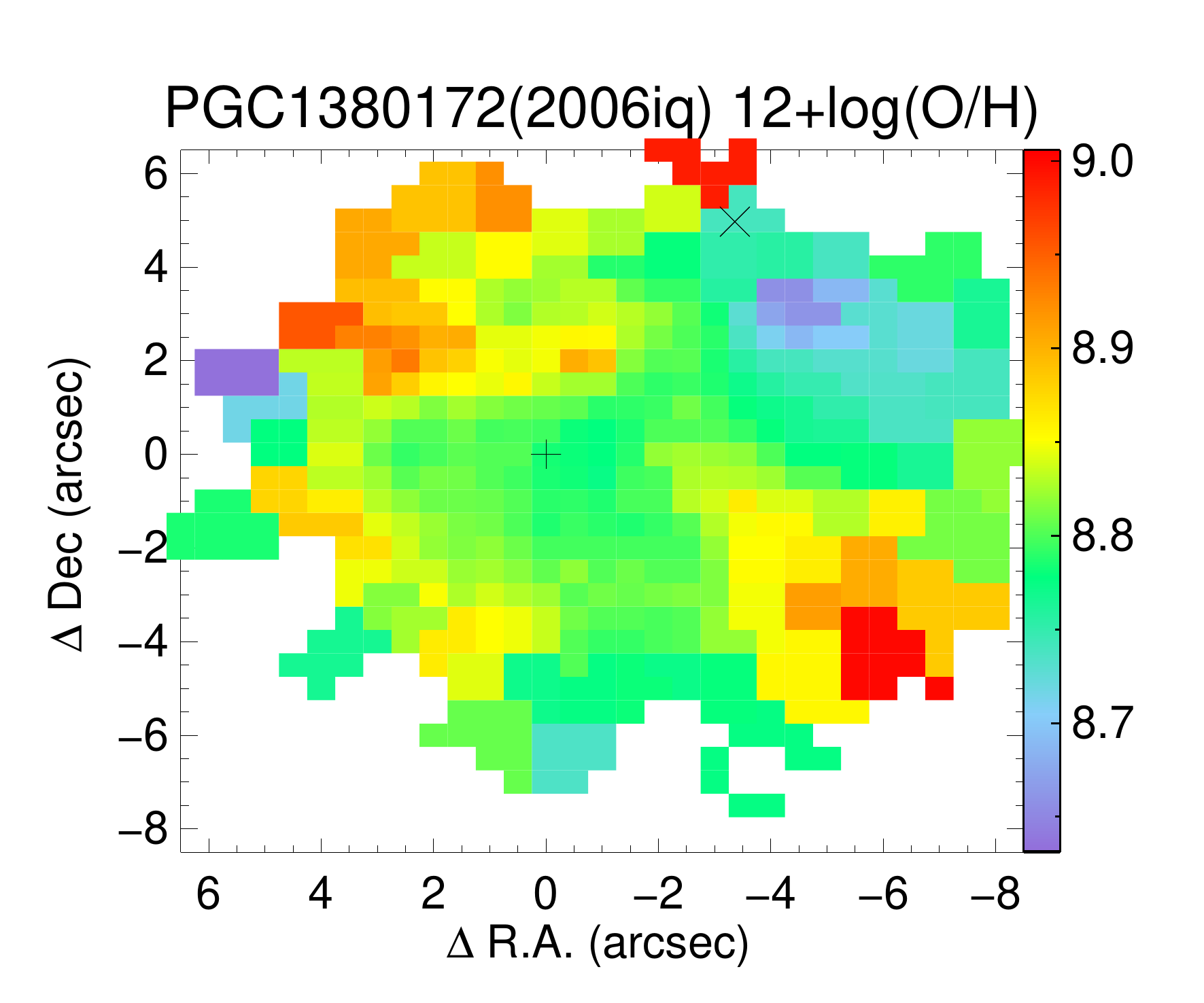}
\hspace{0.2cm}
\includegraphics[width=30mm]{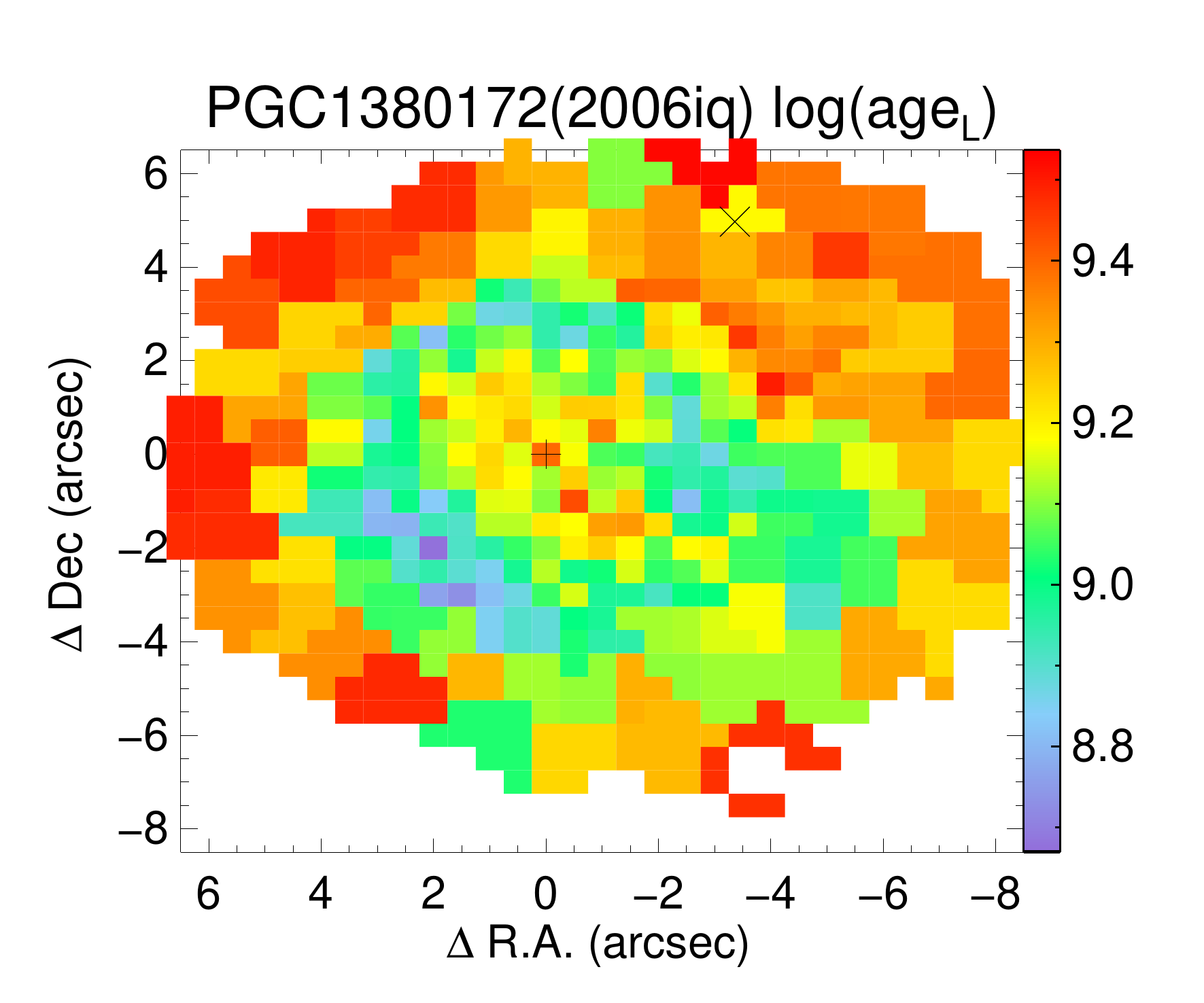}
\end{minipage}
\begin{minipage}{1.0\textwidth}
\hspace{1.5cm}
\includegraphics[width=20mm]{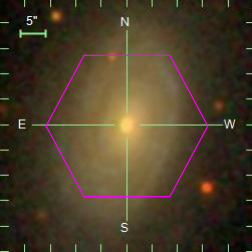}
\hspace{0.2cm}
\includegraphics[width=30mm]{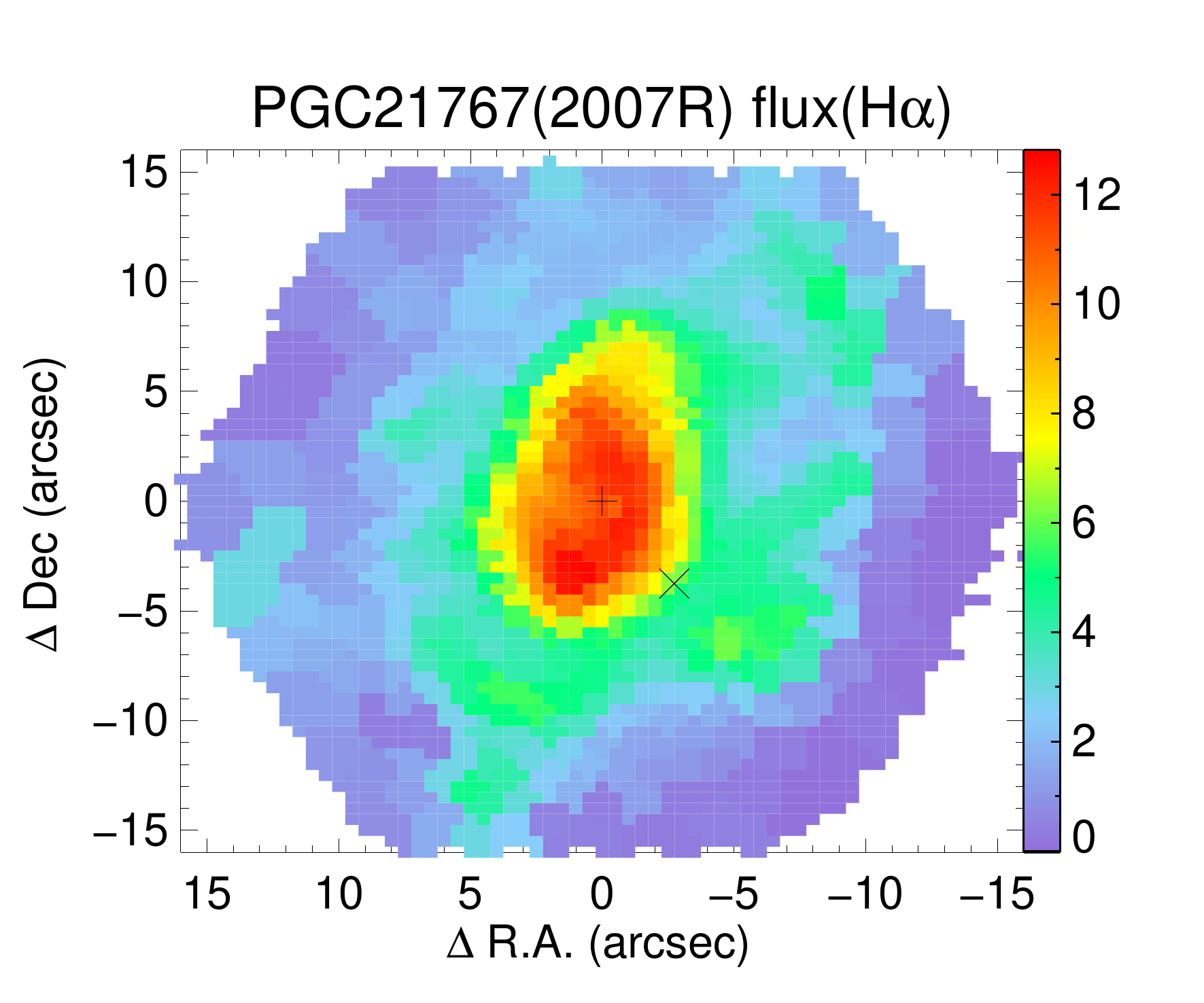}
\hspace{0.2cm}
\includegraphics[width=30mm]{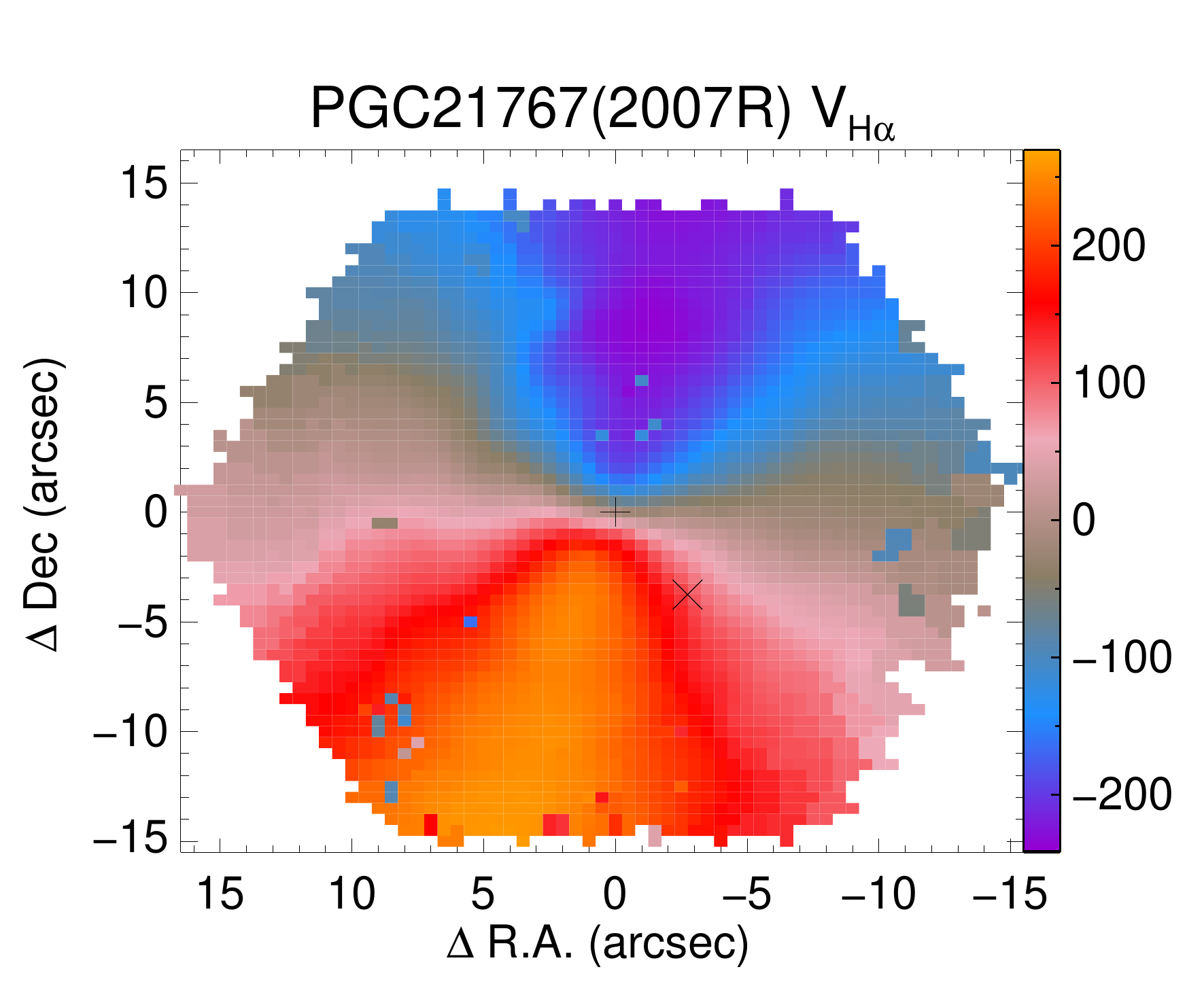}
\hspace{0.2cm}
\includegraphics[width=30mm]{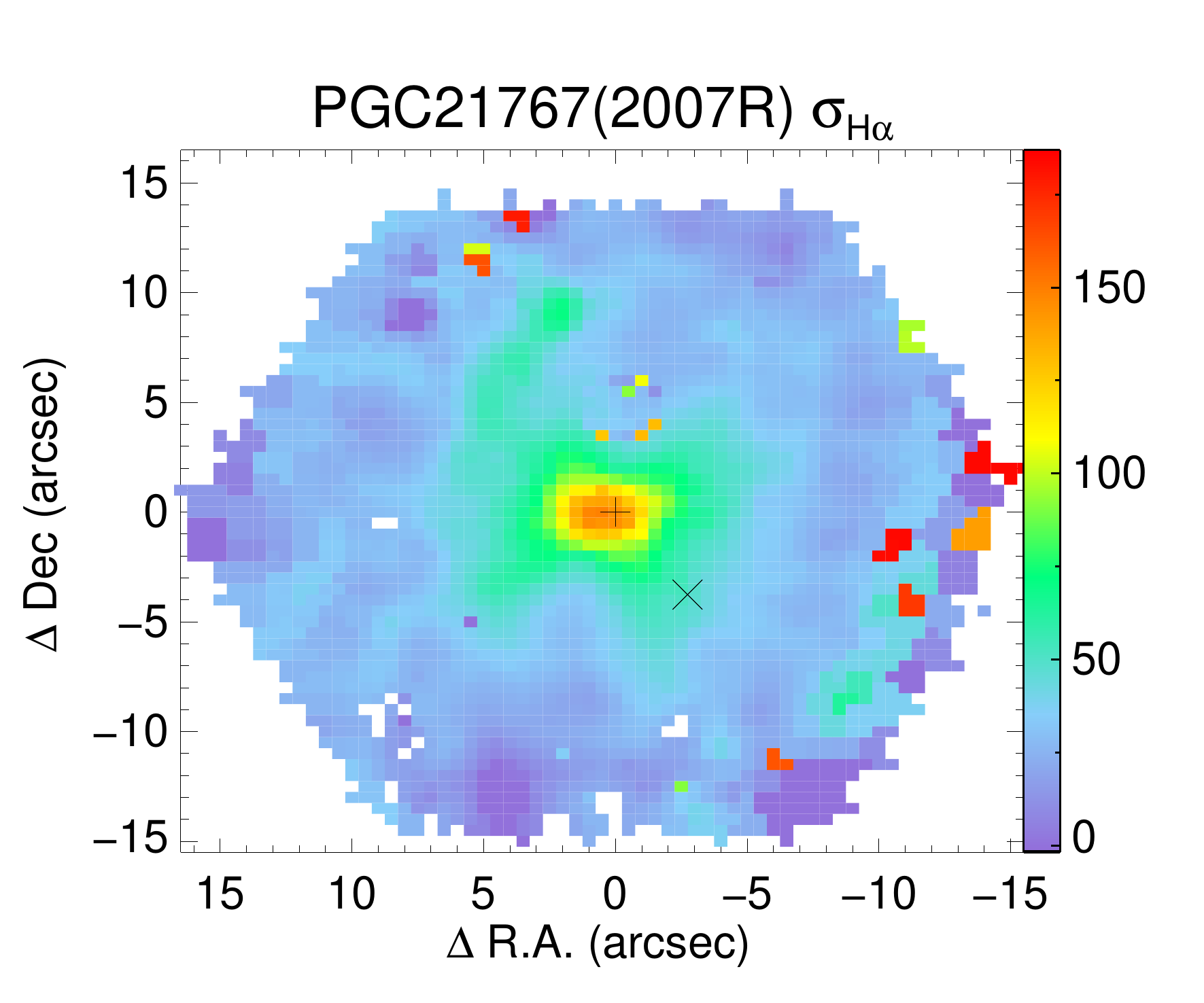}
\end{minipage}
\begin{minipage}{\textwidth}
\hspace{0.9cm}
\includegraphics[width=30mm]{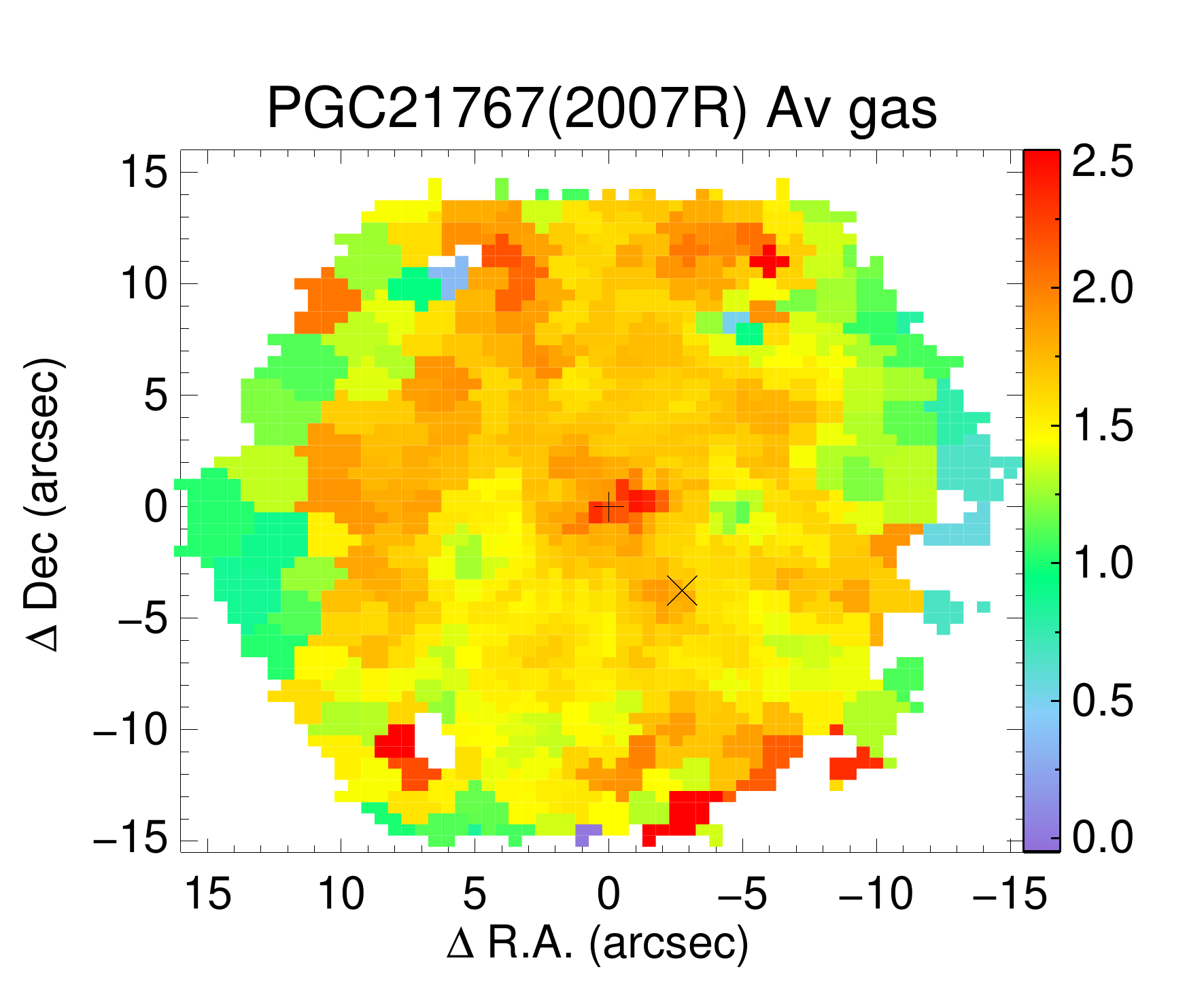}
\hspace{-0.2cm}
\includegraphics[width=30mm]{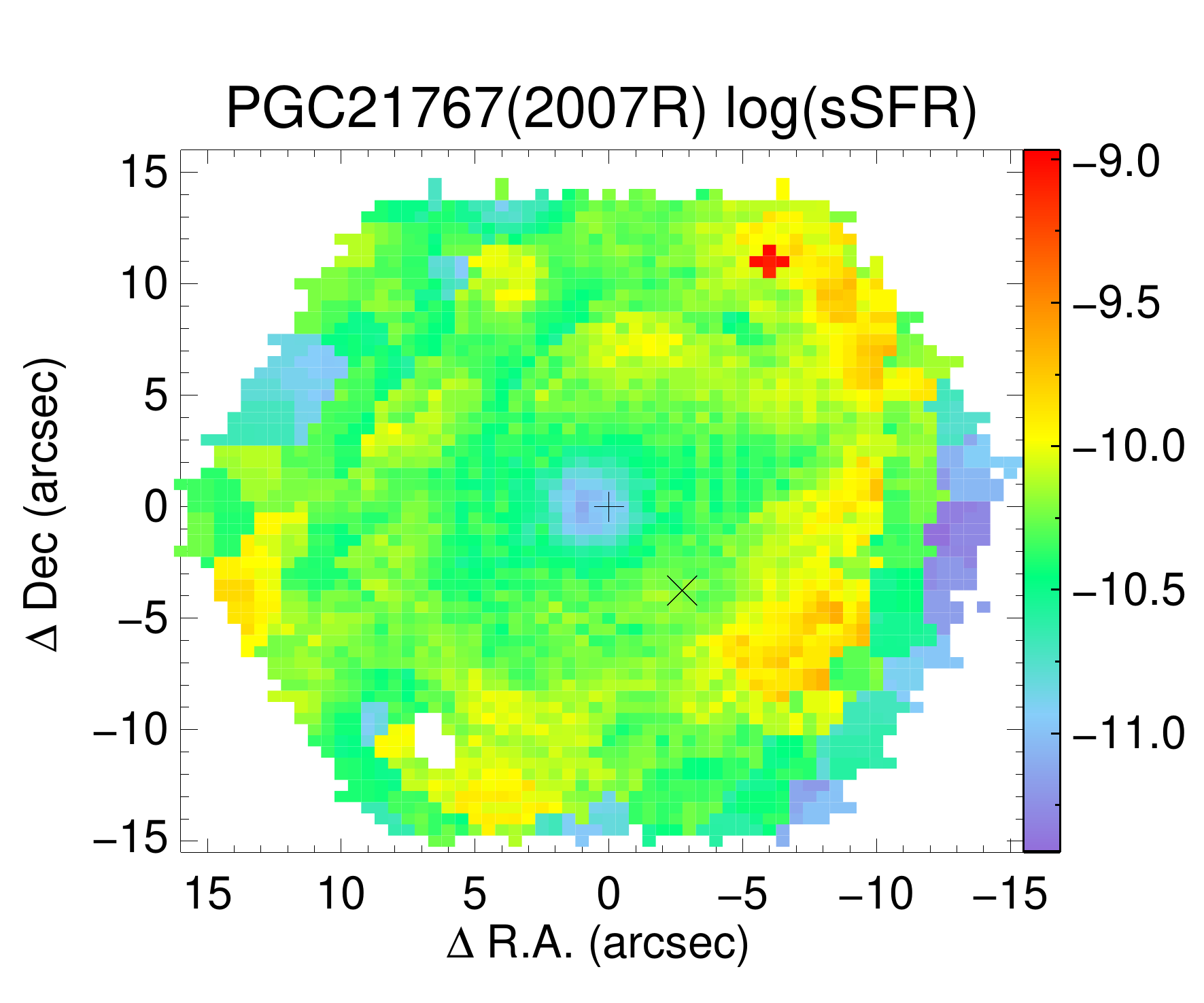}
\hspace{0.2cm}
\includegraphics[width=30mm]{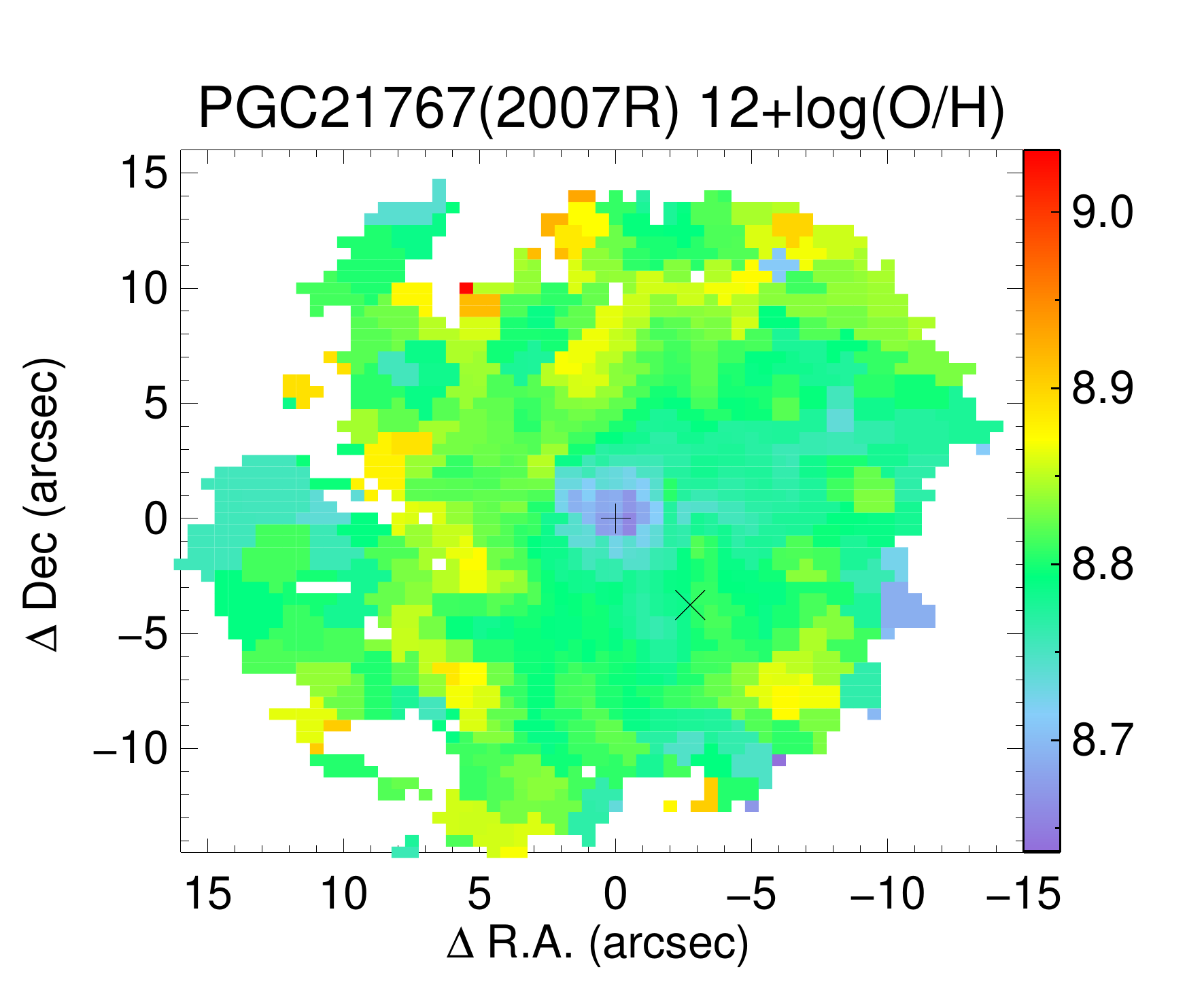}
\hspace{0.2cm}
\includegraphics[width=30mm]{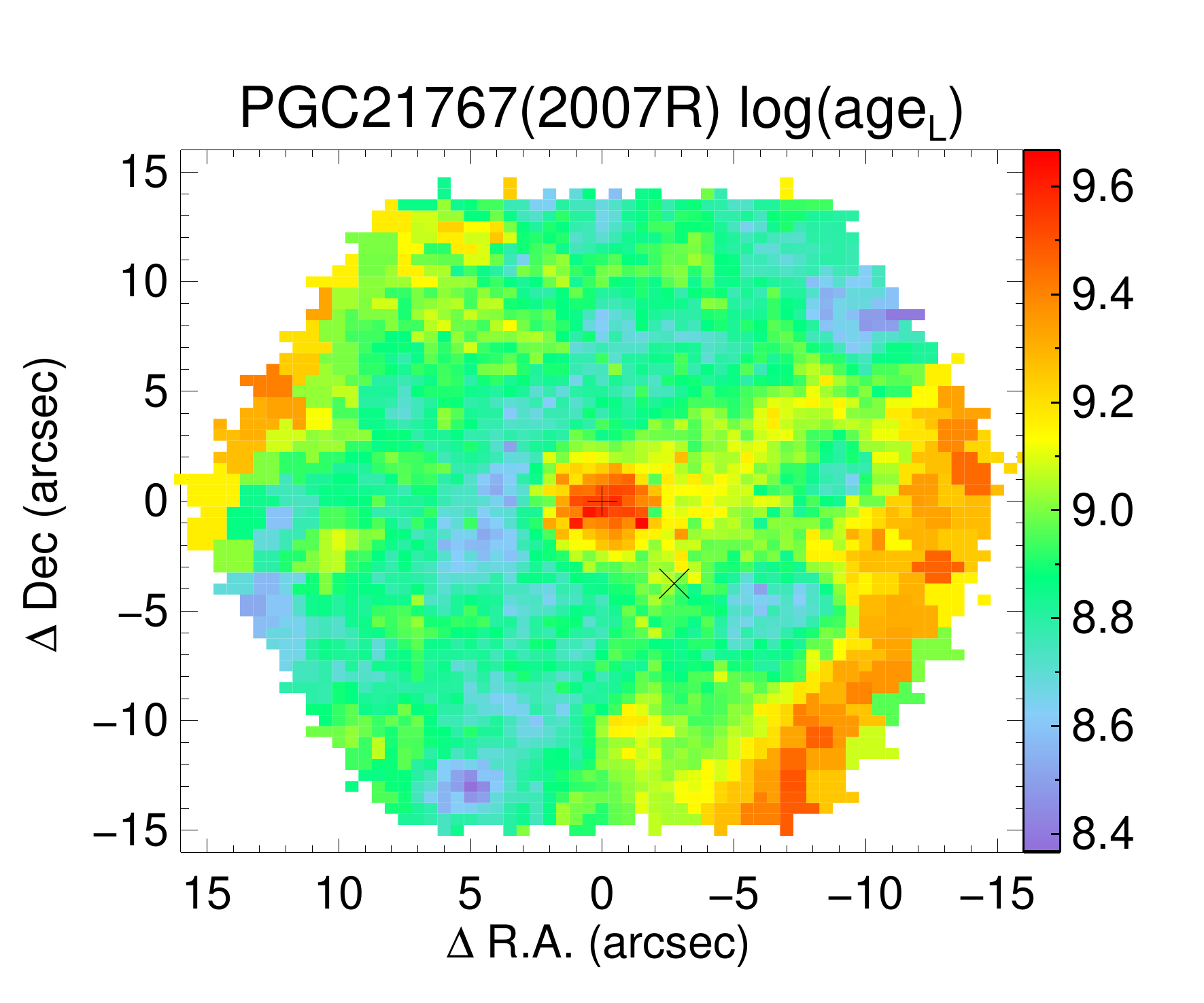}
\end{minipage}
\begin{minipage}{1.0\textwidth}
\hspace{1.5cm}
\includegraphics[width=20mm]{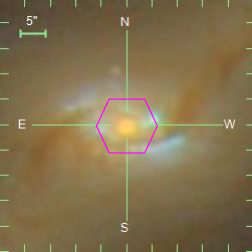}
\hspace{0.2cm}
\includegraphics[width=30mm]{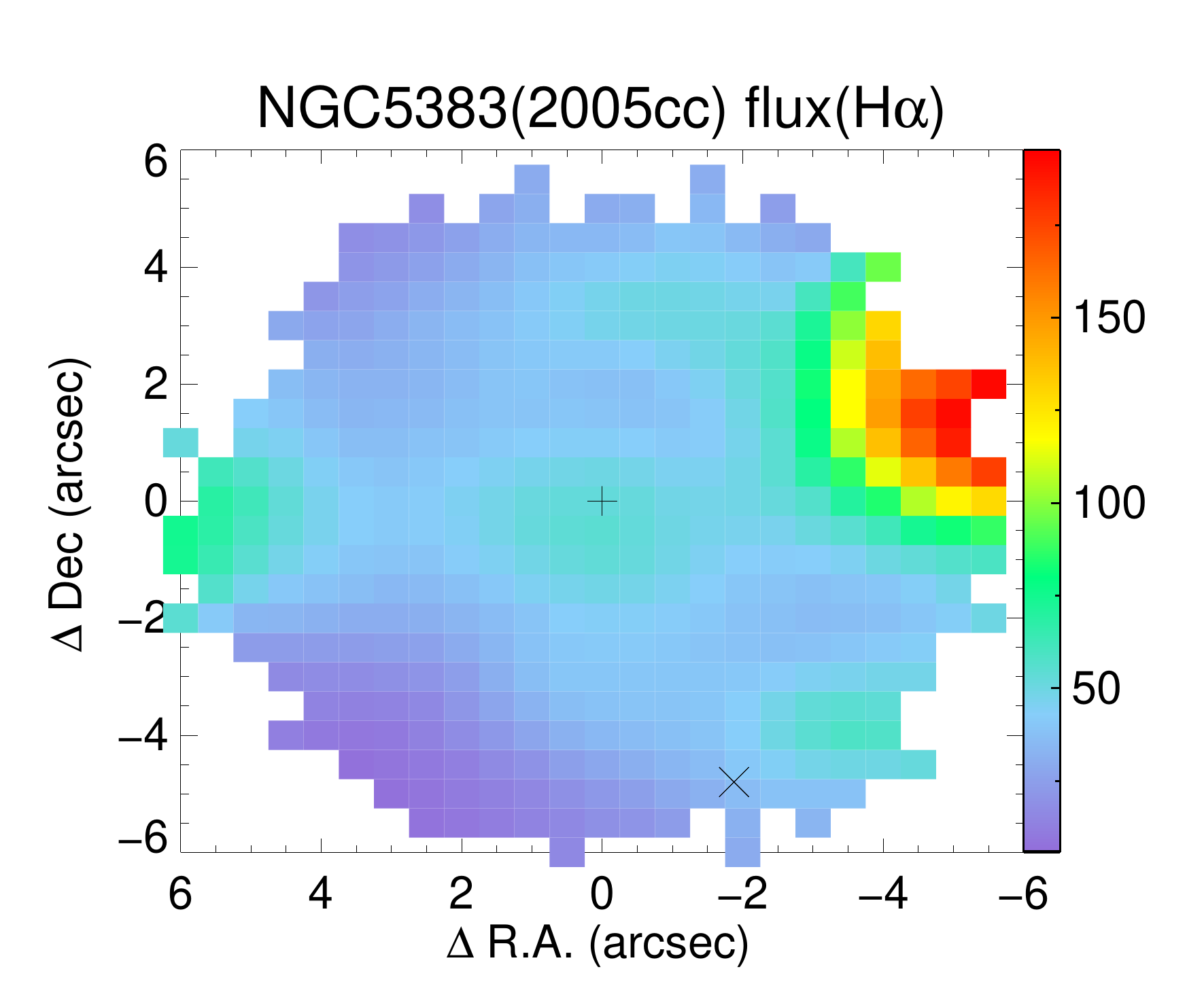}
\hspace{0.2cm}
\includegraphics[width=30mm]{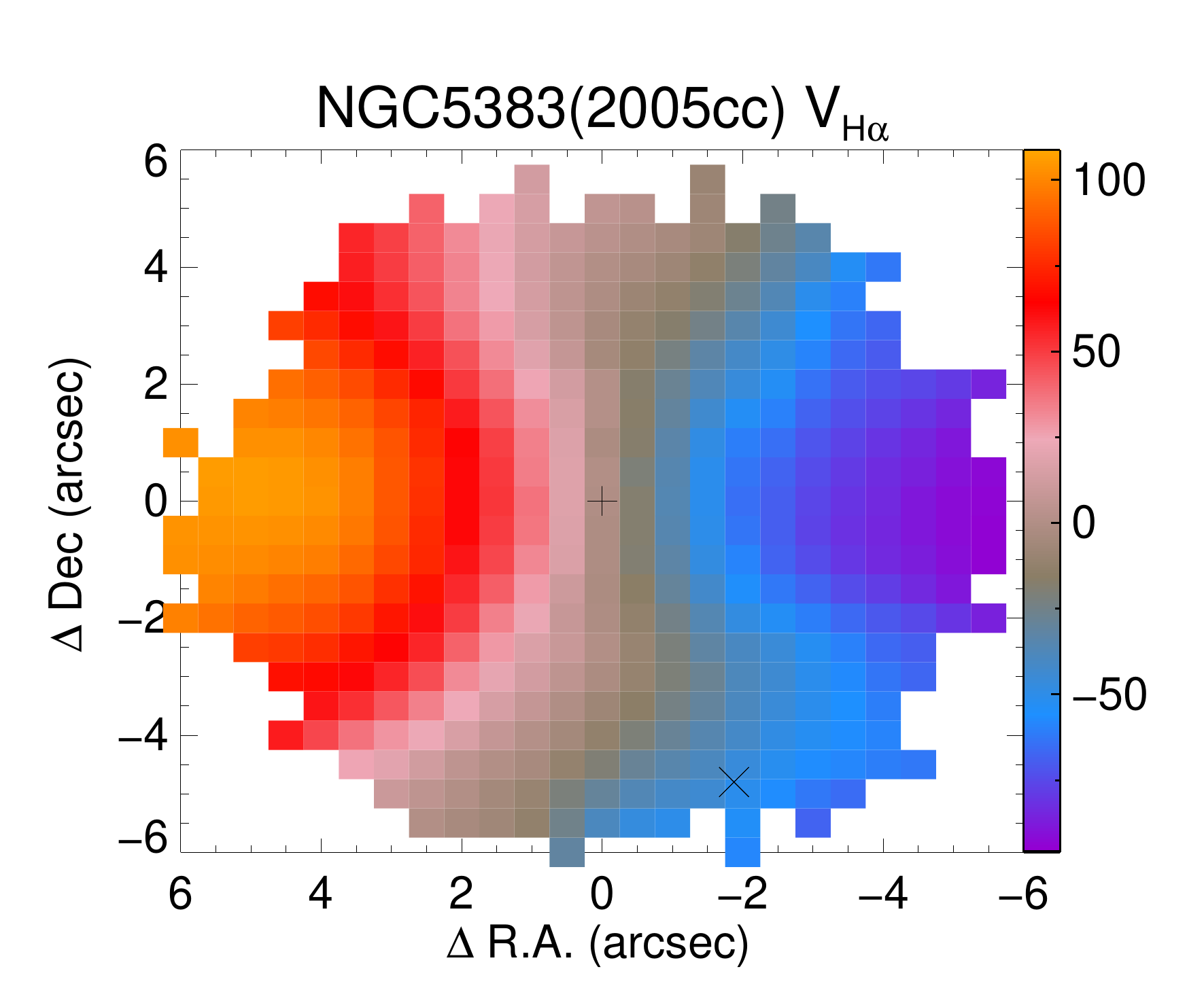}
\hspace{0.2cm}
\includegraphics[width=30mm]{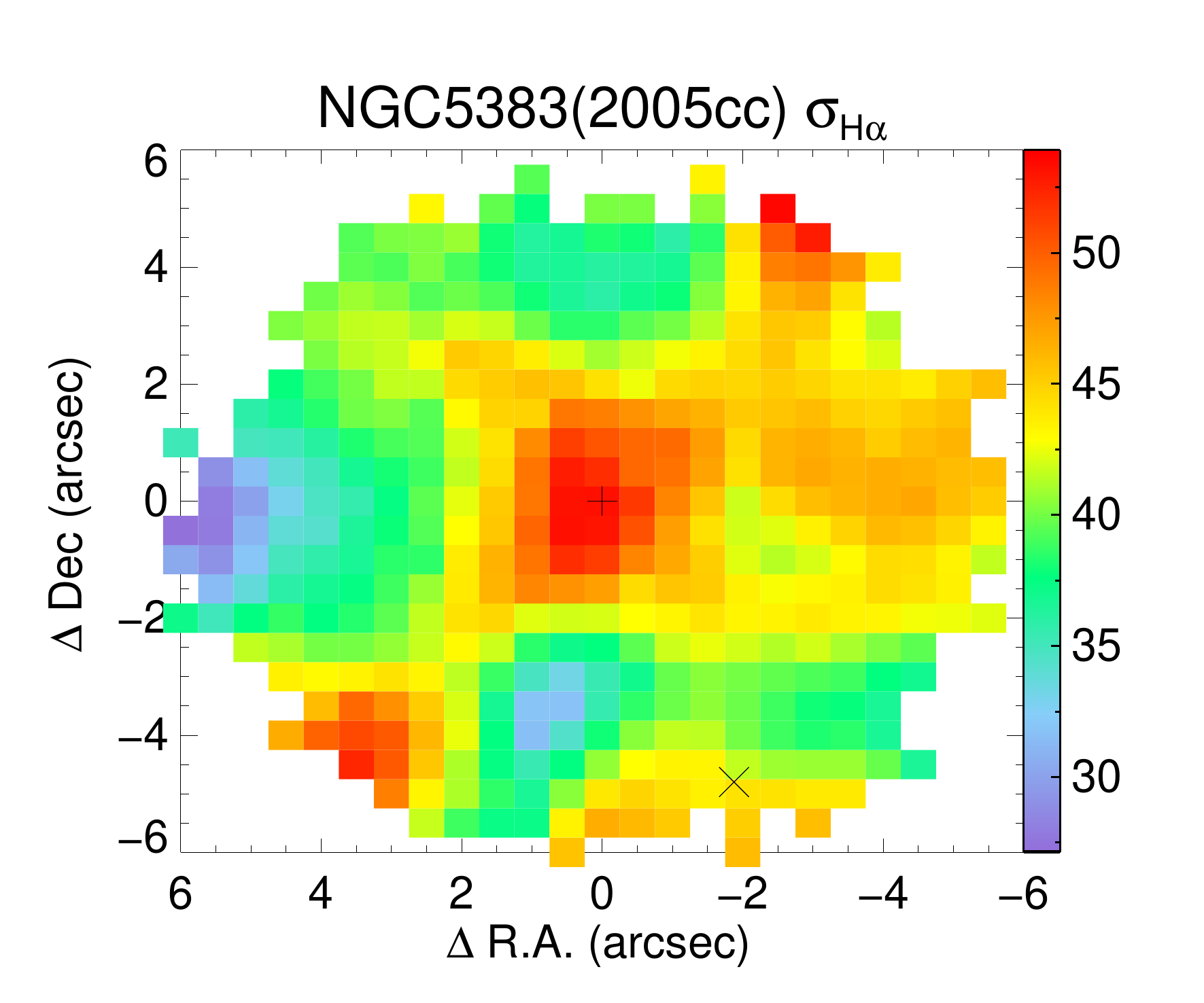}
\end{minipage}
\begin{minipage}{\textwidth}
\hspace{0.9cm}
\includegraphics[width=30mm]{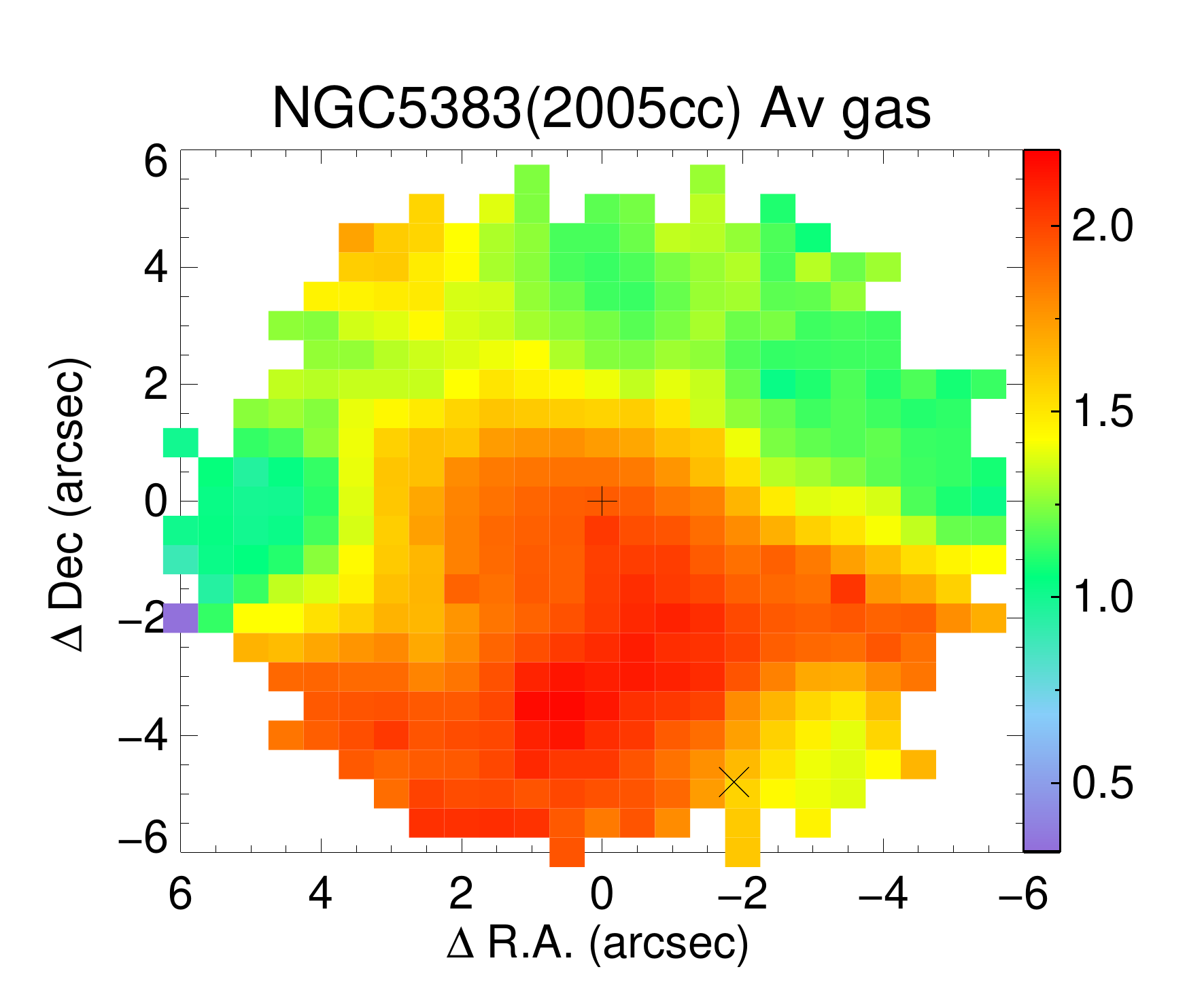}
\hspace{-0.2cm}
\includegraphics[width=30mm]{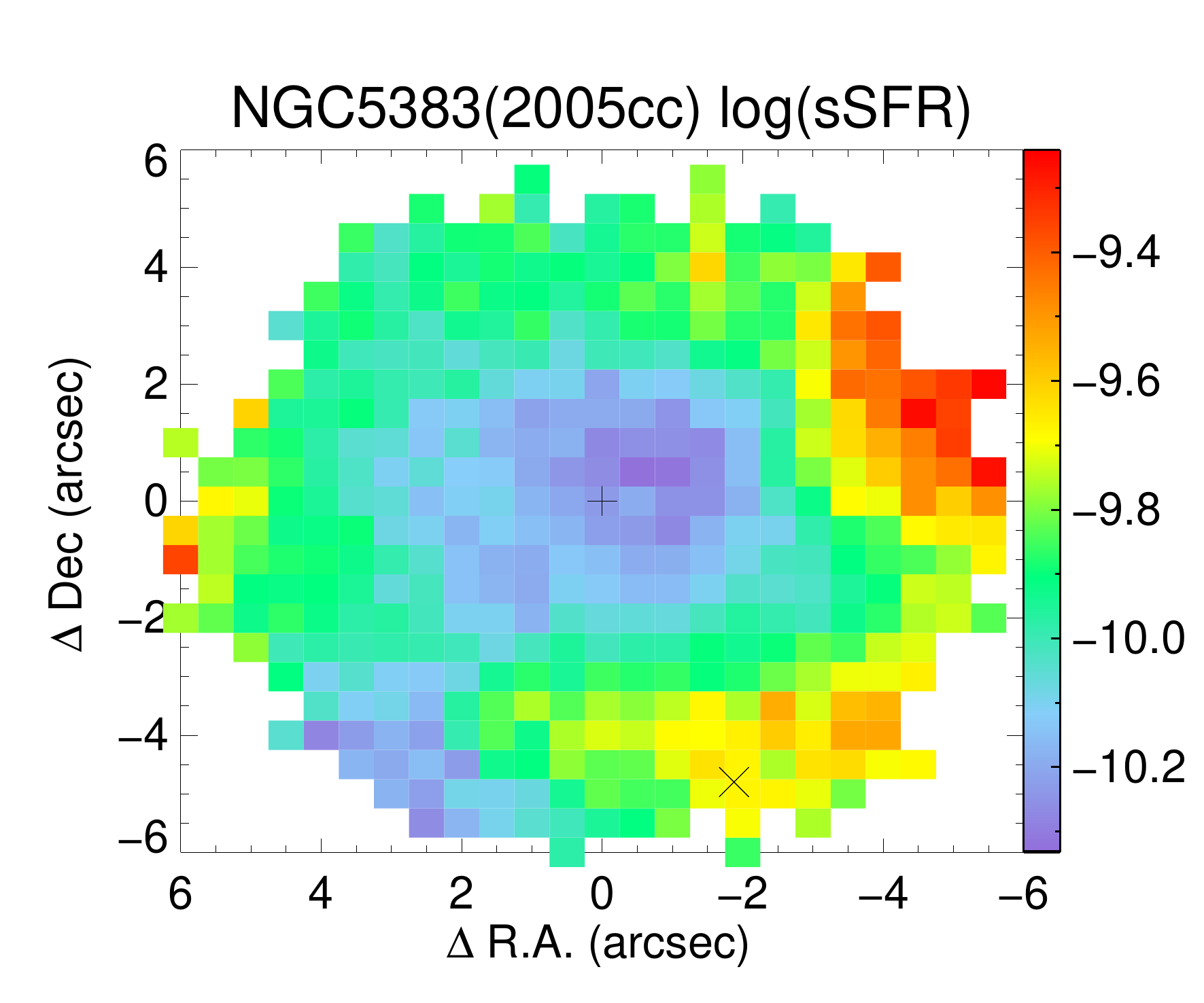}
\hspace{0.2cm}
\includegraphics[width=30mm]{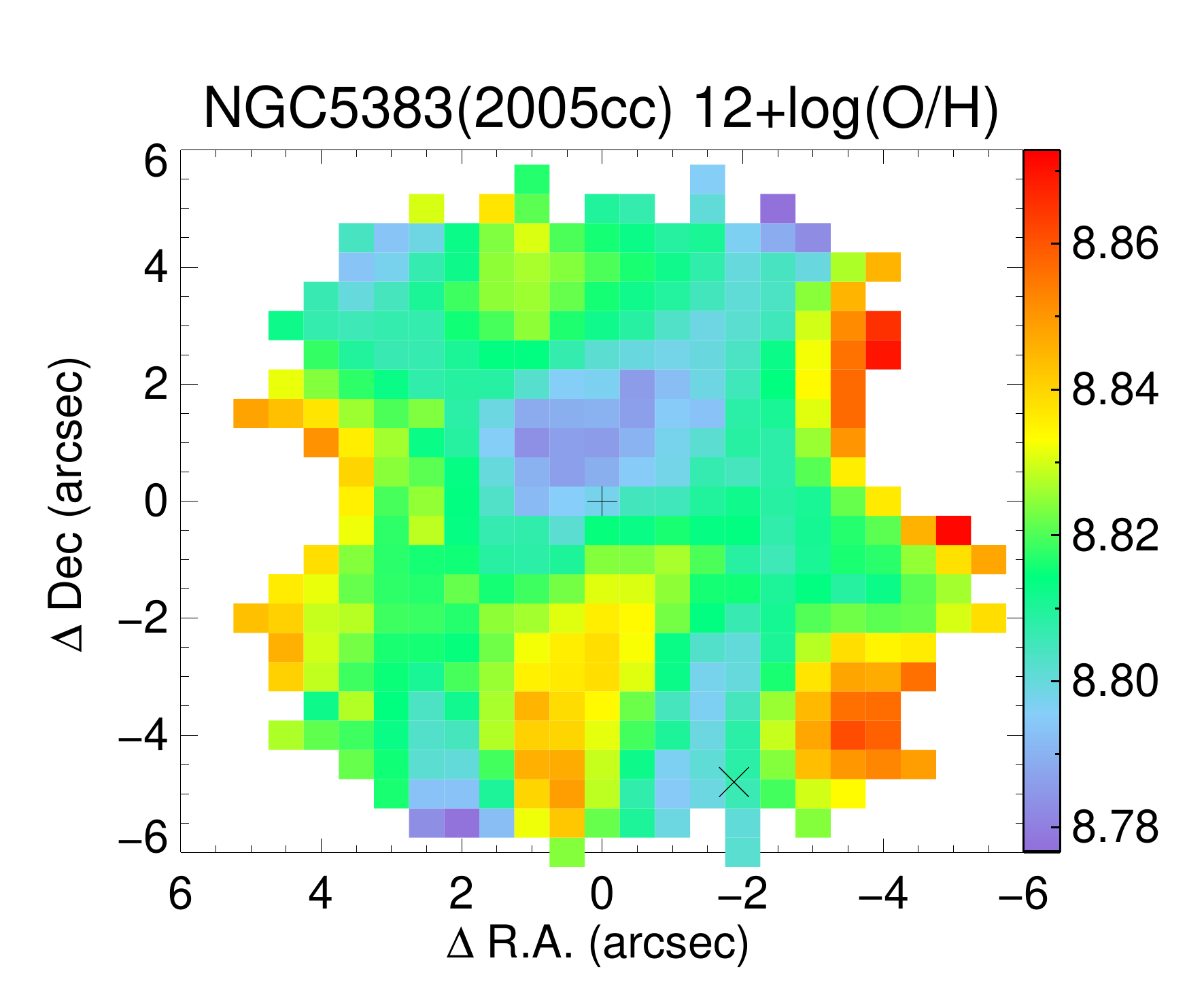}
\hspace{0.2cm}
\includegraphics[width=30mm]{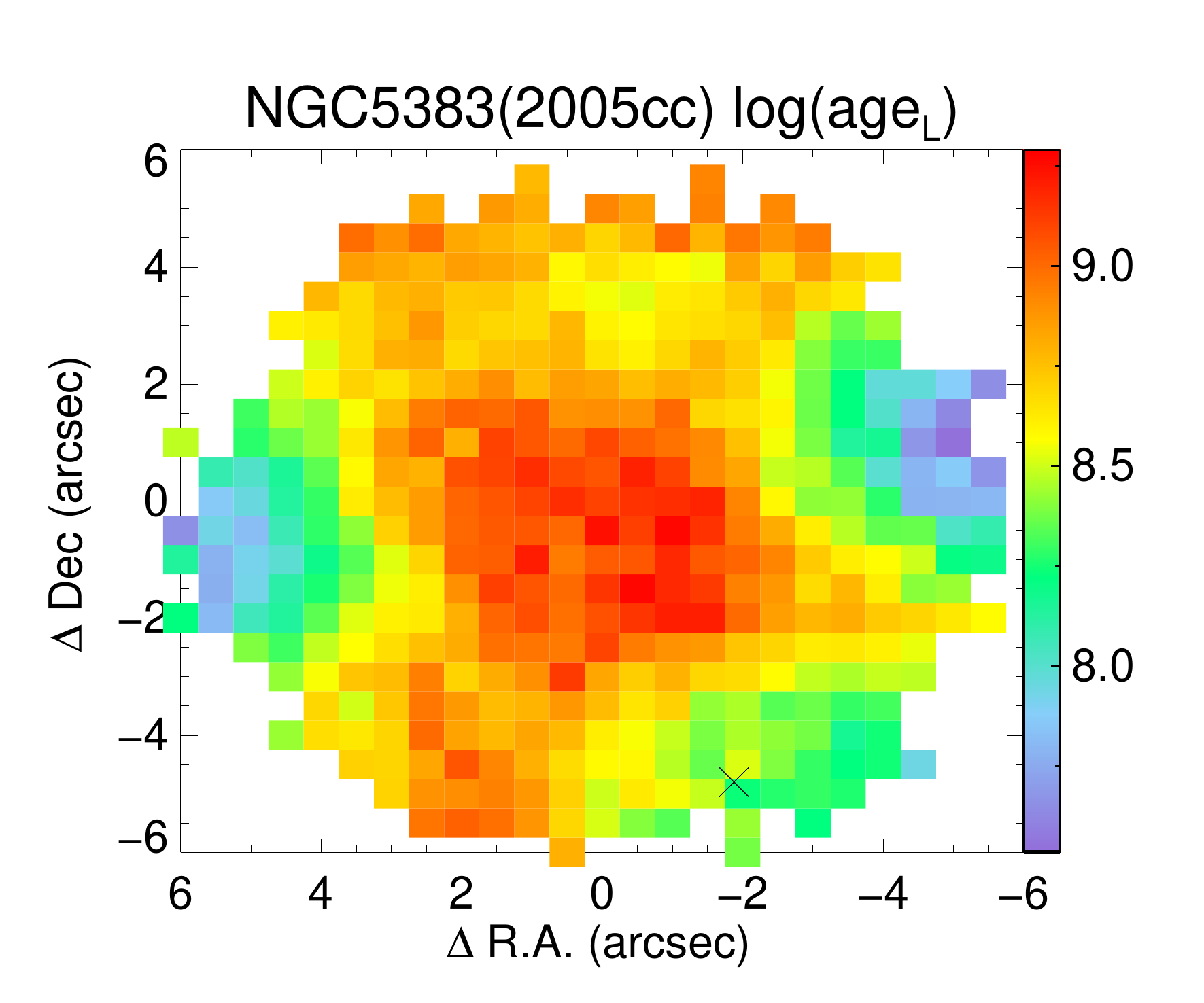}
\end{minipage}
\caption{The 2D maps of SN Ia host galaxies, including H$\alpha$ flux in units of $erg s^{-1} cm^{-2}$, H$\alpha$ velocity in units of $km s^{-1}$, velocity dispersion in units of $km s^{-1}$, gas dust extinction in units of $mag$, sSFR in units of $yr^{-1}$, oxygen abundance estimated using O3N2 method and the logarithm of light-weighted stellar age estimated using {\tt STARLIGHT} in units of $log[yr]$. The coordinates in the $X$, $Y$ axes are in units of $arcsec$ with respect to the 2D map centers. The plus marks the position of the galaxy center and the cross marks the location of the supernovae. There is only one spaxel left by limiting the S/N $\geq$ 3 for the SN 2006iq explosion site in the host galaxy. So, we do not cut spaxels with S/N $\leq$ 3 in the metallicity 2D map for this SN host galaxy.}
\label{fig.SNIa-havd}
\end{figure*}

\begin{figure*}[]
\centering
\begin{minipage}{\textwidth}
\hspace{1.5cm}
\includegraphics[width=20mm]{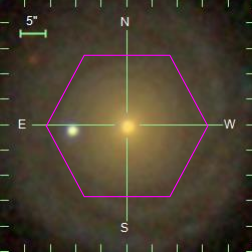}
\hspace{0.2cm}
\includegraphics[width=30mm]{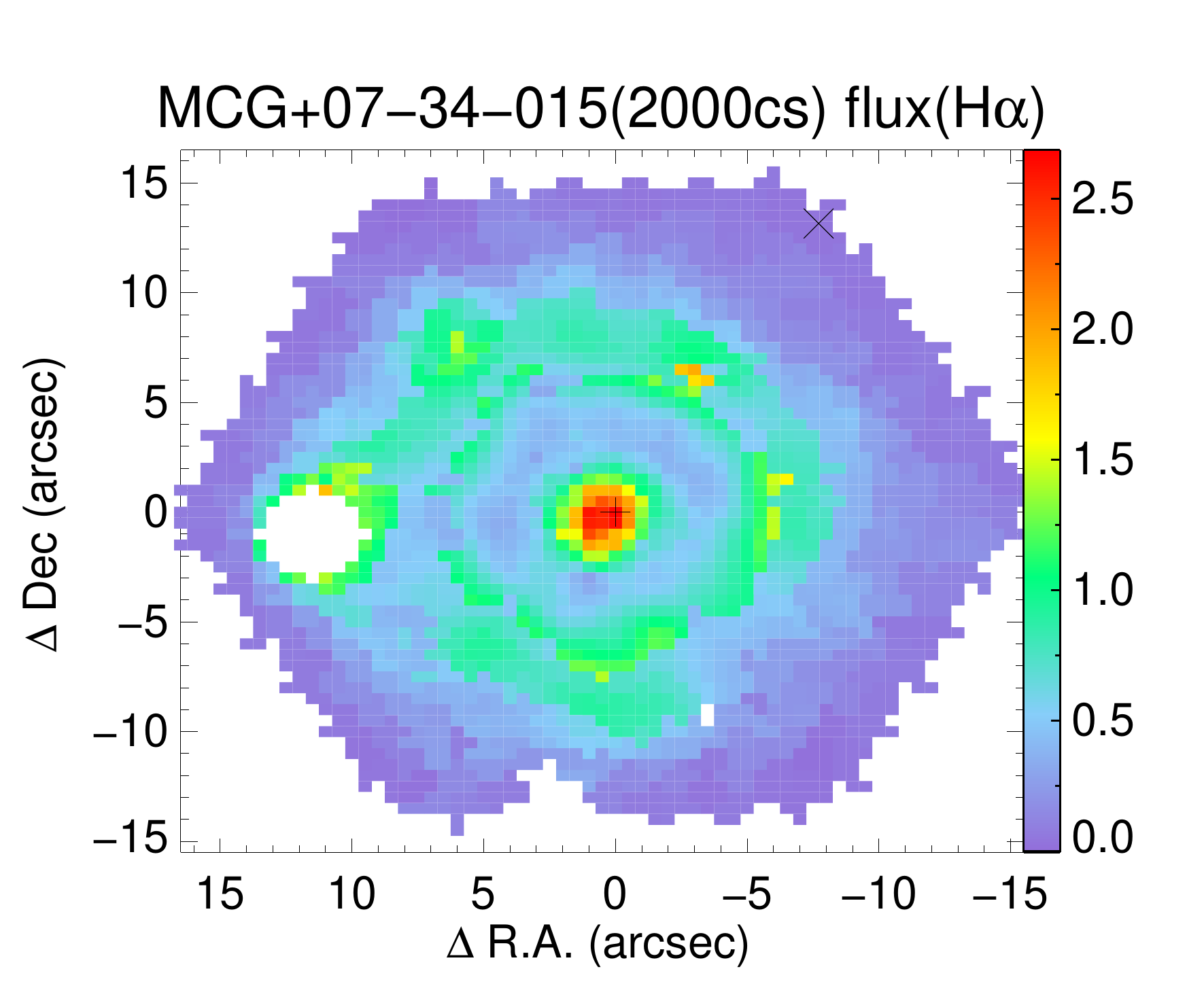}
\hspace{0.2cm}
\includegraphics[width=30mm]{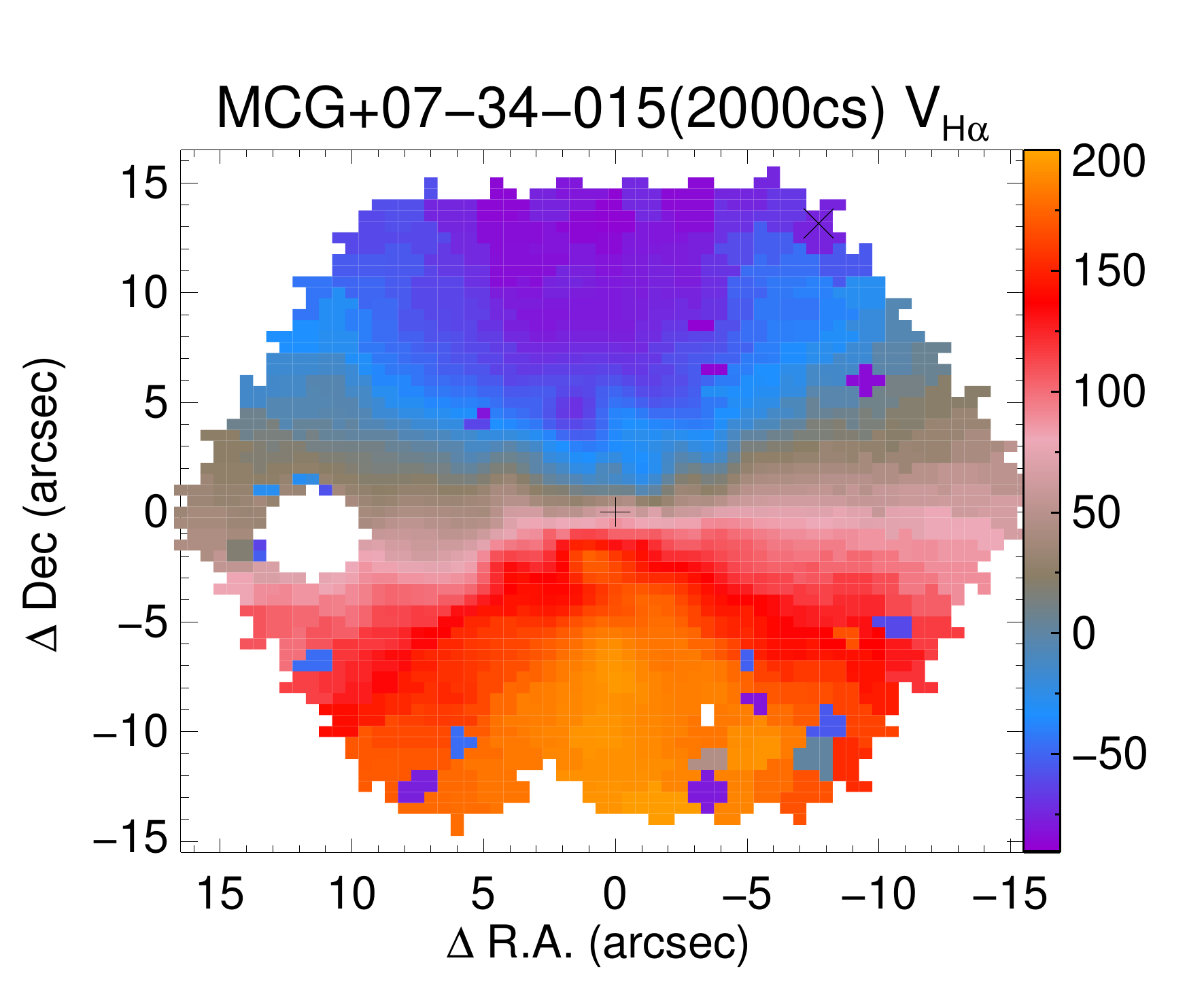}
\hspace{0.2cm}
\includegraphics[width=30mm]{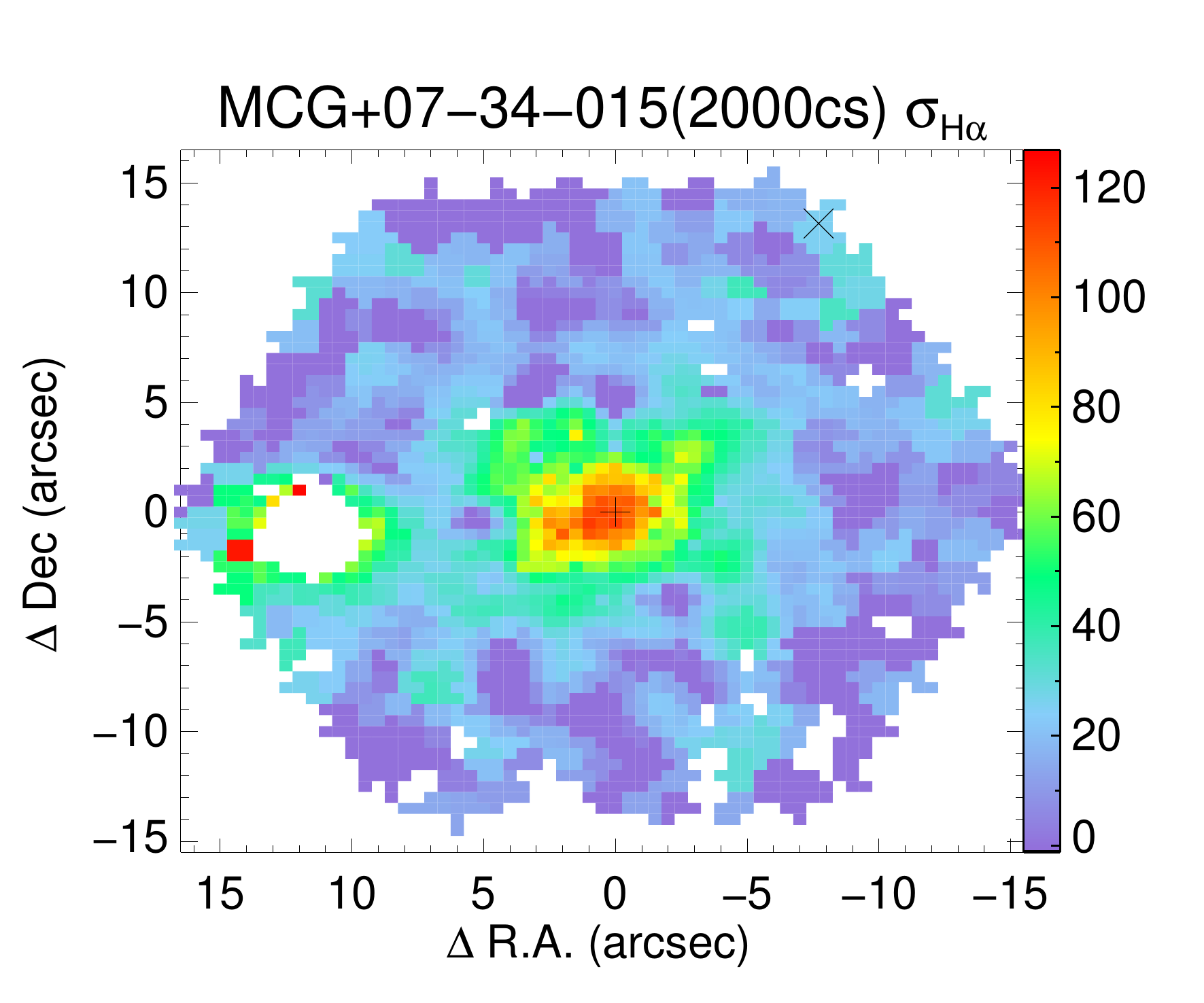}
\end{minipage}
\begin{minipage}{\textwidth}
\hspace{0.9cm}
\includegraphics[width=30mm]{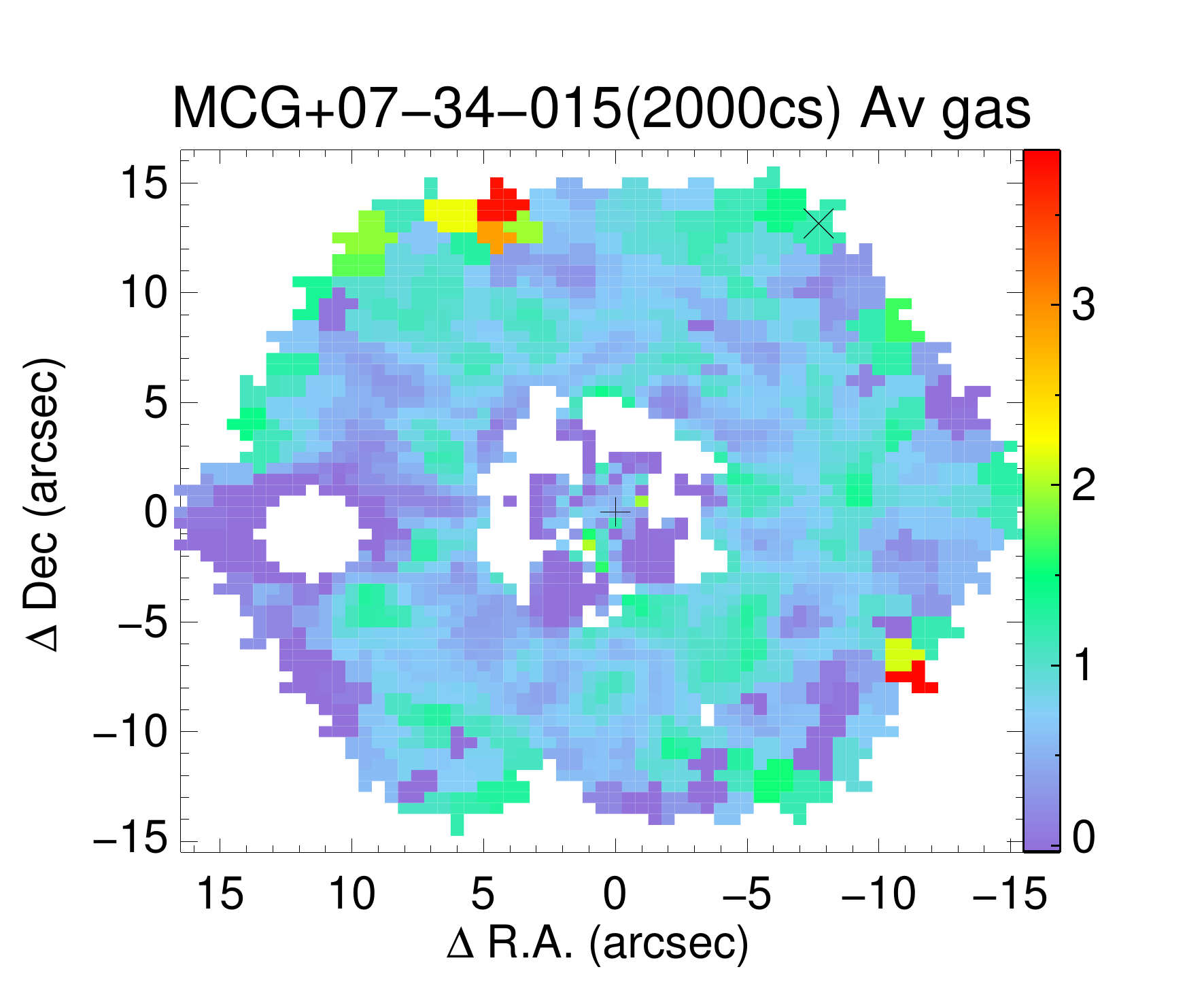}
\hspace{-0.2cm}
\includegraphics[width=30mm]{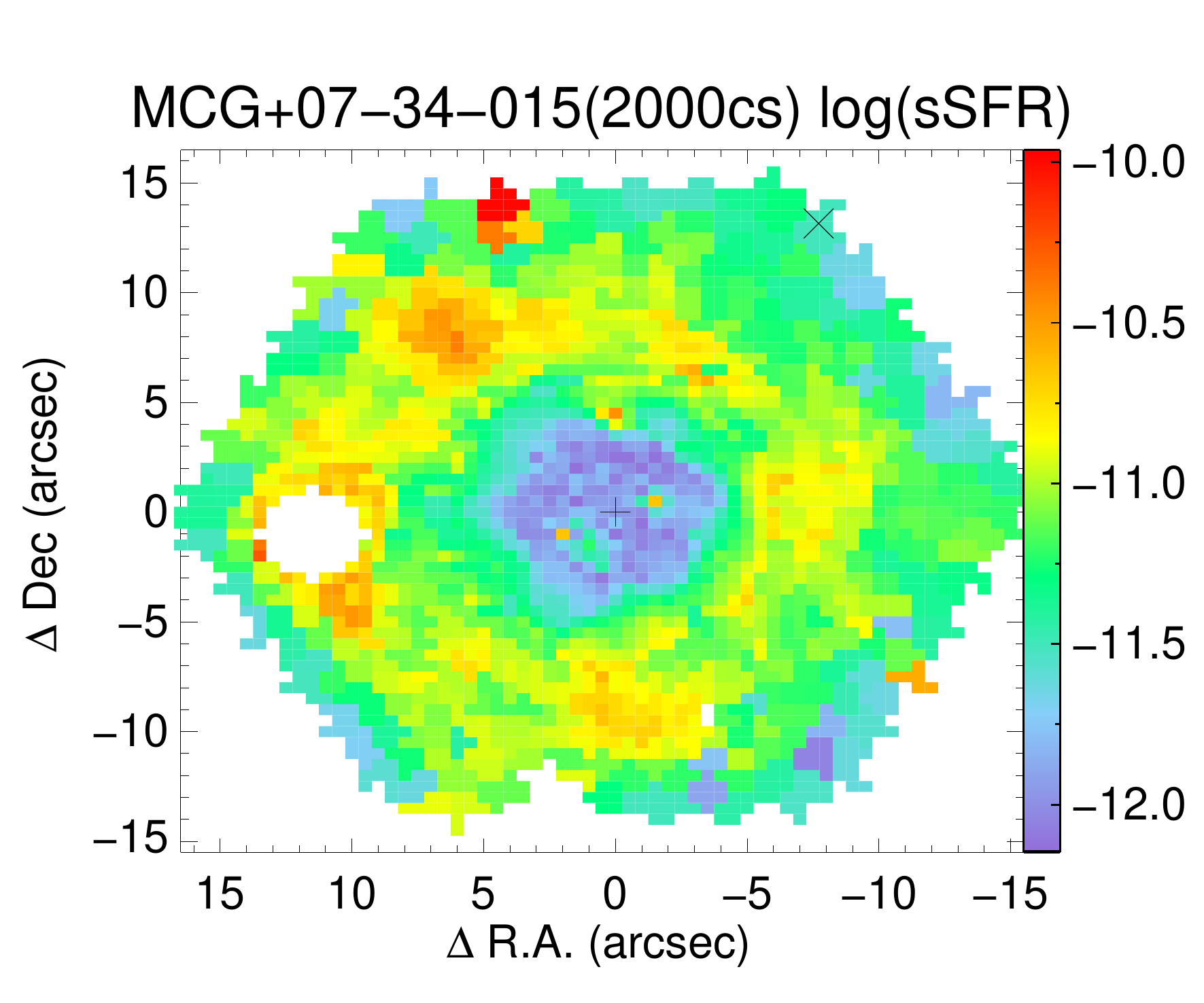}
\hspace{0.2cm}
\includegraphics[width=30mm]{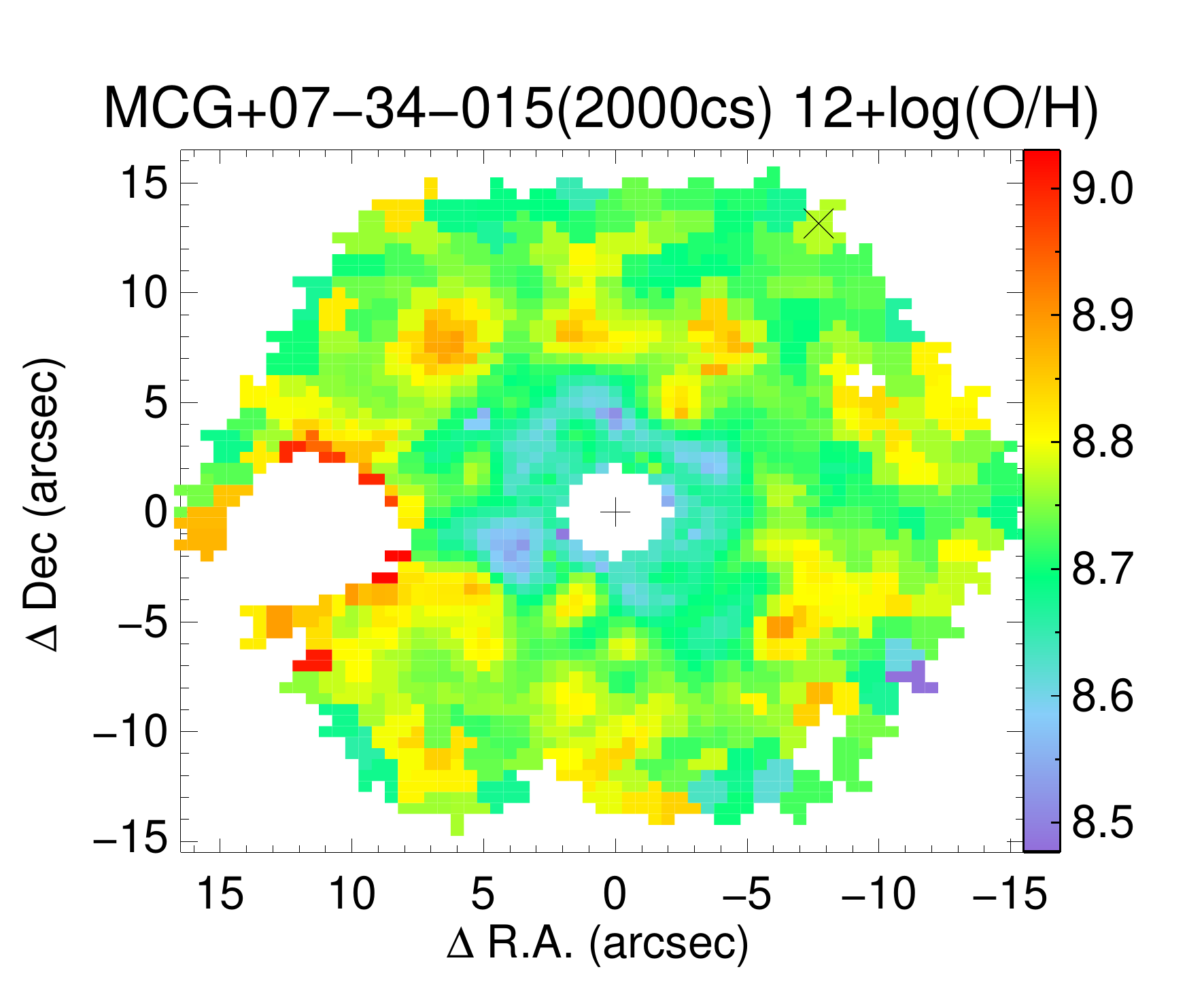}
\hspace{0.2cm}
\includegraphics[width=30mm]{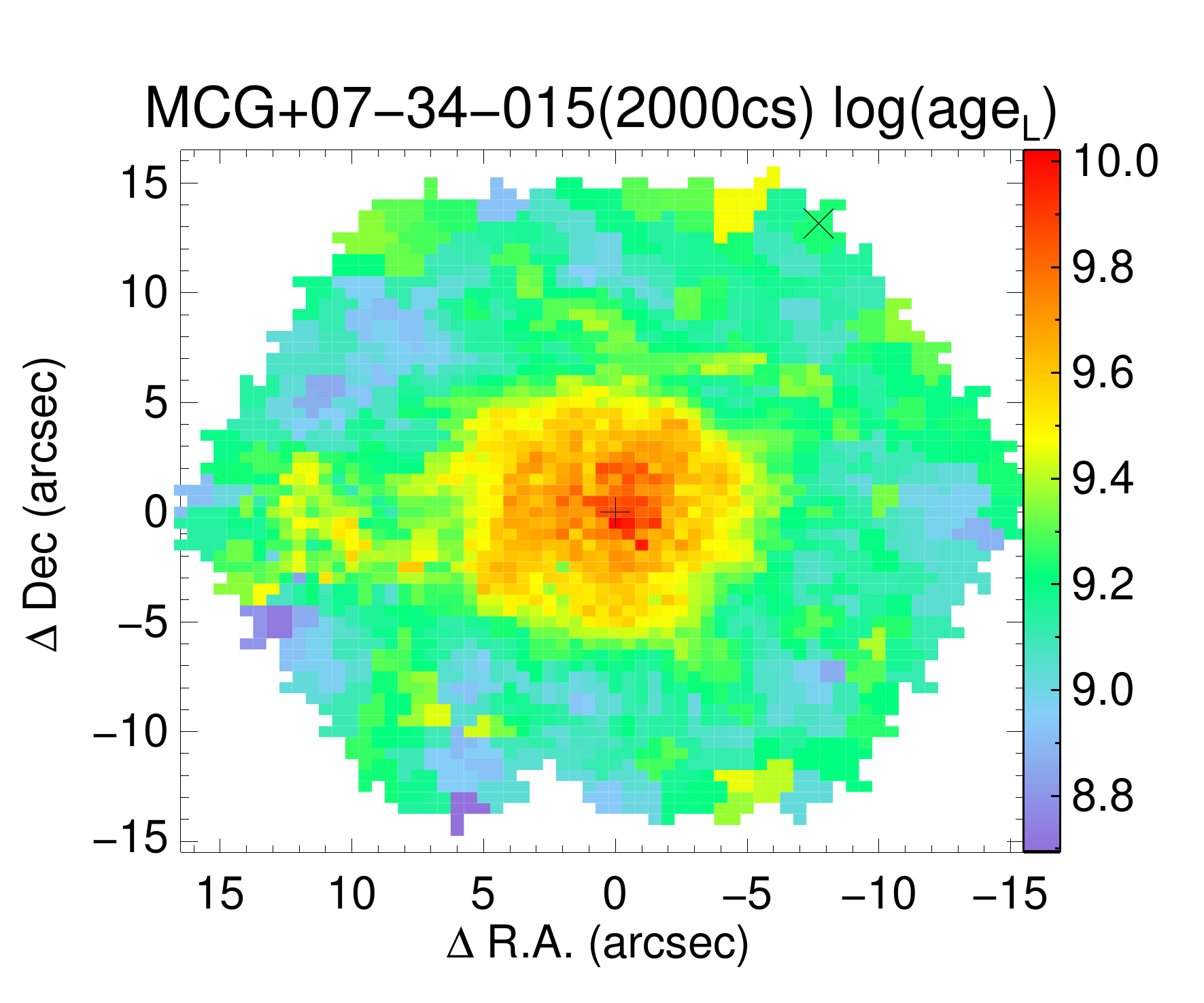}
\end{minipage}
\begin{minipage}{\textwidth}
\hspace{1.5cm}
\includegraphics[width=20mm]{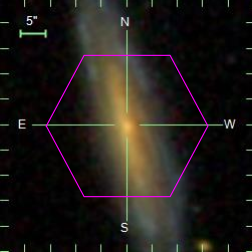}
\hspace{0.2cm}
\includegraphics[width=30mm]{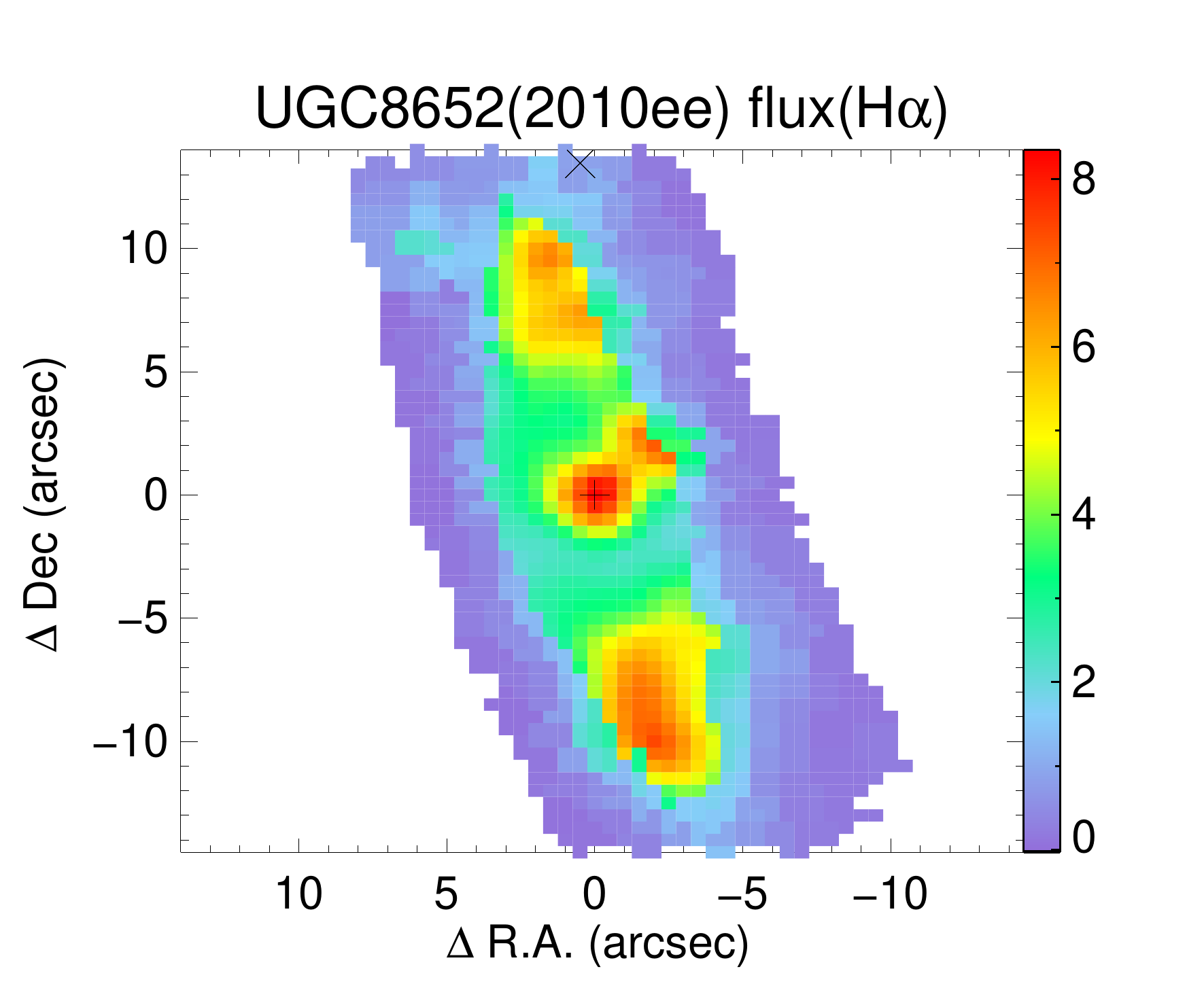}
\hspace{0.2cm}
\includegraphics[width=30mm]{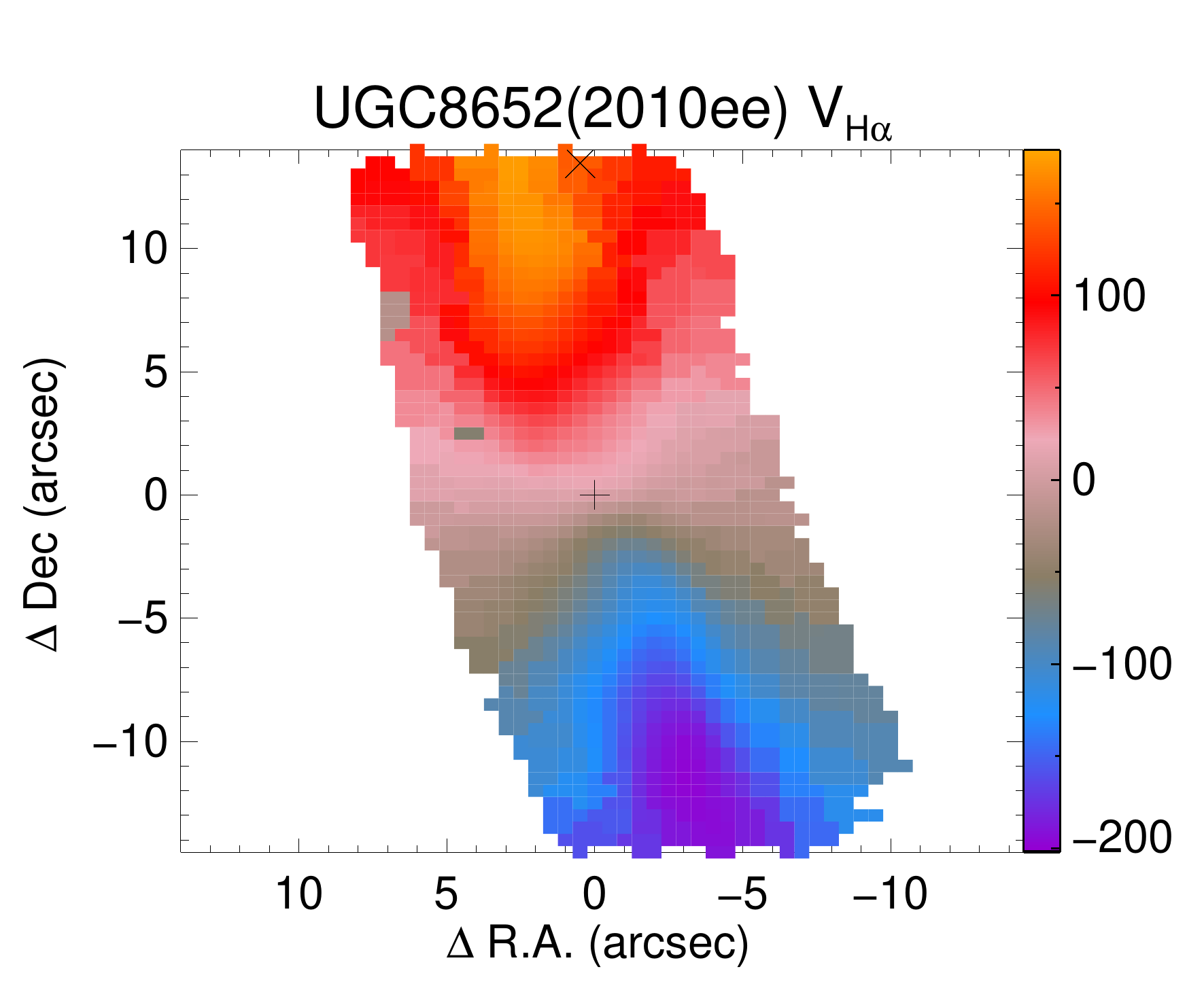}
\hspace{0.2cm}
\includegraphics[width=30mm]{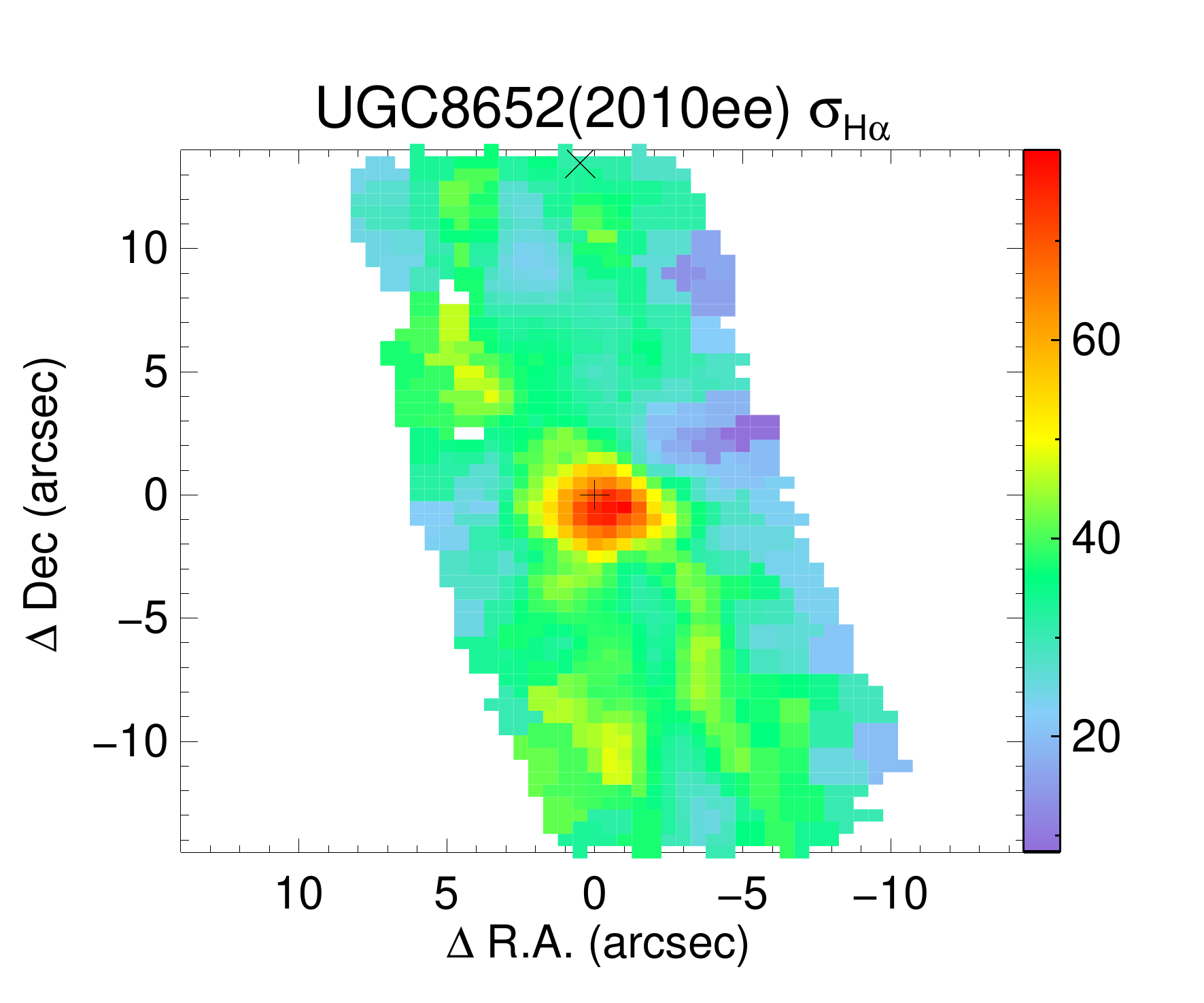}
\end{minipage}
\begin{minipage}{\textwidth}
\hspace{0.9cm}
\includegraphics[width=30mm]{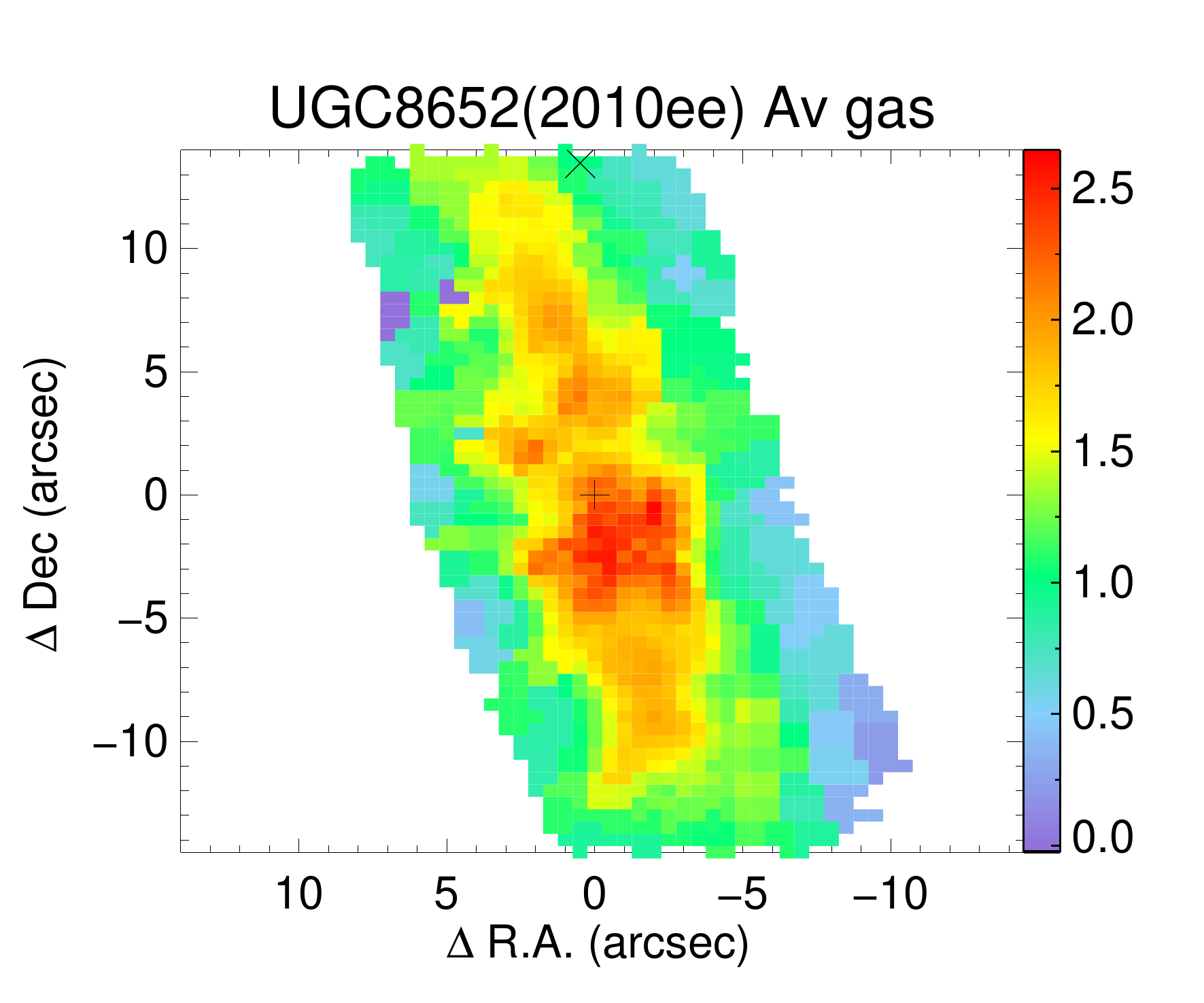}
\hspace{-0.2cm}
\includegraphics[width=30mm]{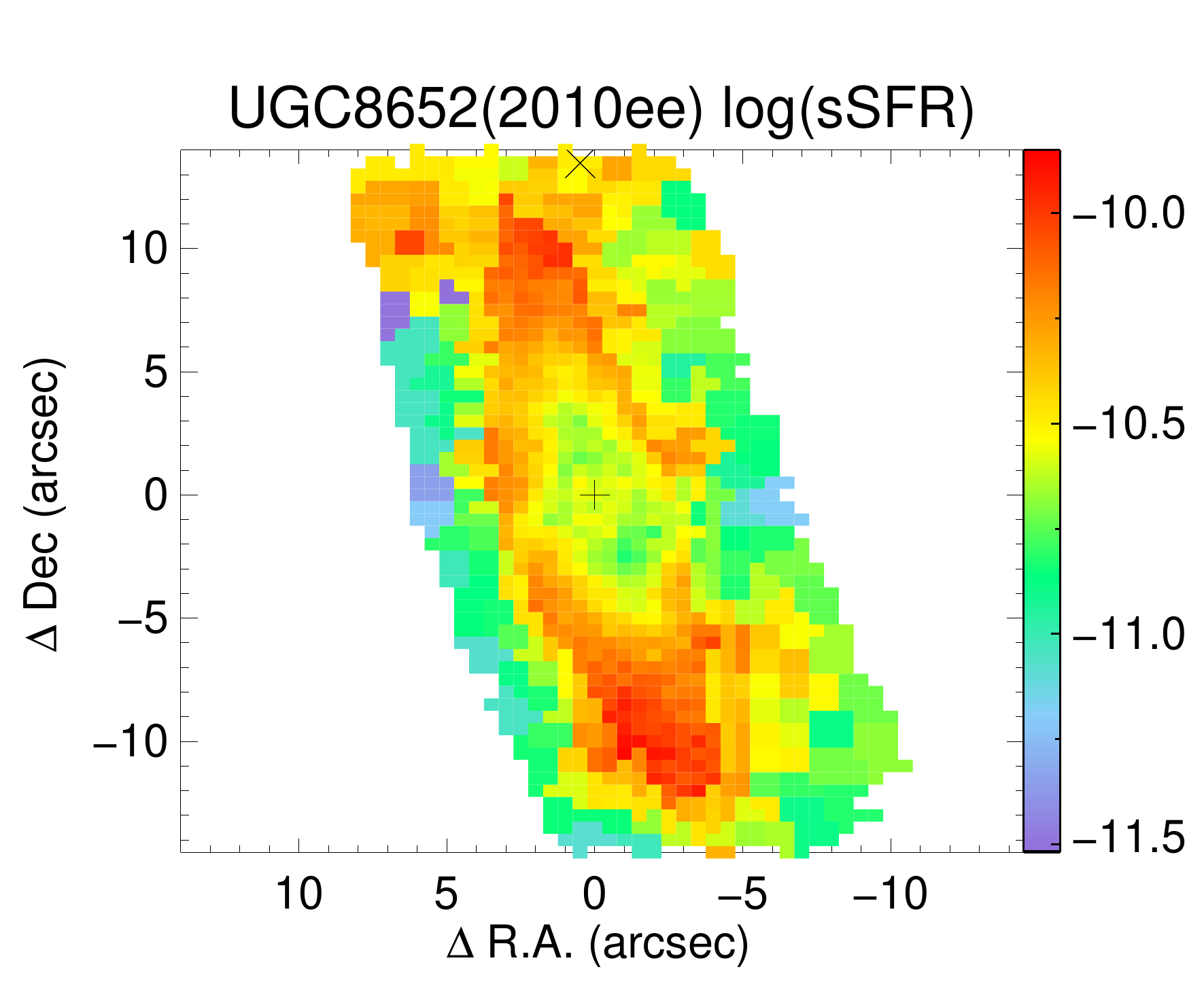}
\hspace{0.2cm}
\includegraphics[width=30mm]{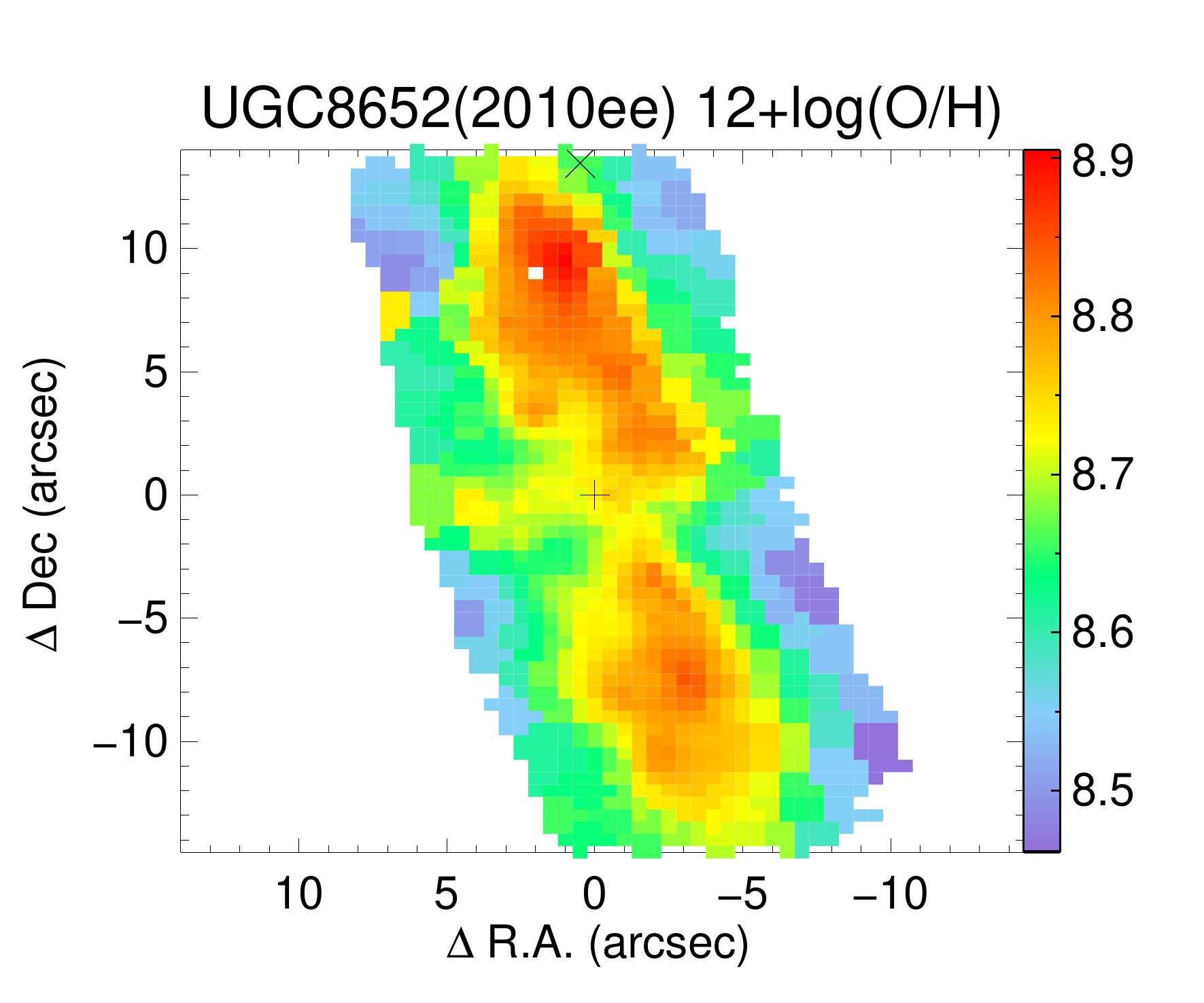}
\hspace{0.2cm}
\includegraphics[width=30mm]{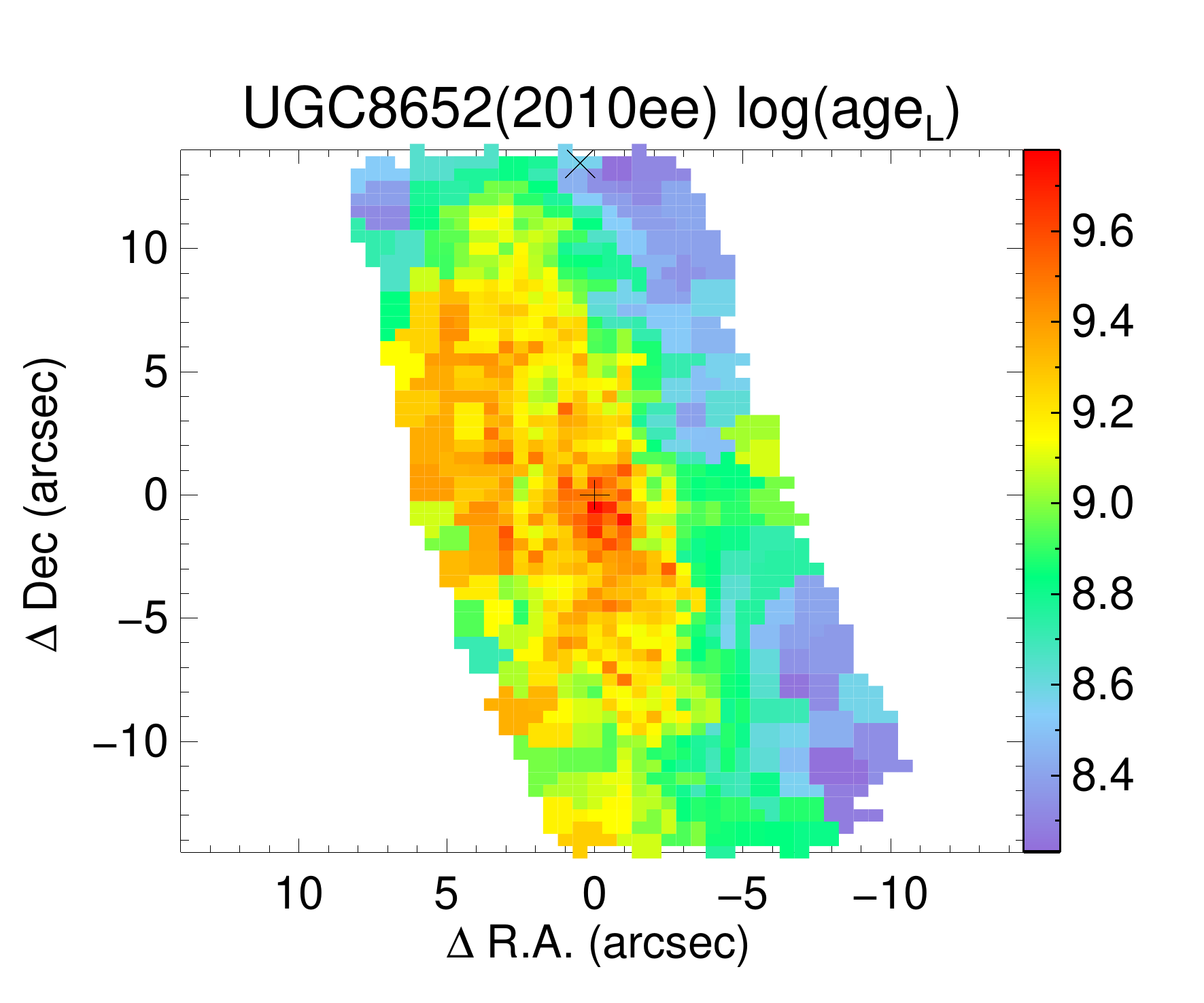}
\end{minipage}
\begin{minipage}{\textwidth}
\hspace{1.5cm}
\includegraphics[width=20mm]{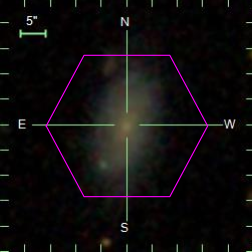}
\hspace{0.2cm}
\includegraphics[width=30mm]{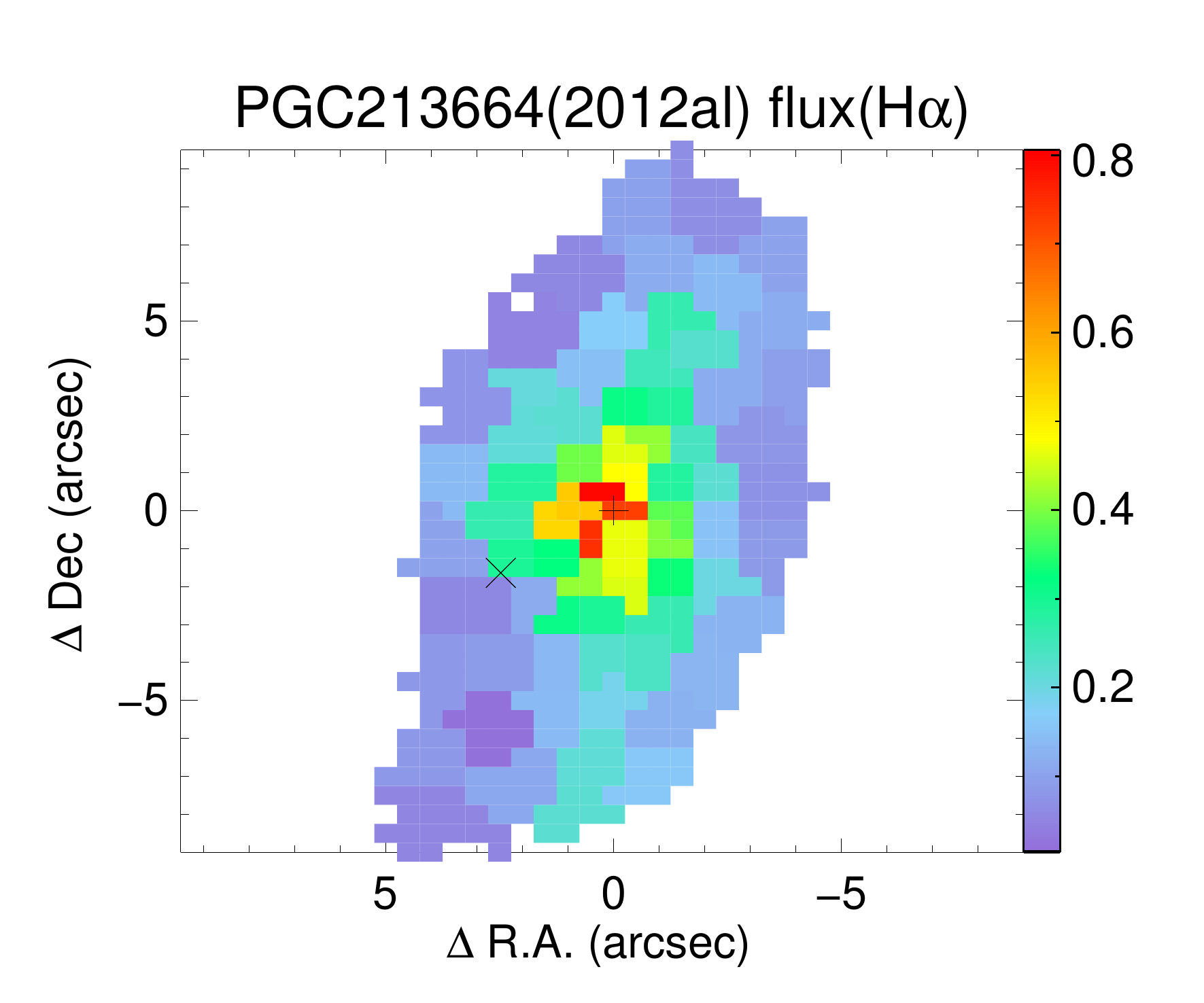}
\hspace{0.2cm}
\includegraphics[width=30mm]{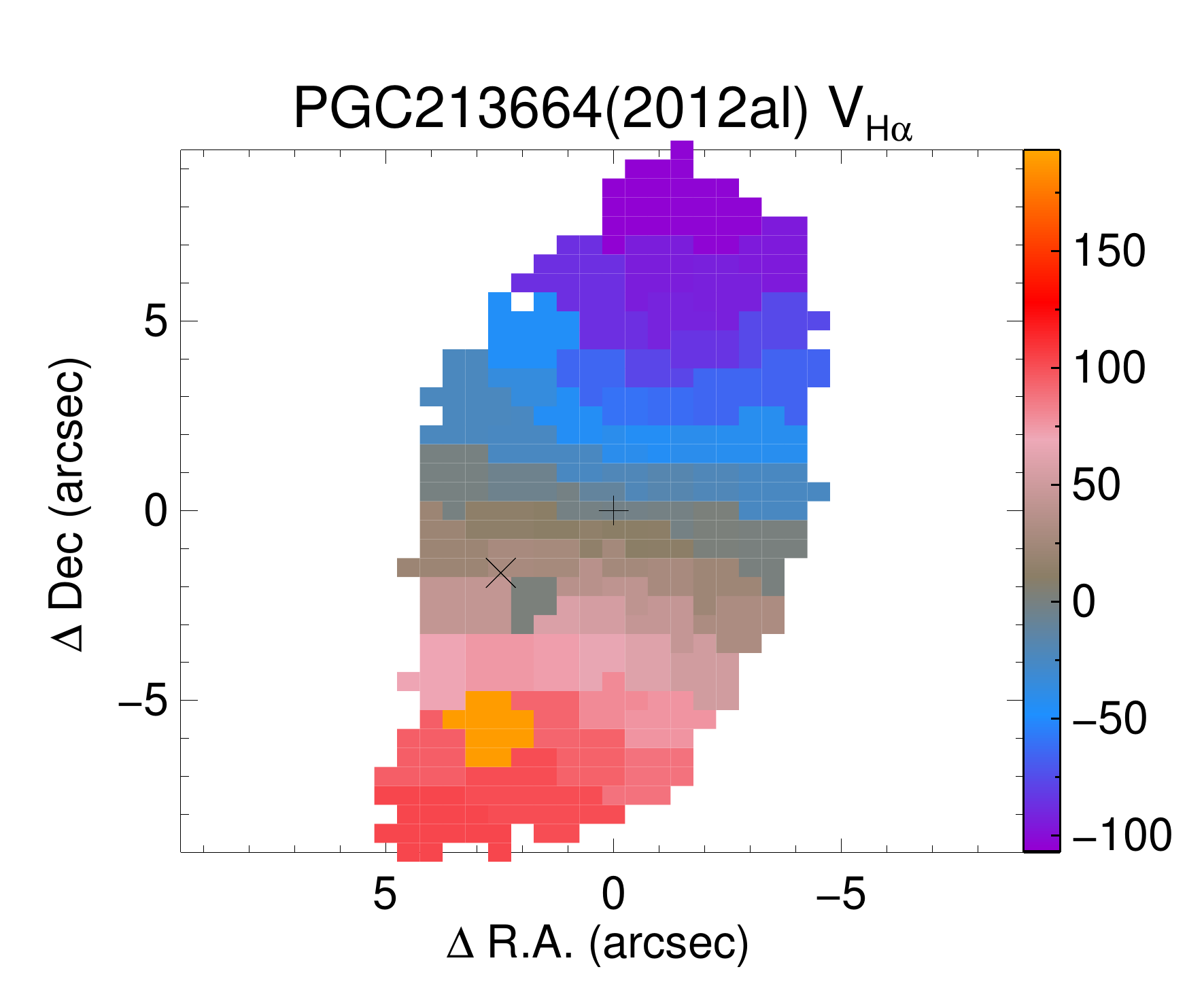}
\hspace{0.2cm}
\includegraphics[width=30mm]{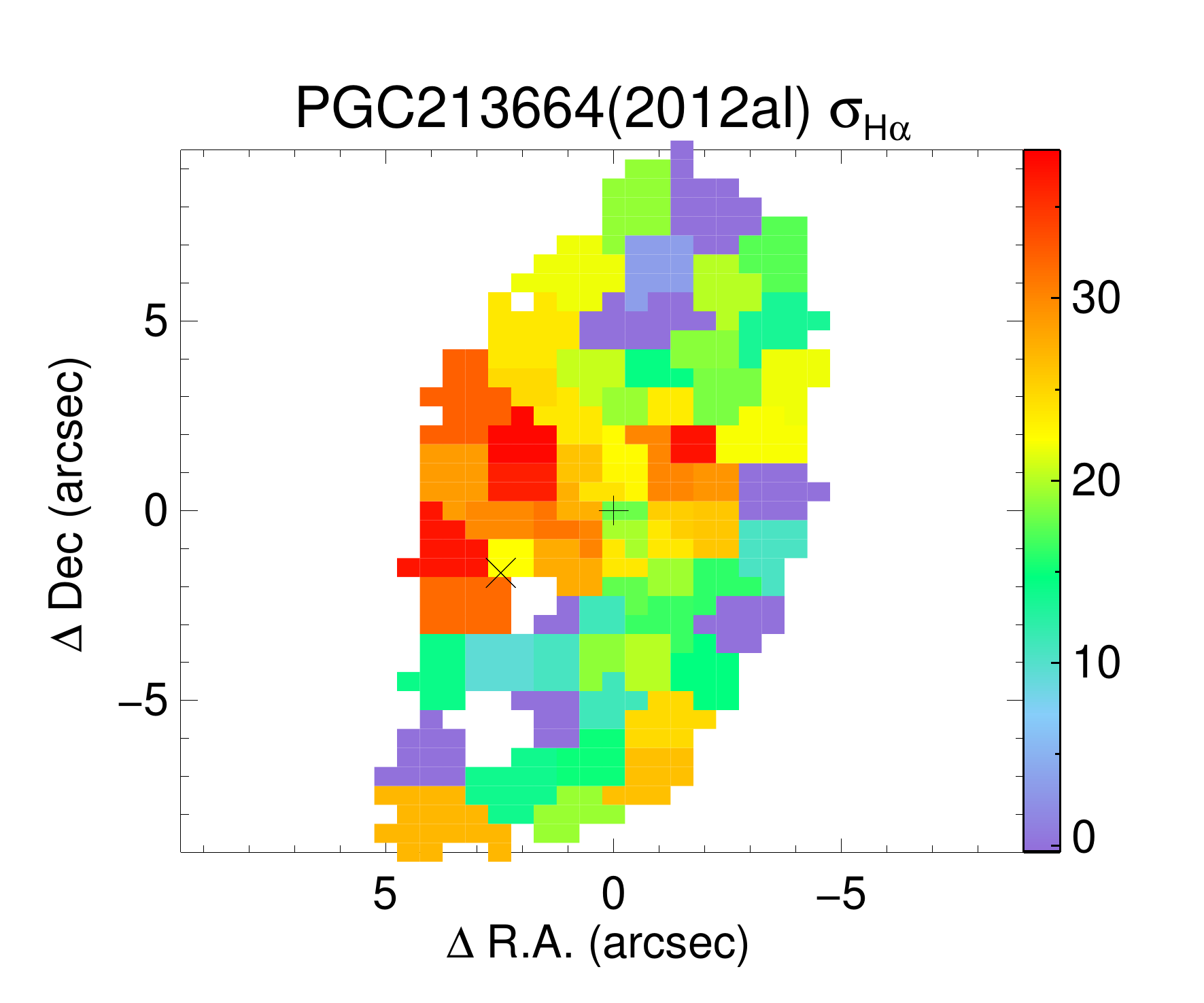}
\end{minipage}
\begin{minipage}{\textwidth}
\hspace{0.9cm}
\includegraphics[width=30mm]{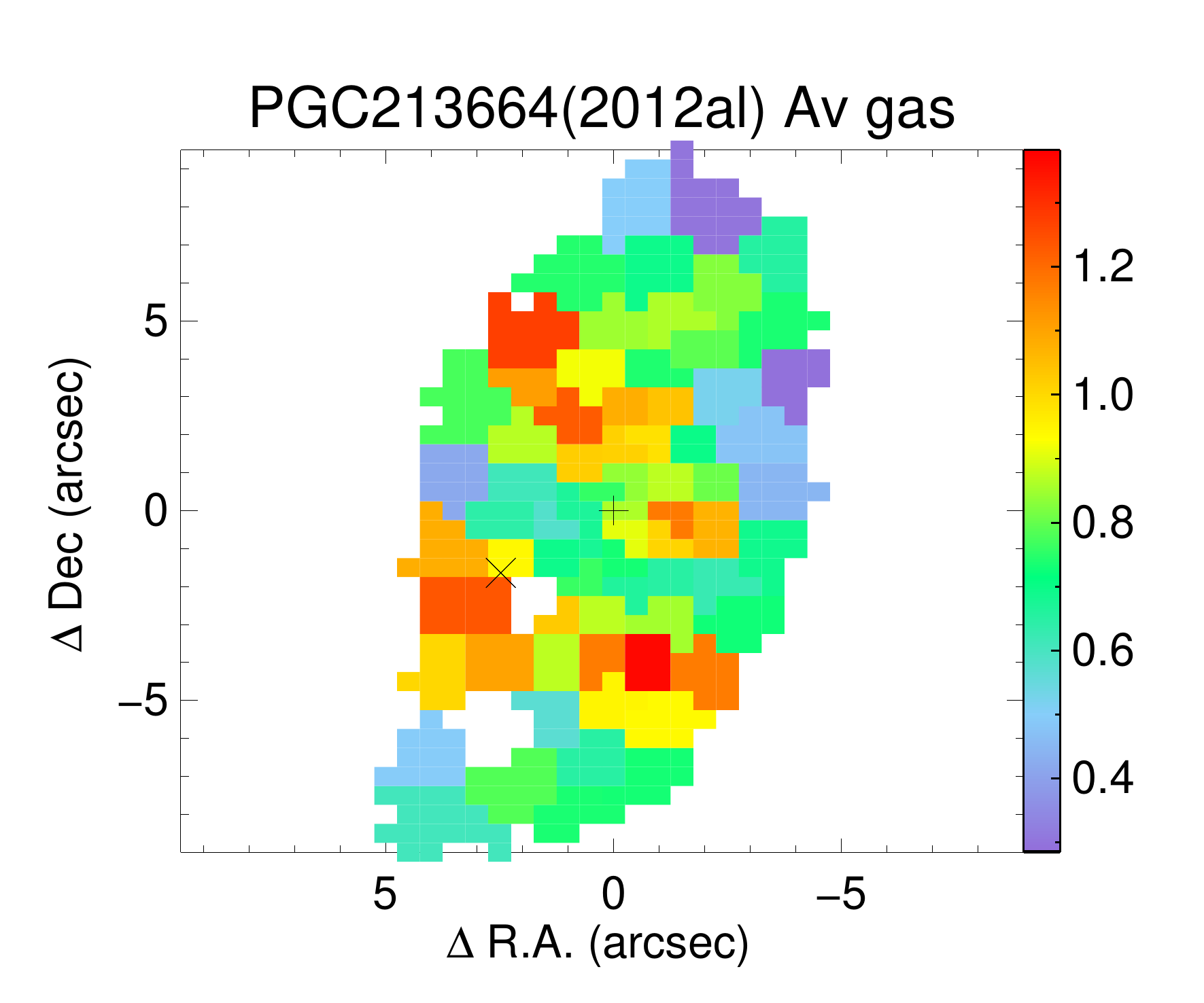}
\hspace{-0.2cm}
\includegraphics[width=30mm]{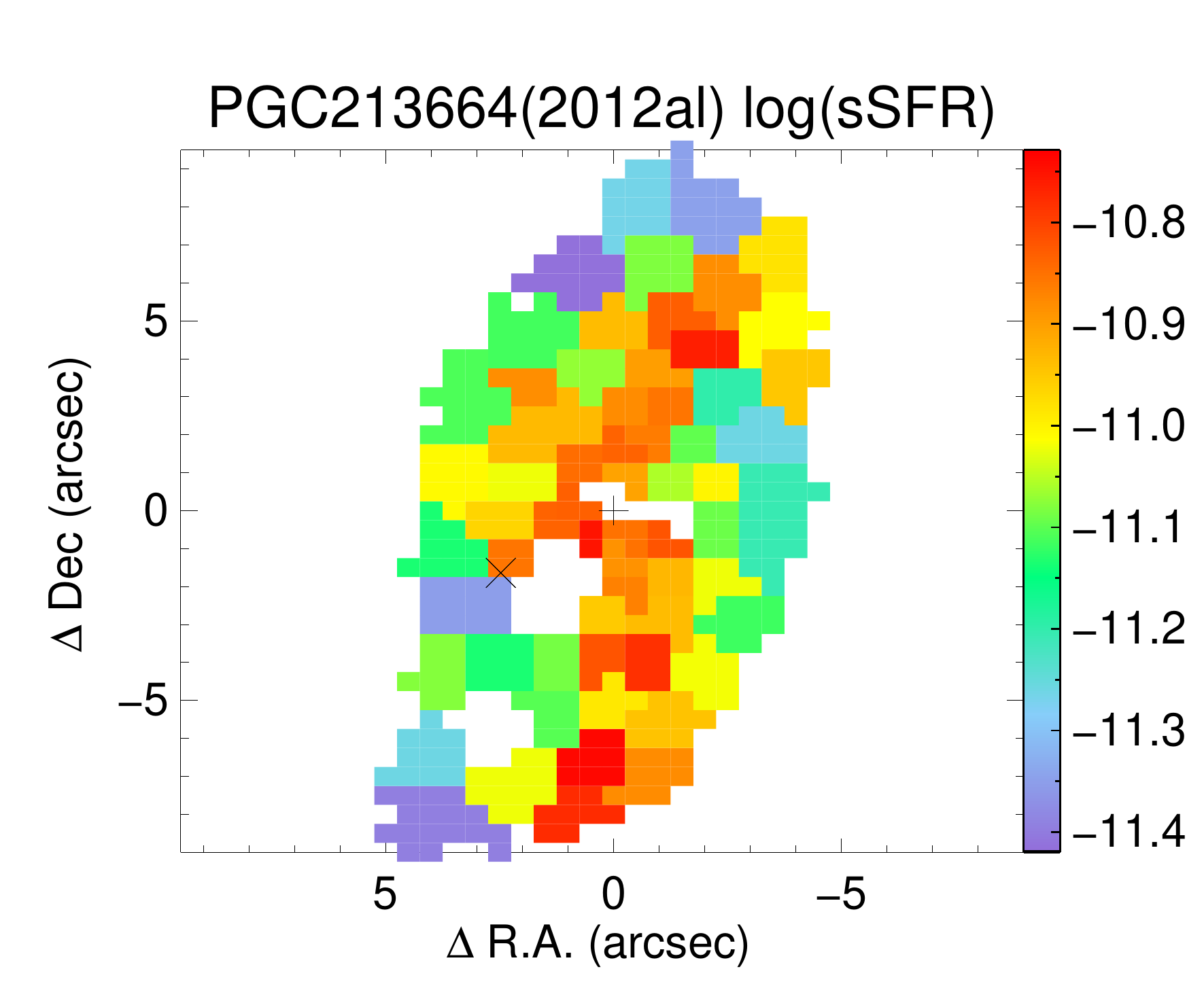}
\hspace{0.2cm}
\includegraphics[width=30mm]{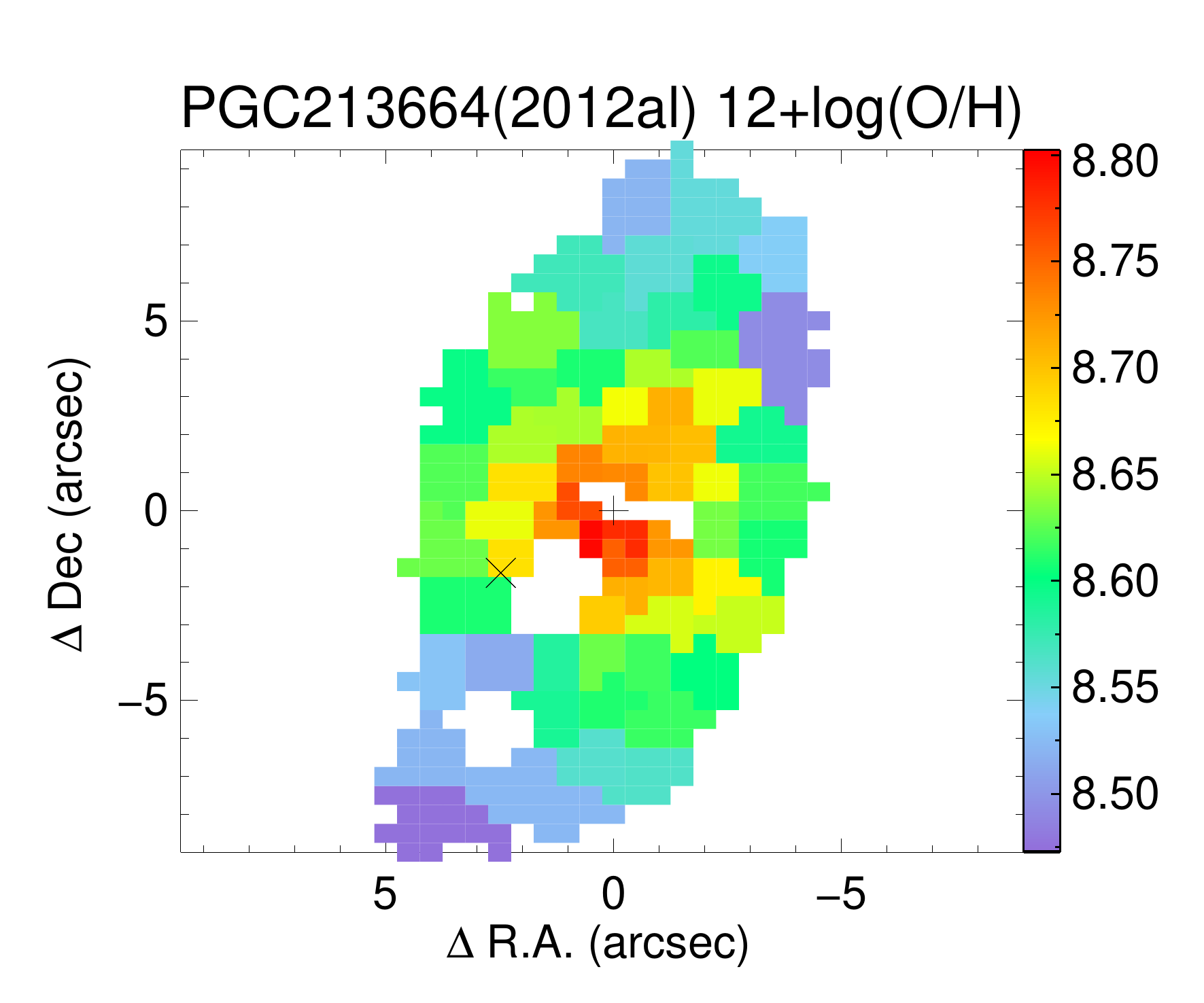}
\hspace{0.2cm}
\includegraphics[width=30mm]{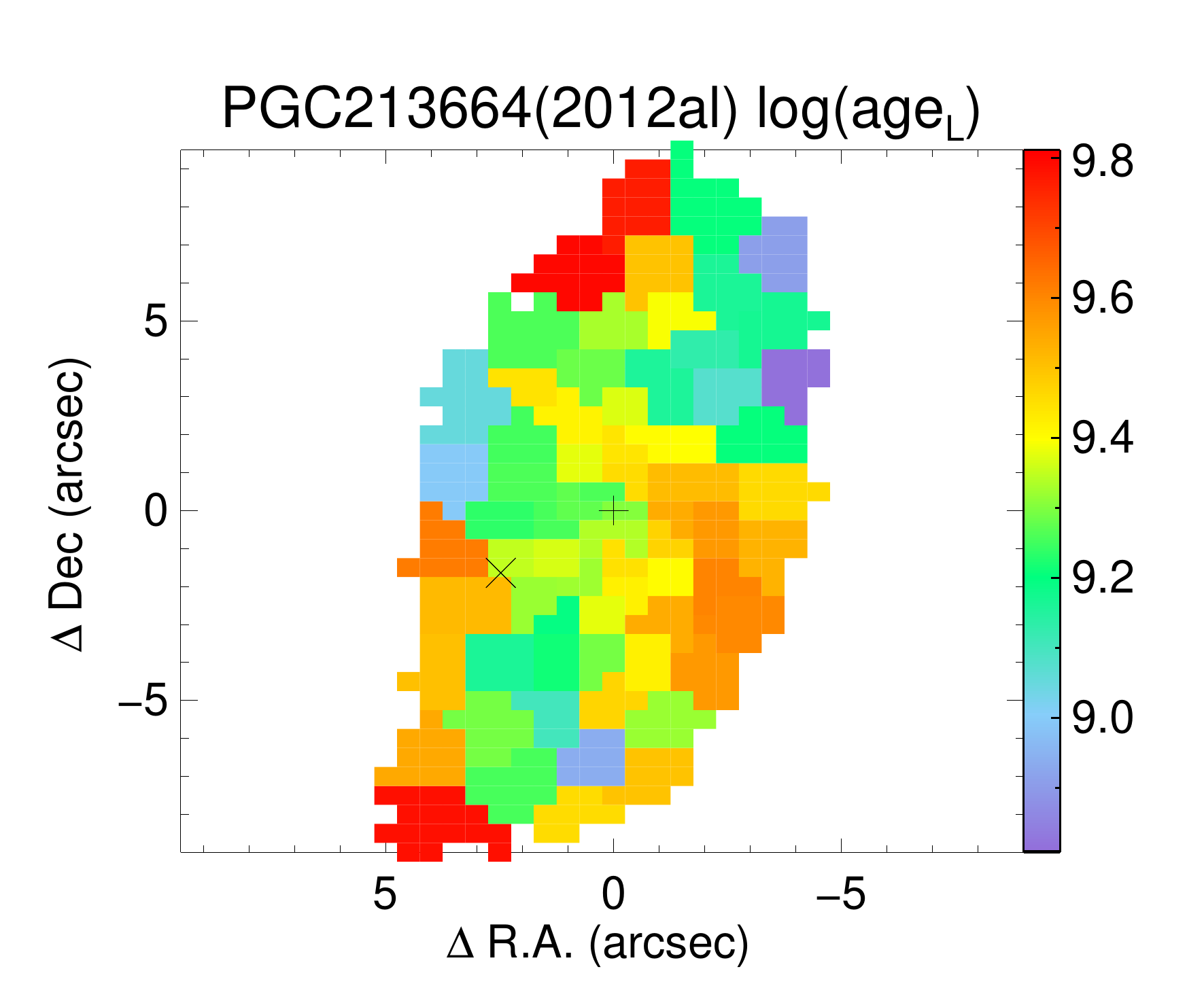}
\end{minipage}
\begin{minipage}{\textwidth}
\hspace{1.5cm}
\includegraphics[width=20mm]{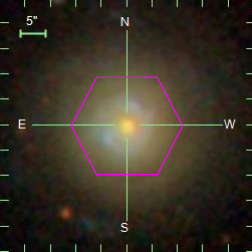}
\hspace{0.2cm}
\includegraphics[width=30mm]{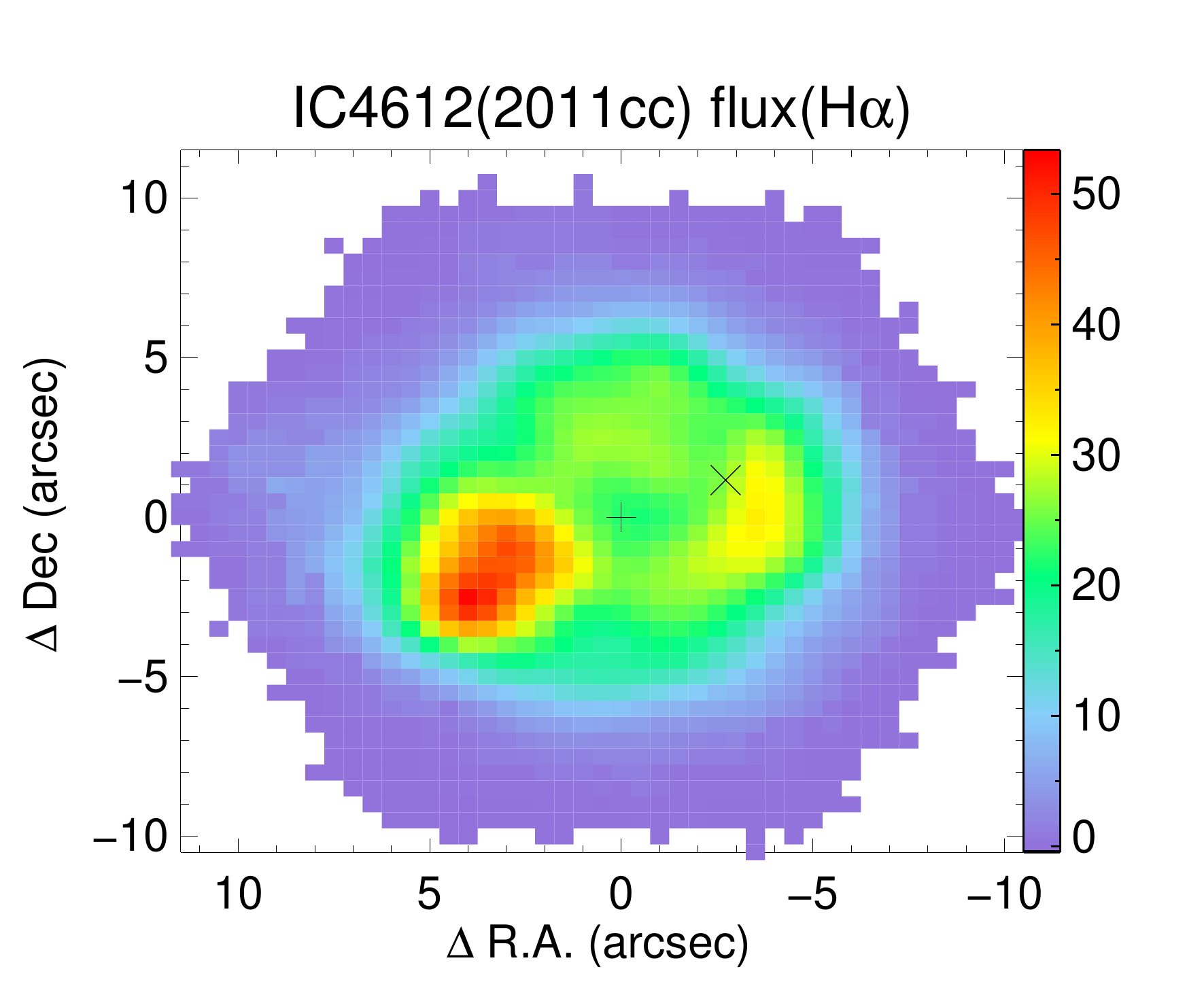}
\hspace{0.2cm}
\includegraphics[width=30mm]{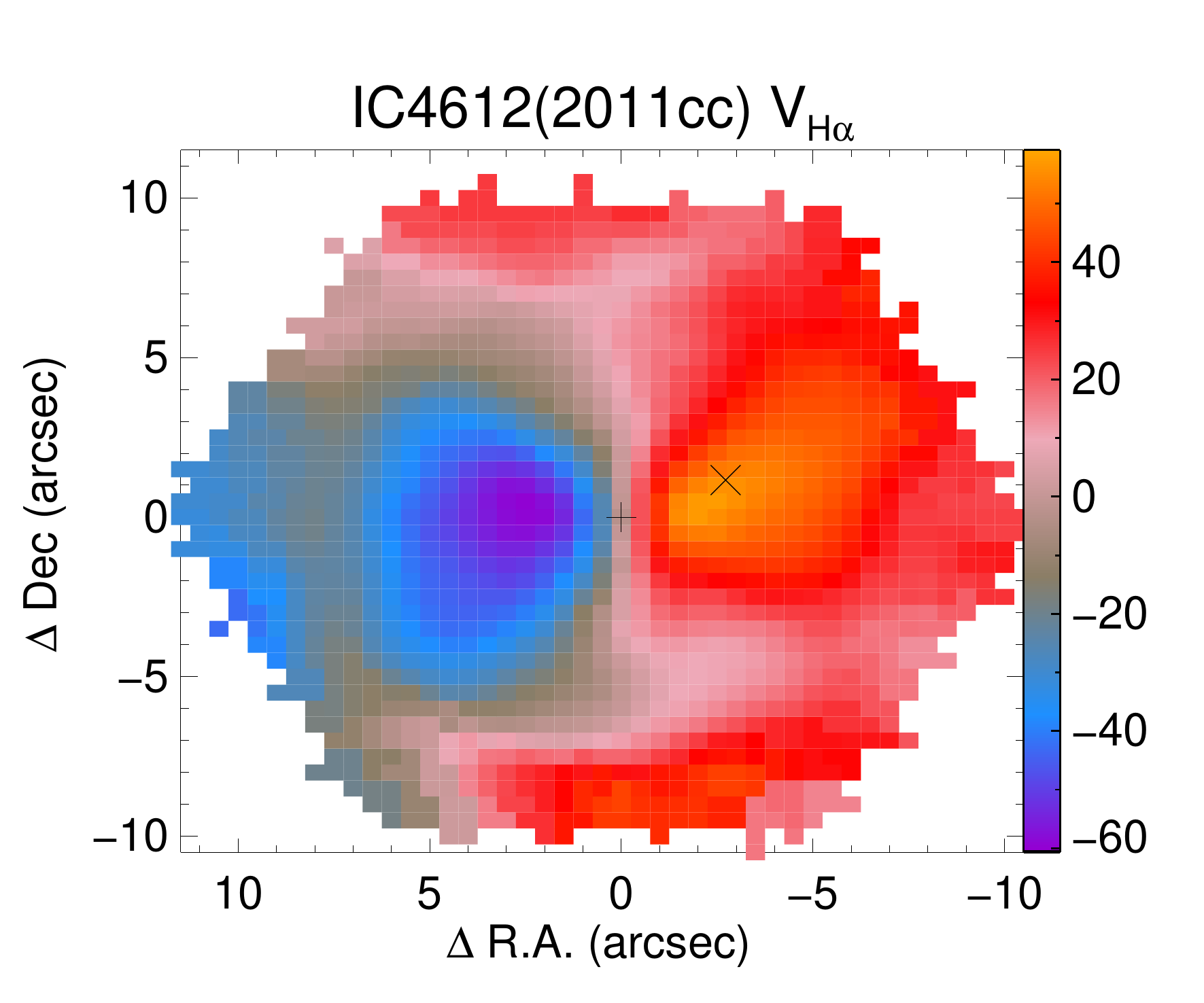}
\hspace{0.2cm}
\includegraphics[width=30mm]{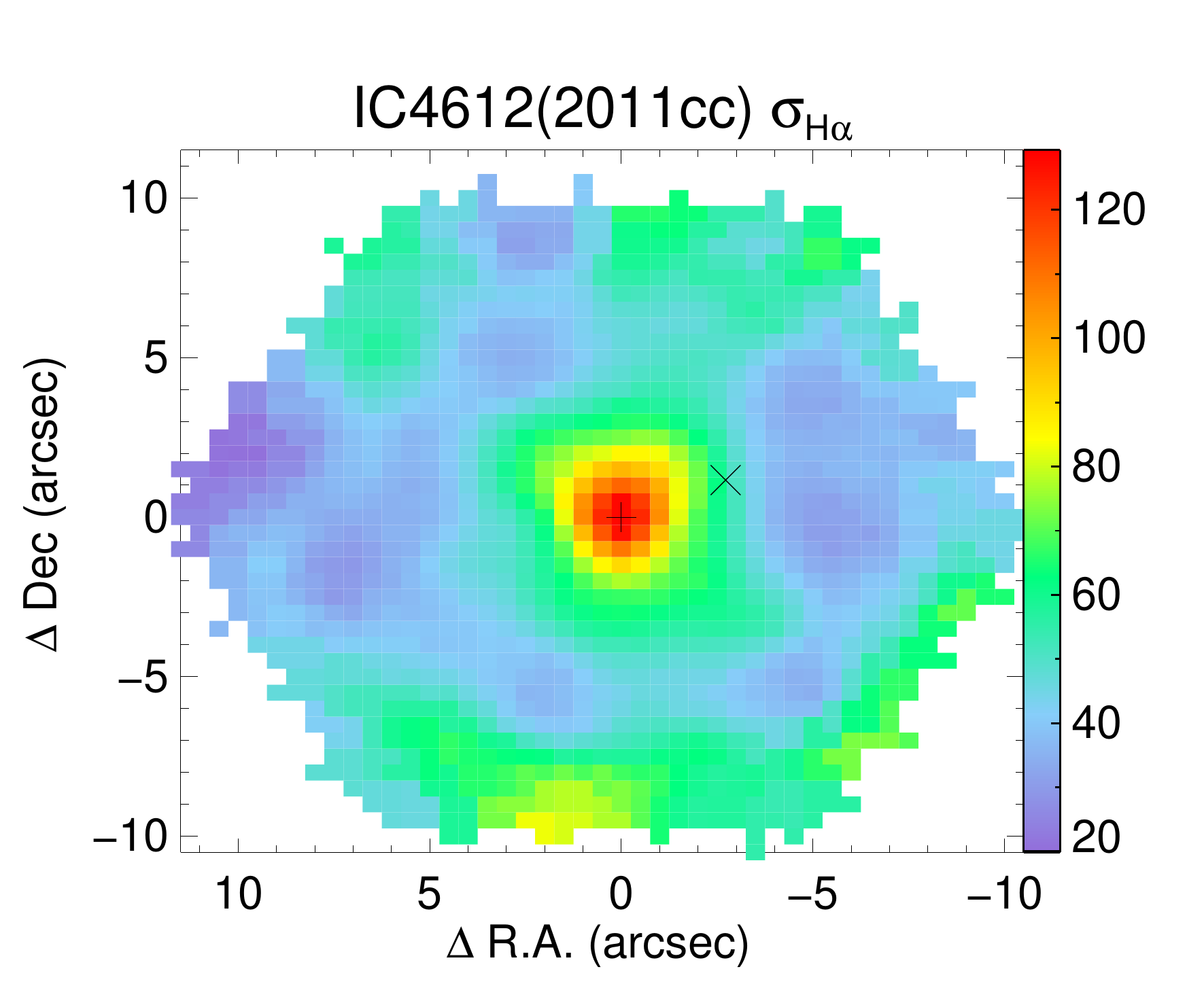}
\end{minipage}
\begin{minipage}{\textwidth}
\hspace{0.9cm}
\includegraphics[width=30mm]{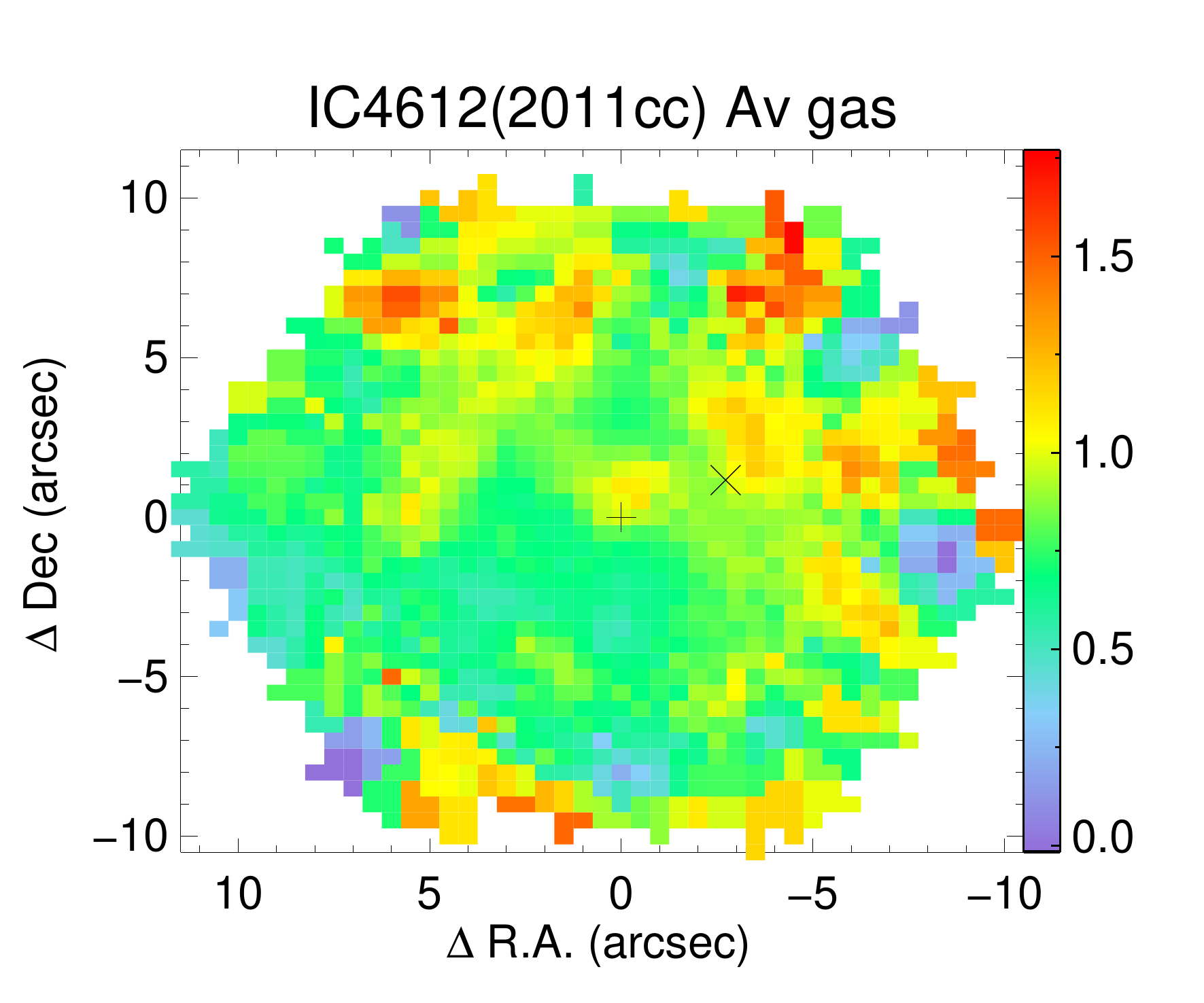}
\hspace{-0.2cm}
\includegraphics[width=30mm]{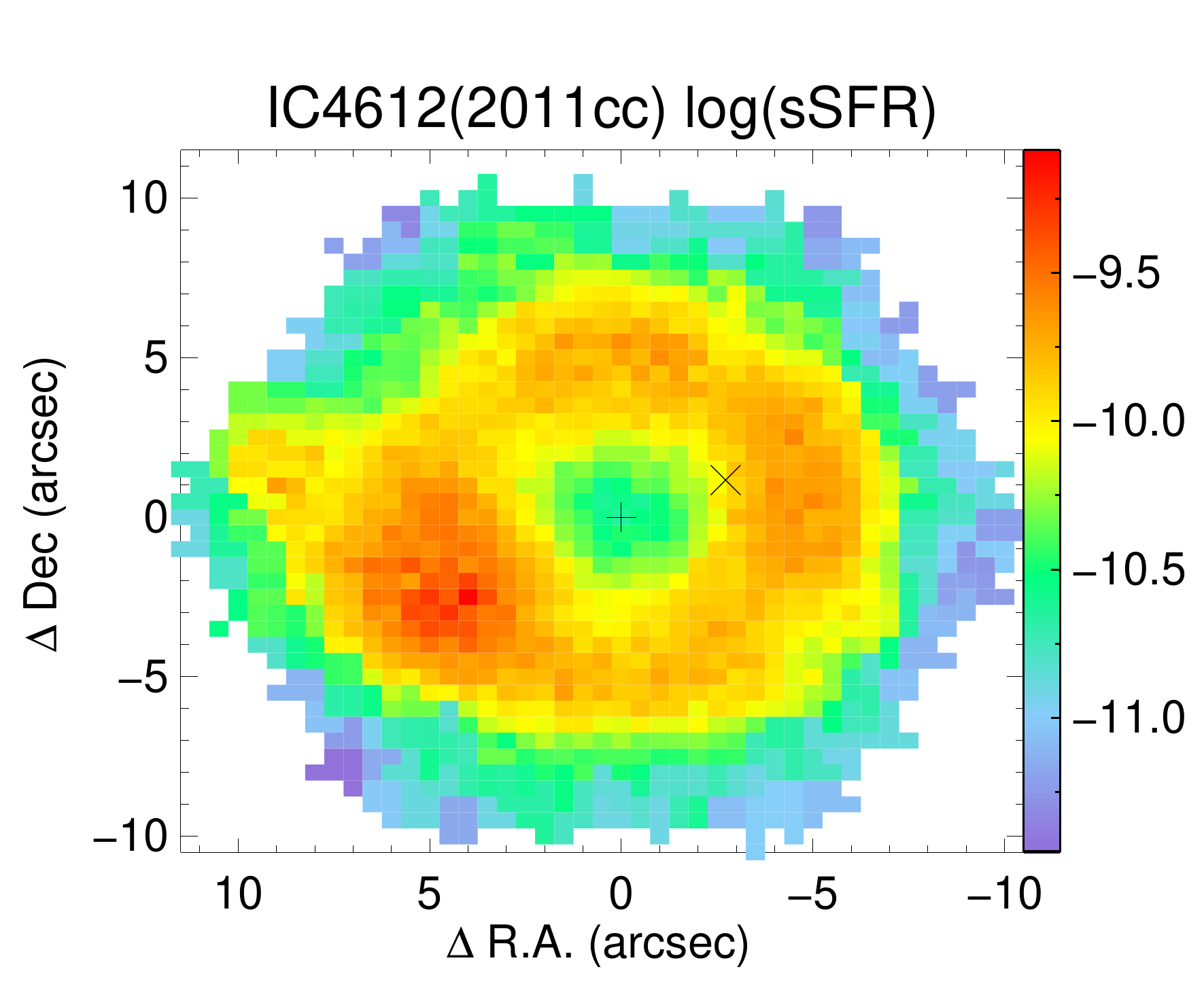}
\hspace{0.2cm}
\includegraphics[width=30mm]{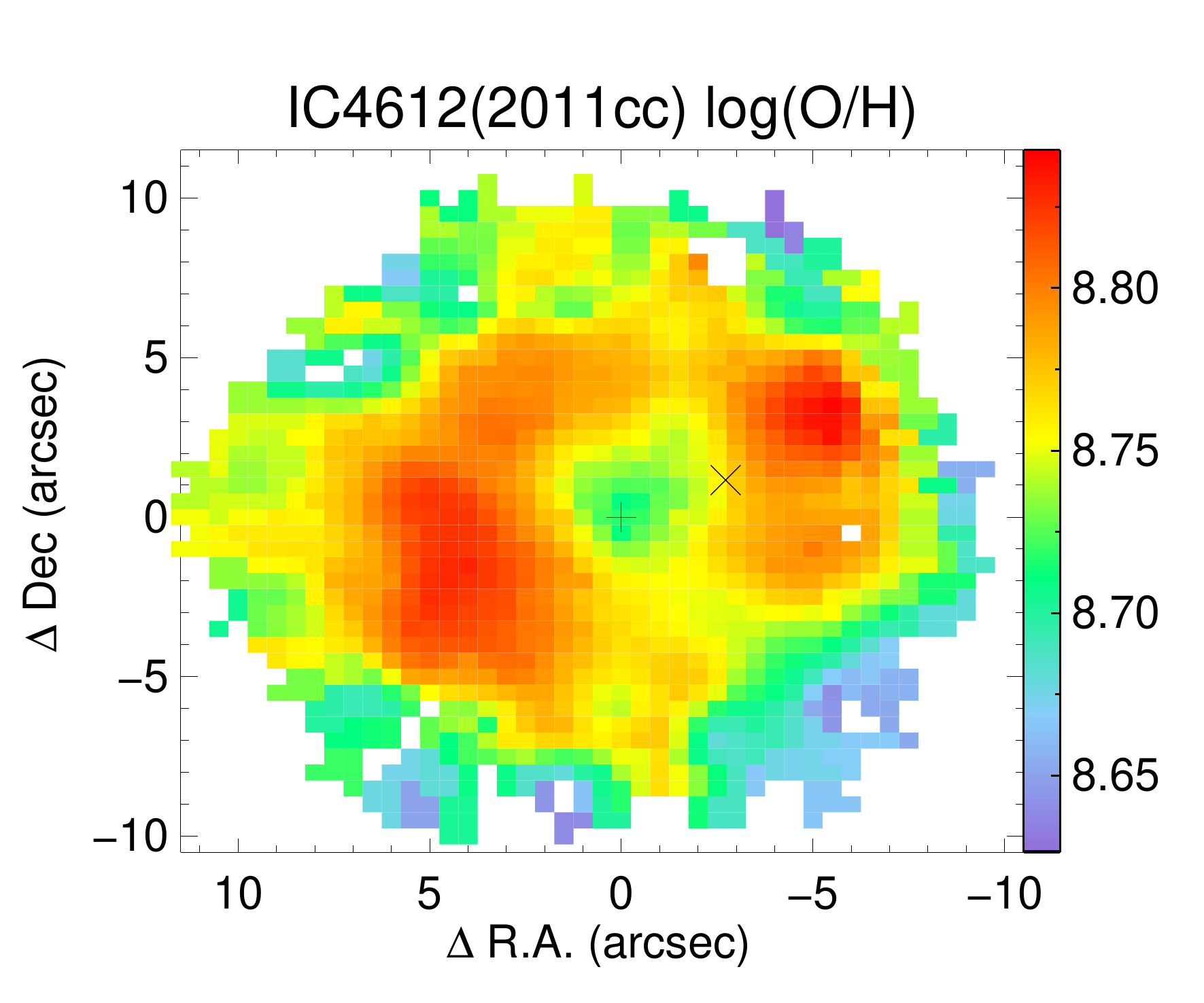}
\hspace{0.2cm}
\includegraphics[width=30mm]{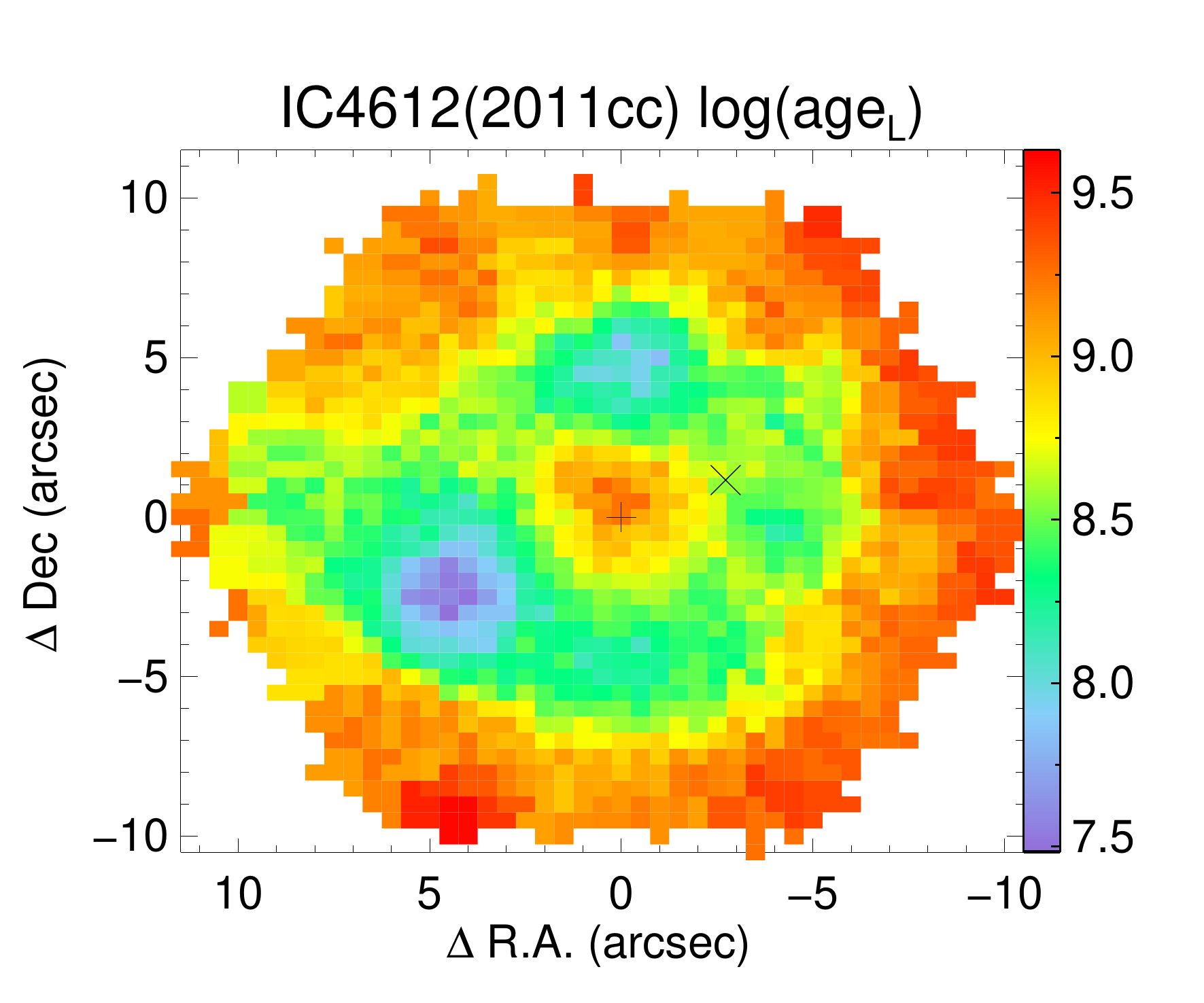}
\end{minipage}
\end{figure*}

\begin{figure*}[]
\centering
\begin{minipage}{\textwidth}
\hspace{1.5cm}
\includegraphics[width=20mm]{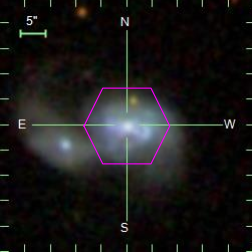}
\hspace{0.2cm}
\includegraphics[width=30mm]{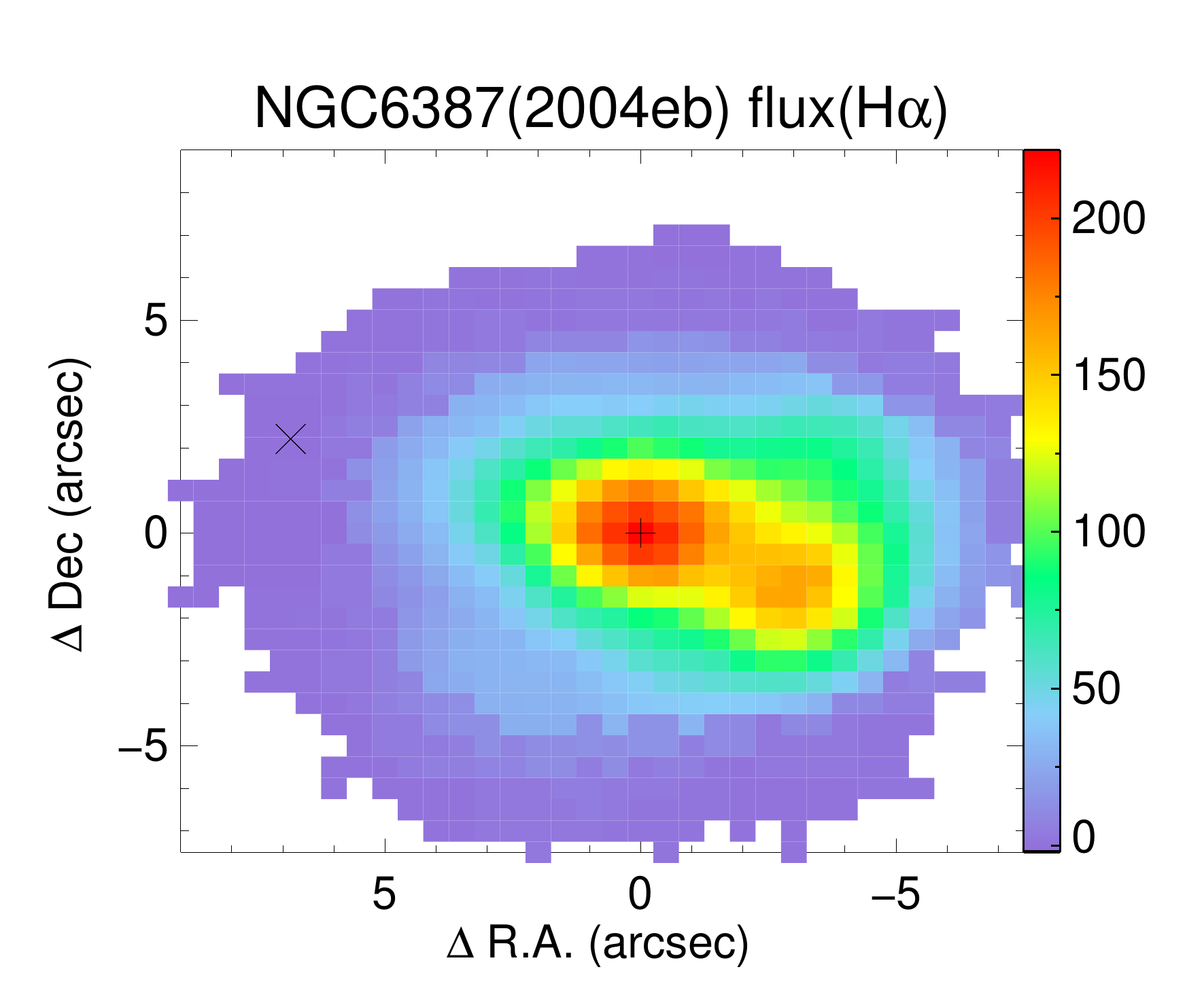}
\hspace{0.2cm}
\includegraphics[width=30mm]{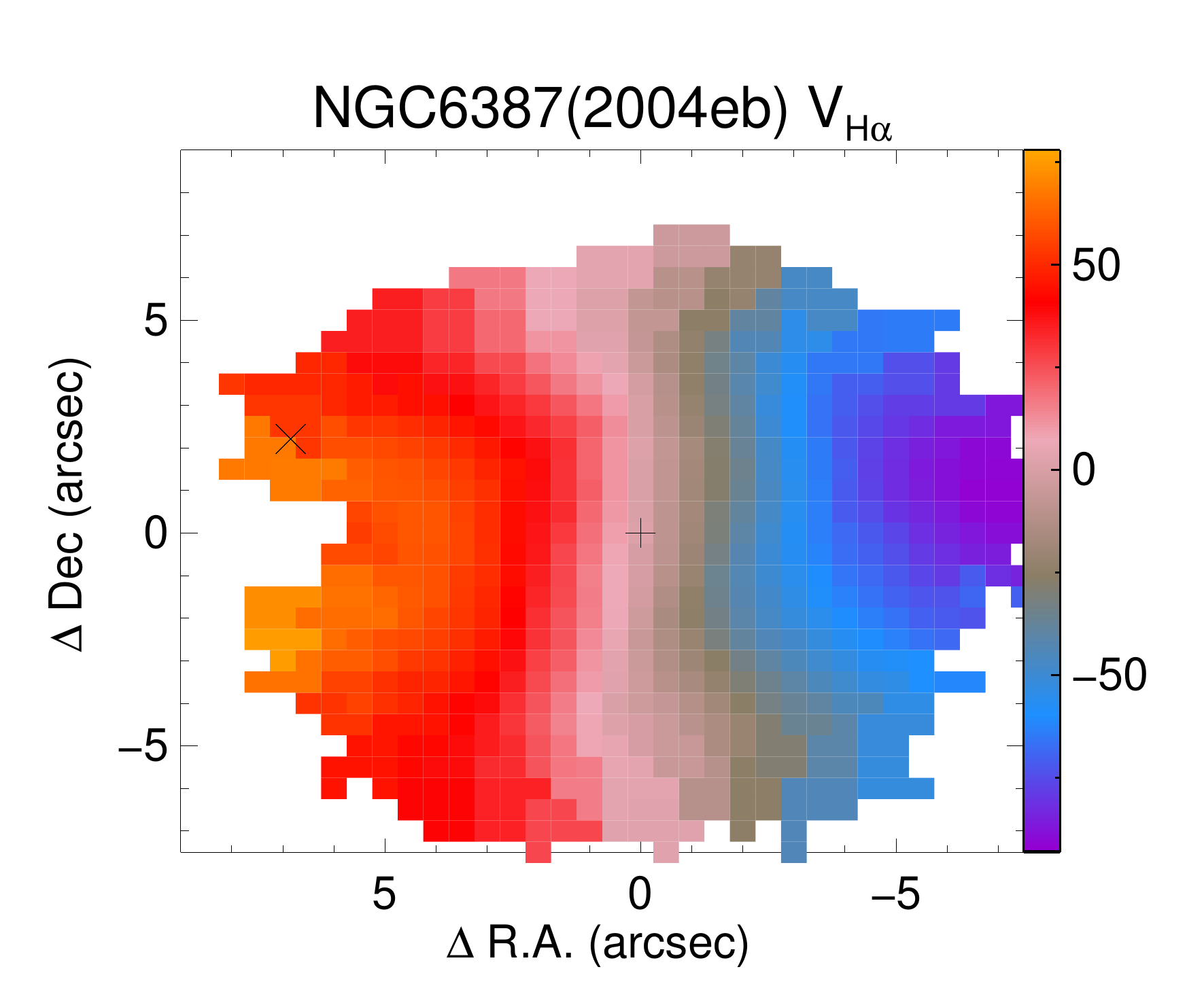}
\hspace{0.2cm}
\includegraphics[width=30mm]{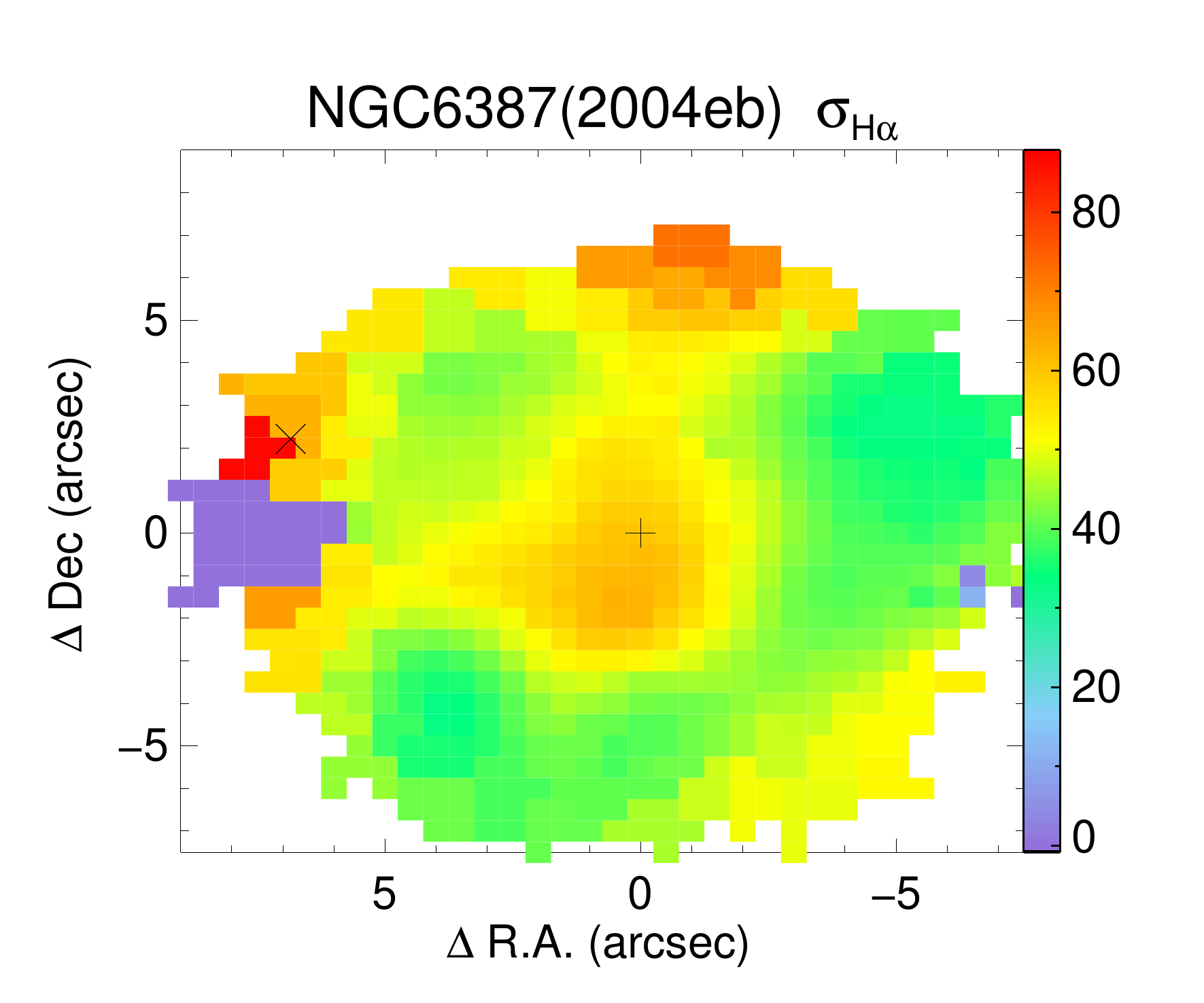}
\end{minipage}
\begin{minipage}{\textwidth}
\hspace{0.9cm}
\includegraphics[width=30mm]{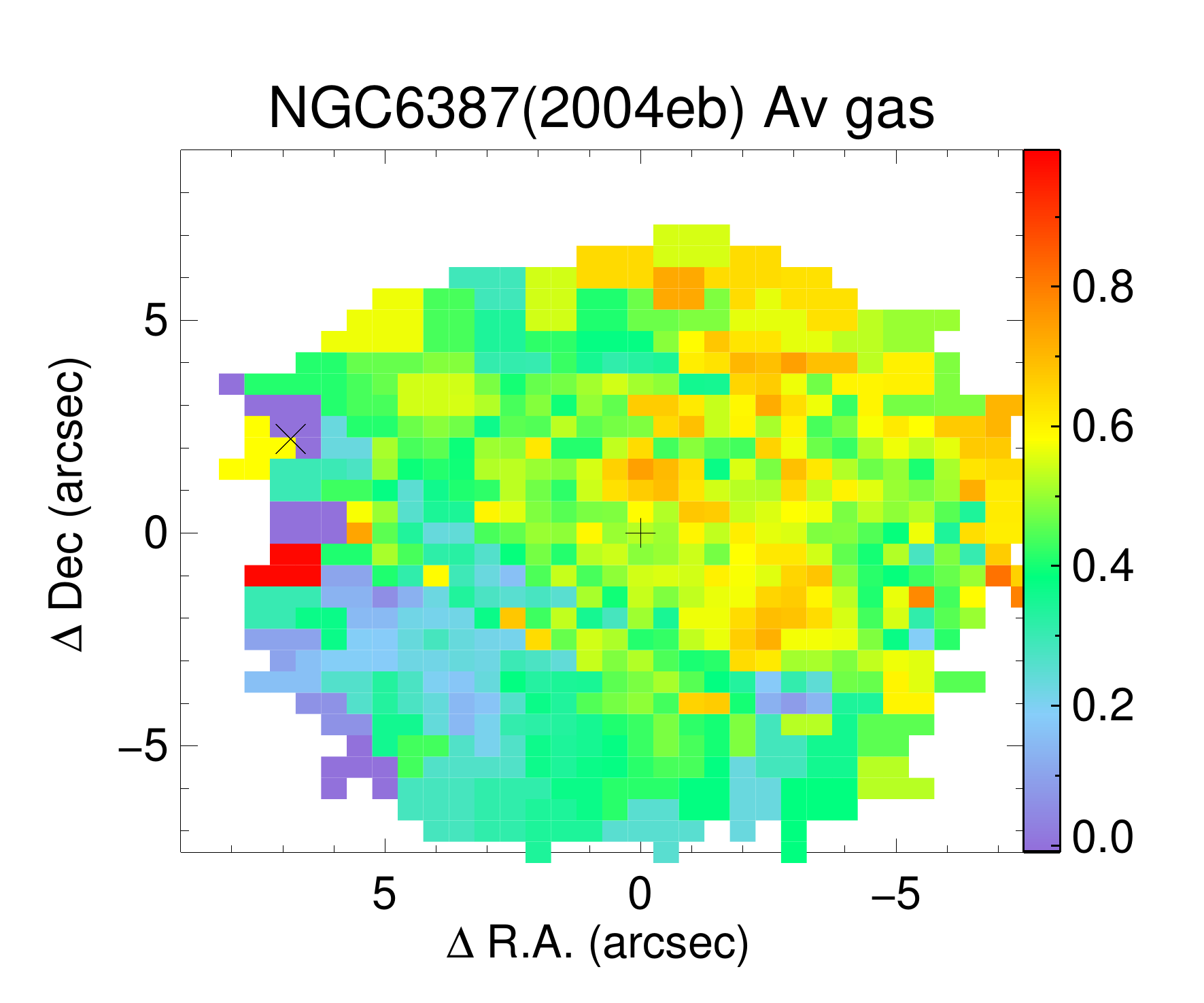}
\hspace{-0.2cm}
\includegraphics[width=30mm]{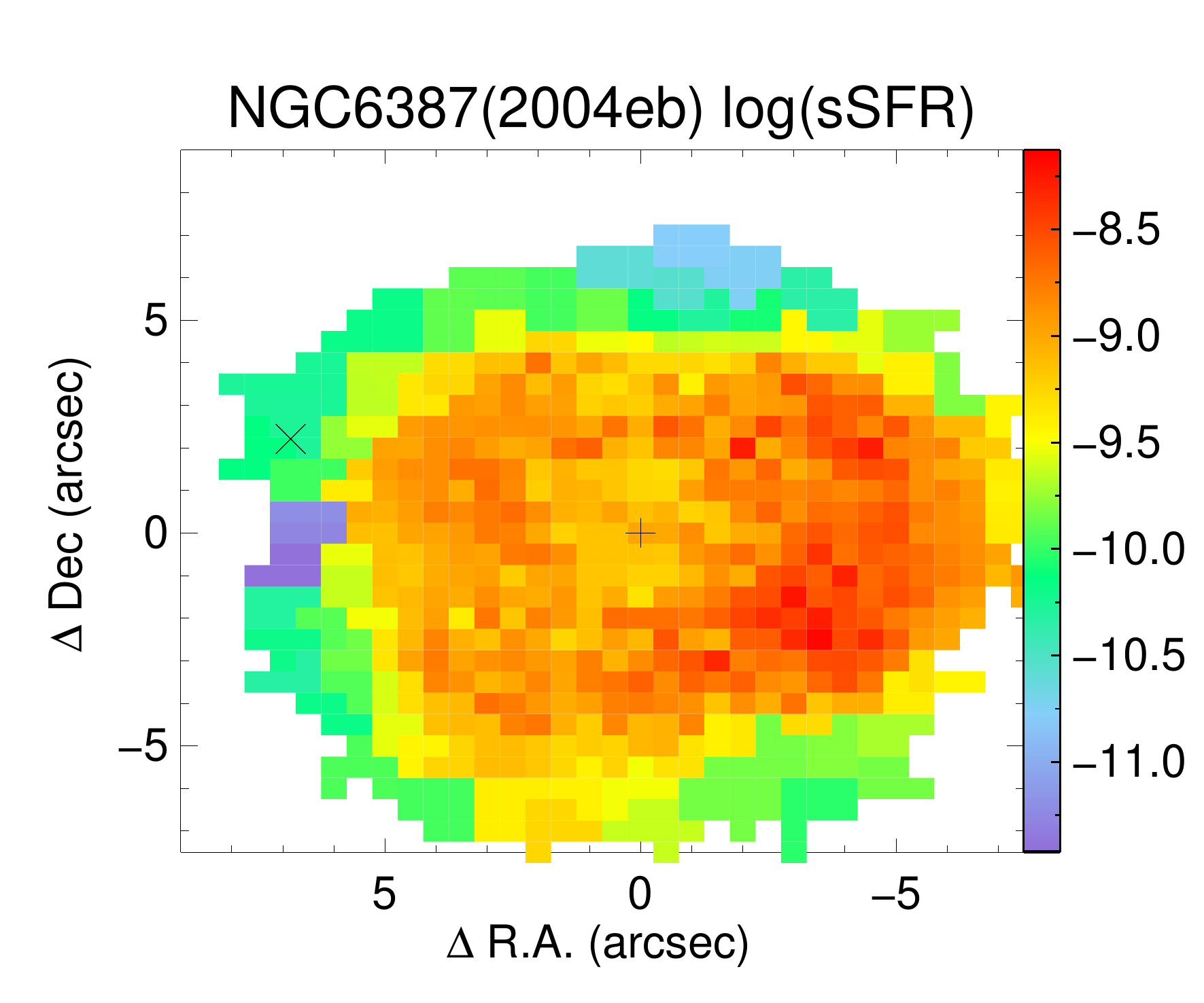}
\hspace{0.2cm}
\includegraphics[width=30mm]{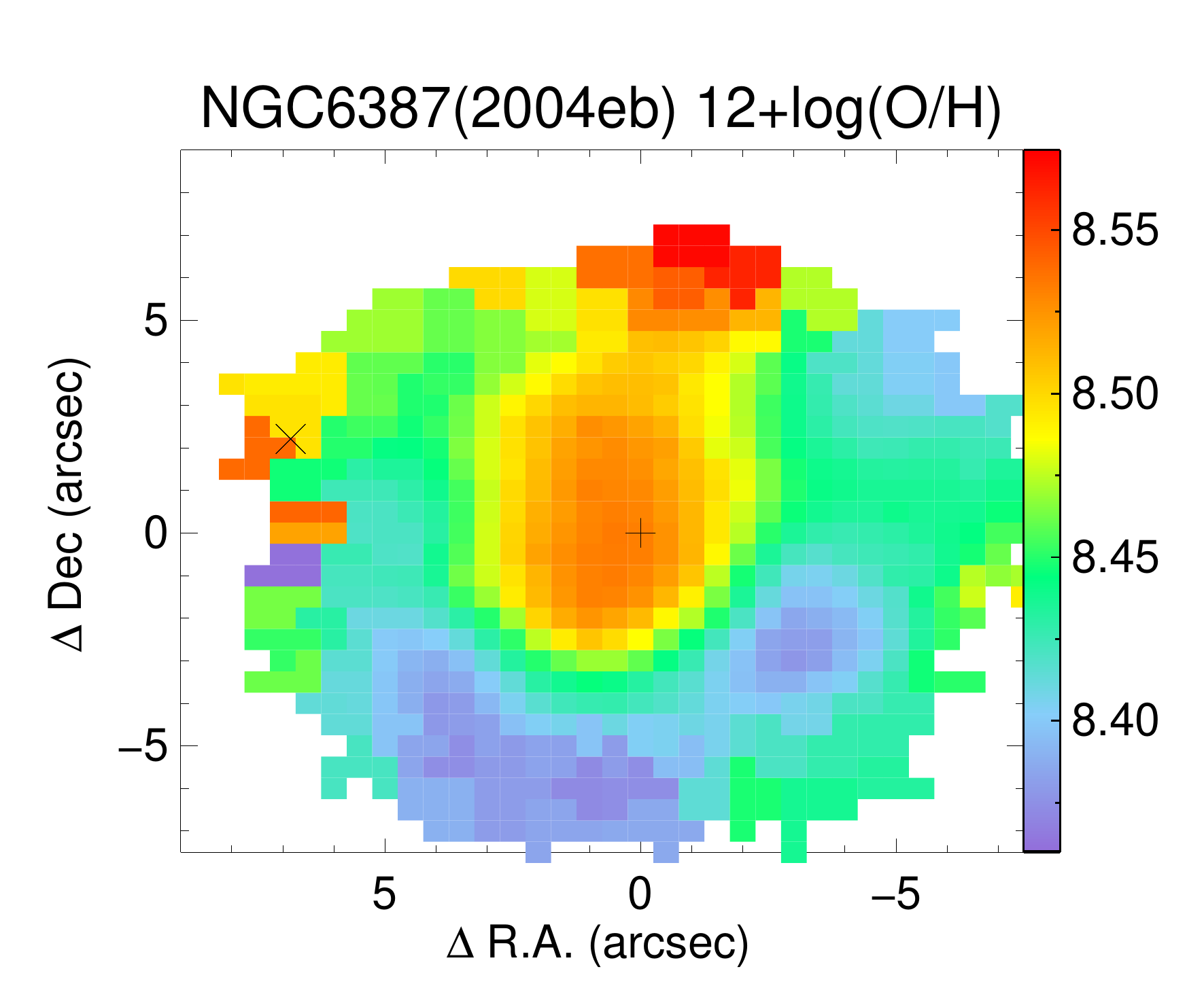}
\hspace{0.2cm}
\includegraphics[width=30mm]{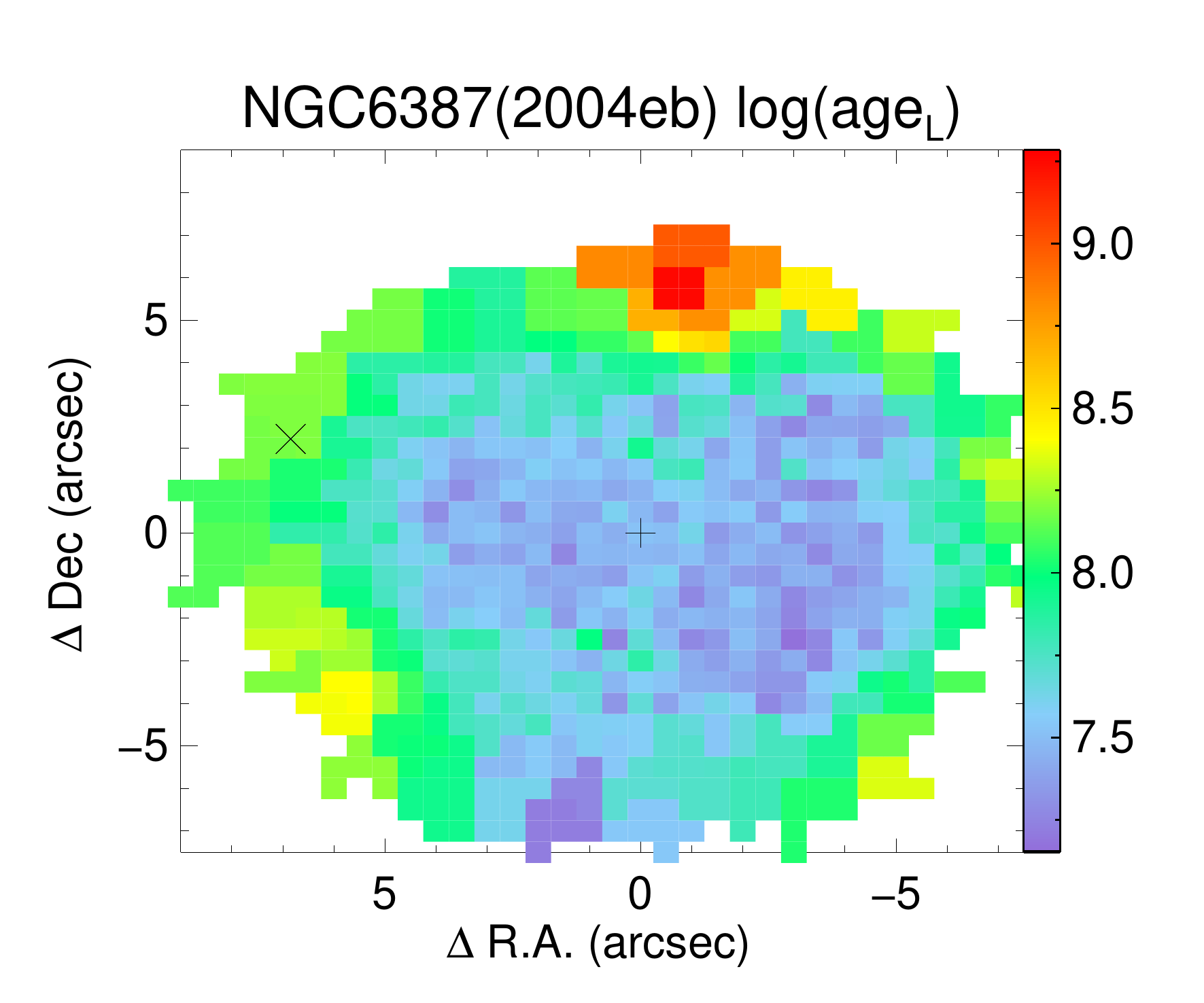}
\end{minipage}
\caption{Same as Fig.~\ref{fig.SNIa-havd}, but for SN II host galaxies. There are only several spaxels left by limiting the S/N $\geq$ 3 for the SN 2000cs  explosion site in the host galaxy. So, we do not cut spaxels with S/N $\leq$ 3 in the sSFR and metallicity 2D maps for this SN host galaxy.}
\label{fig.SNII-havd}
\end{figure*}

\begin{figure*}[]
\centering
\begin{minipage}{\textwidth}
\hspace{1.5cm}
\includegraphics[width=20mm]{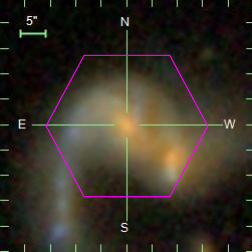}
\hspace{0.2cm}
\includegraphics[width=30mm]{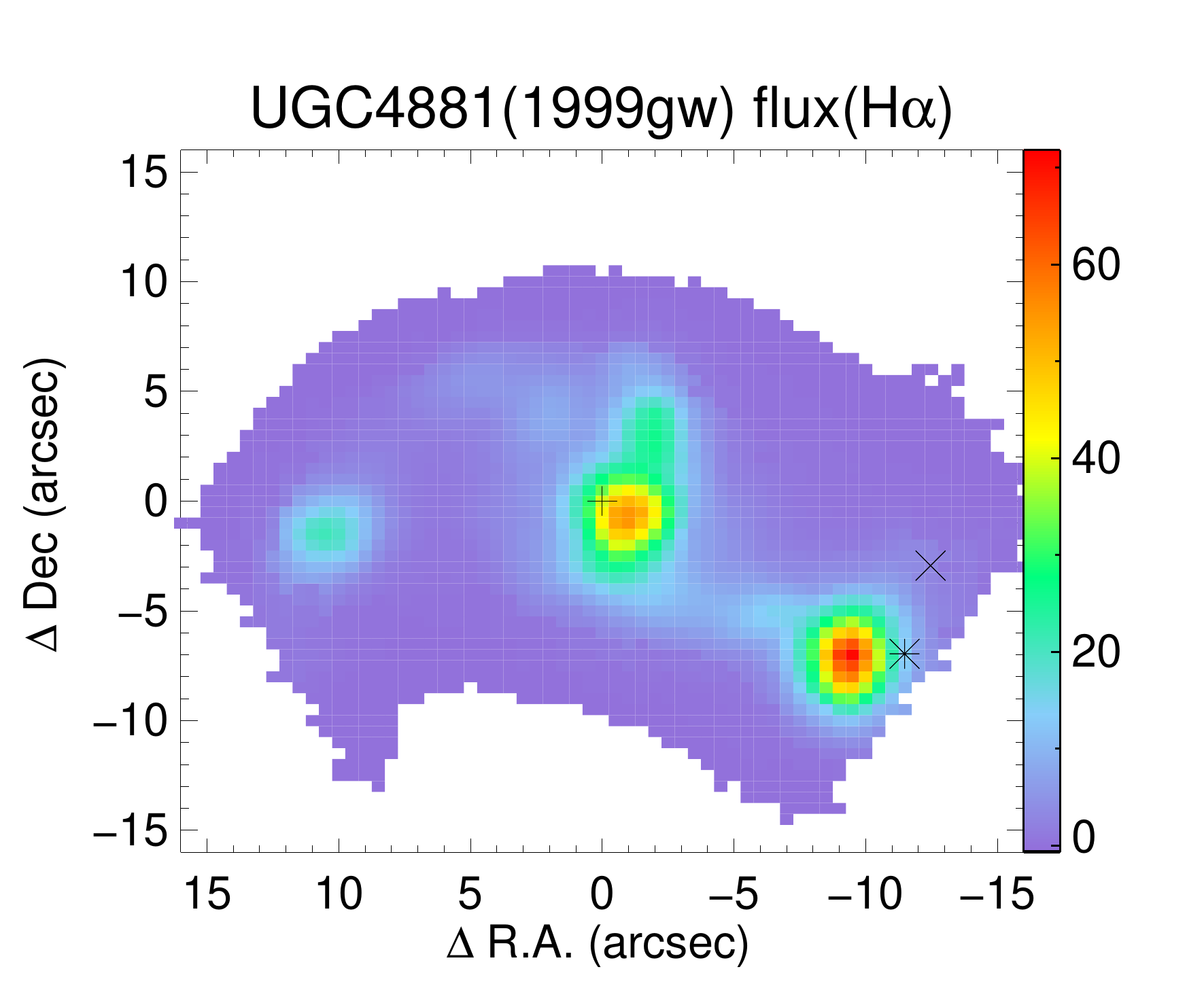}
\hspace{0.2cm}
\includegraphics[width=30mm]{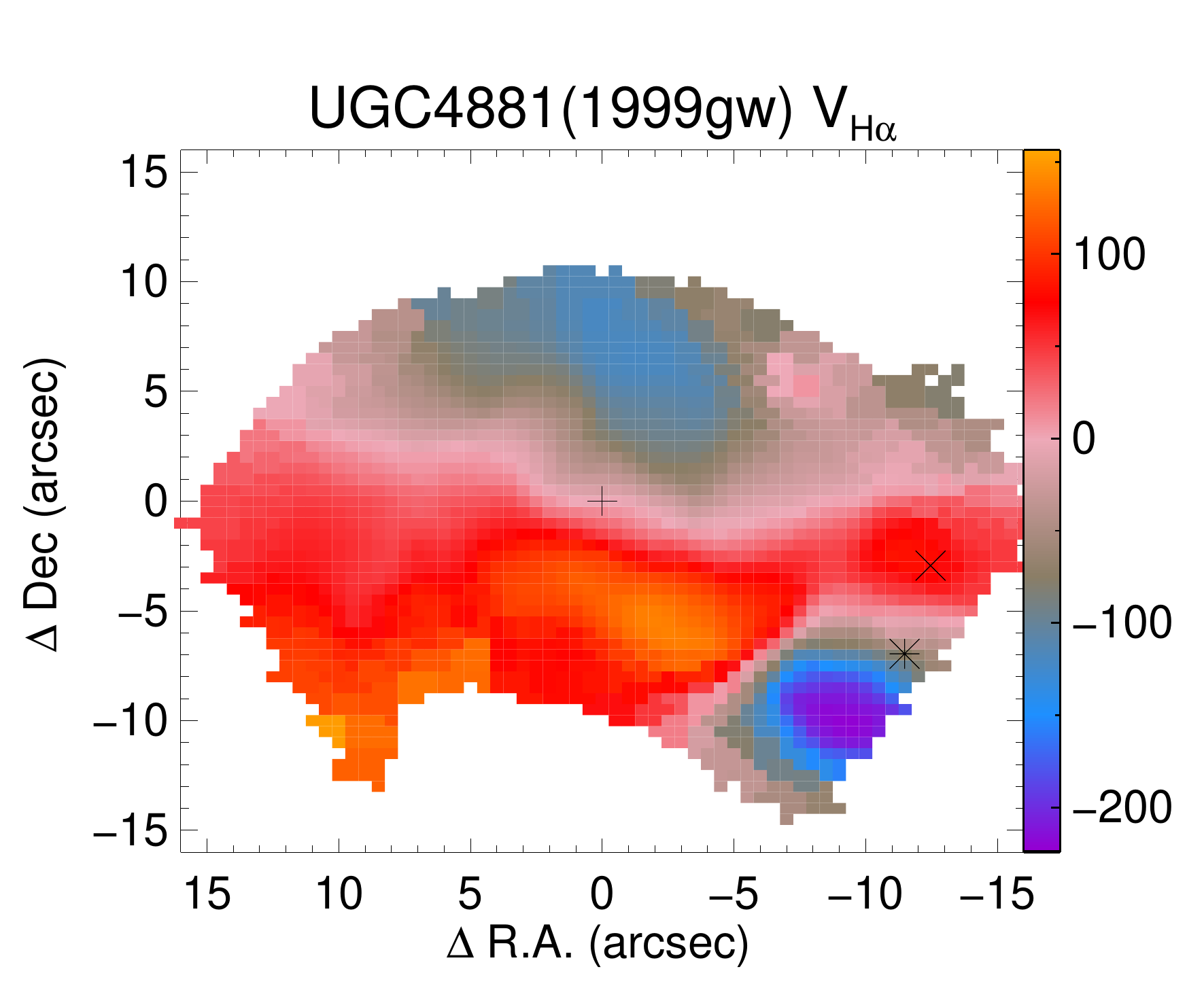}
\hspace{0.2cm}
\includegraphics[width=30mm]{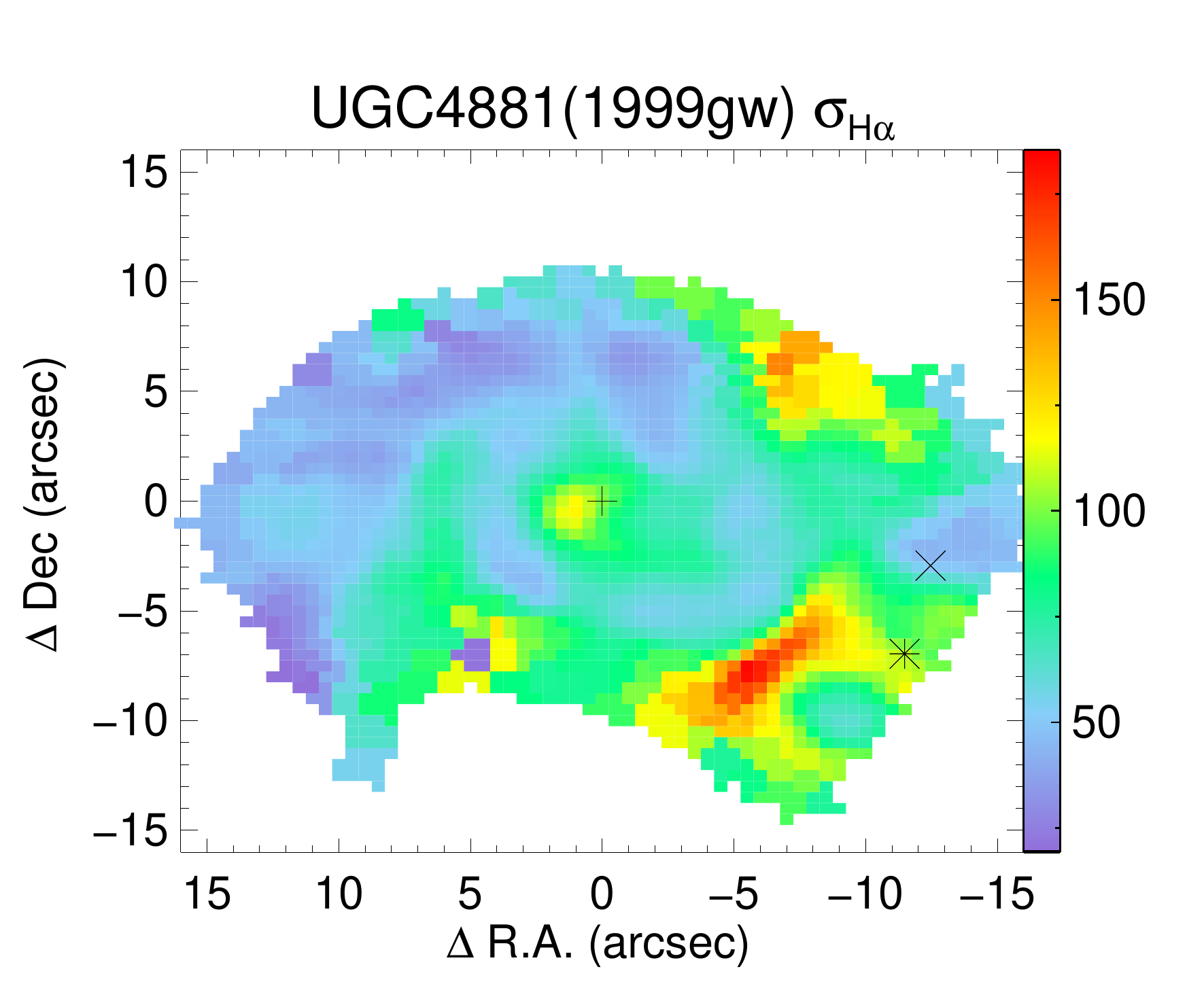}
\end{minipage}
\begin{minipage}{\textwidth}
\hspace{0.9cm}
\includegraphics[width=30mm]{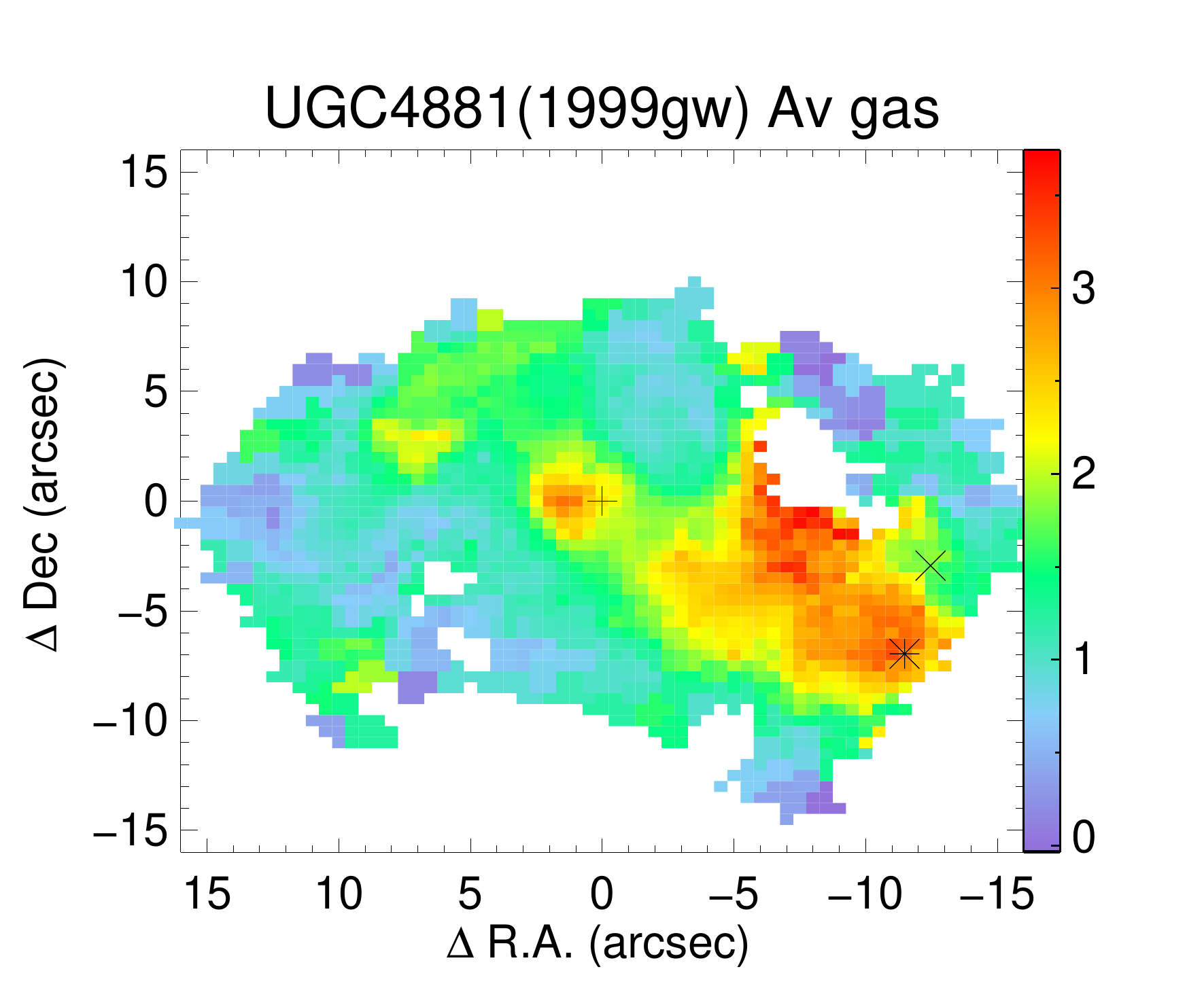}
\hspace{-0.2cm}
\includegraphics[width=30mm]{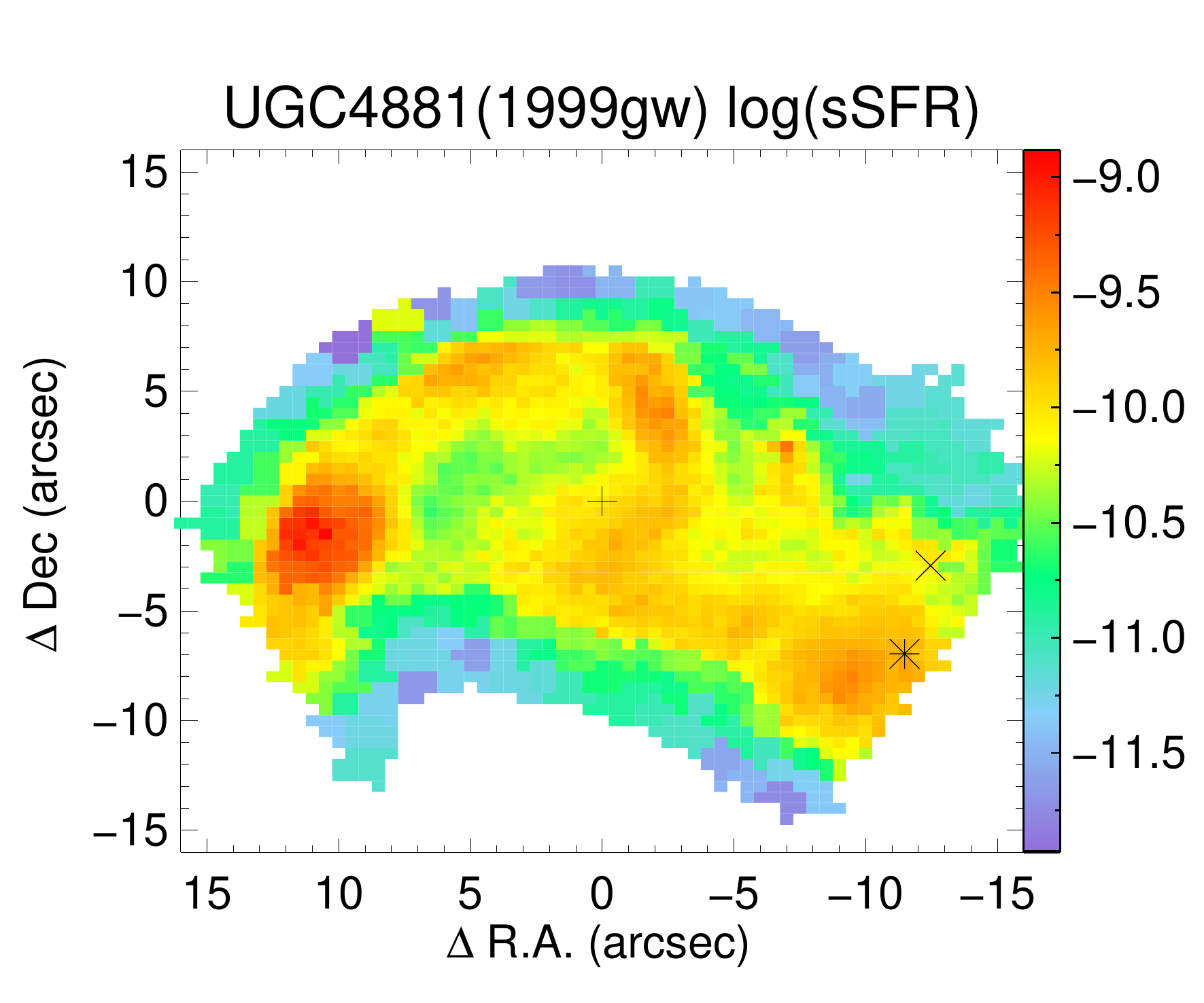}
\hspace{0.2cm}
\includegraphics[width=30mm]{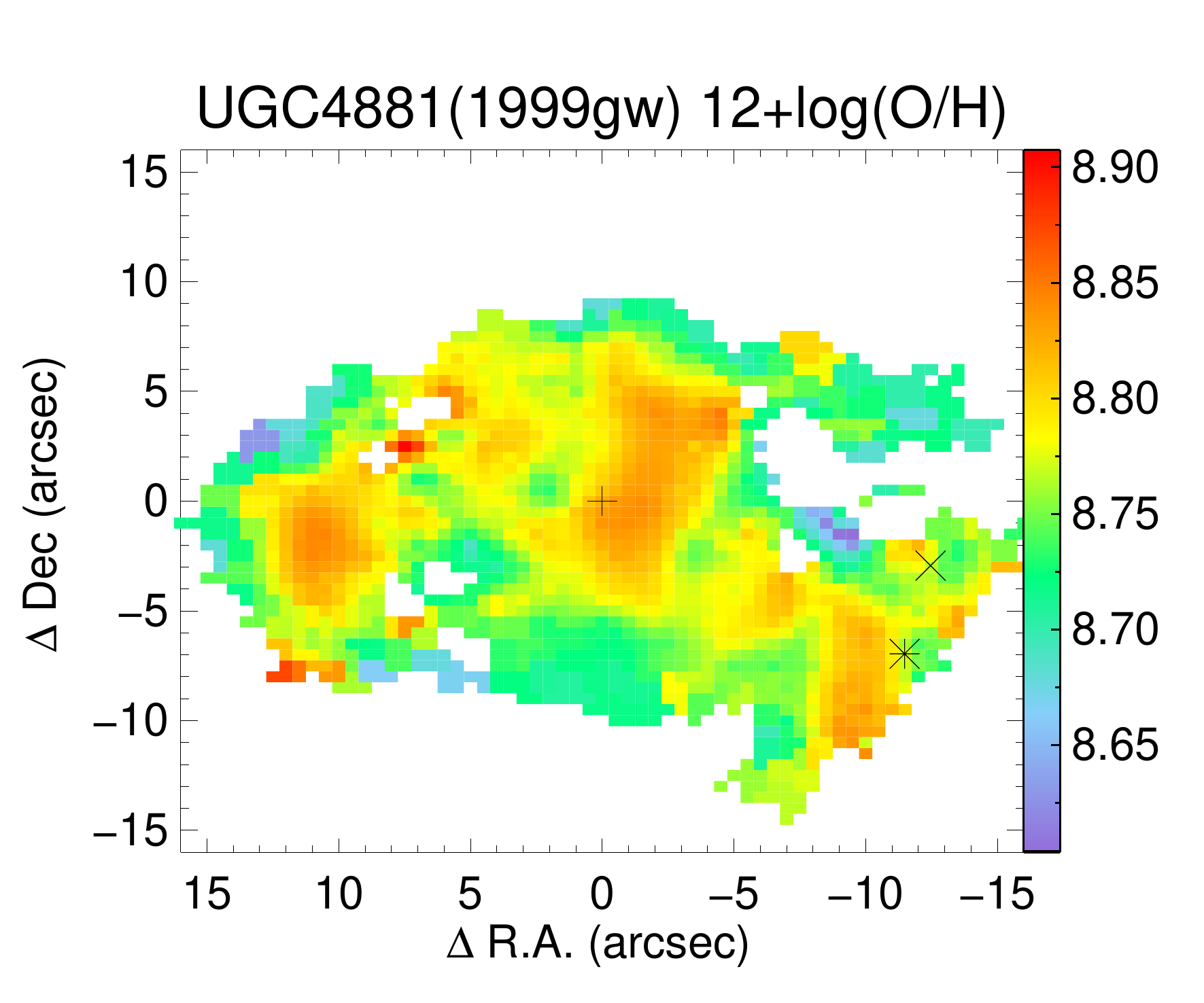}
\hspace{0.2cm}
\includegraphics[width=30mm]{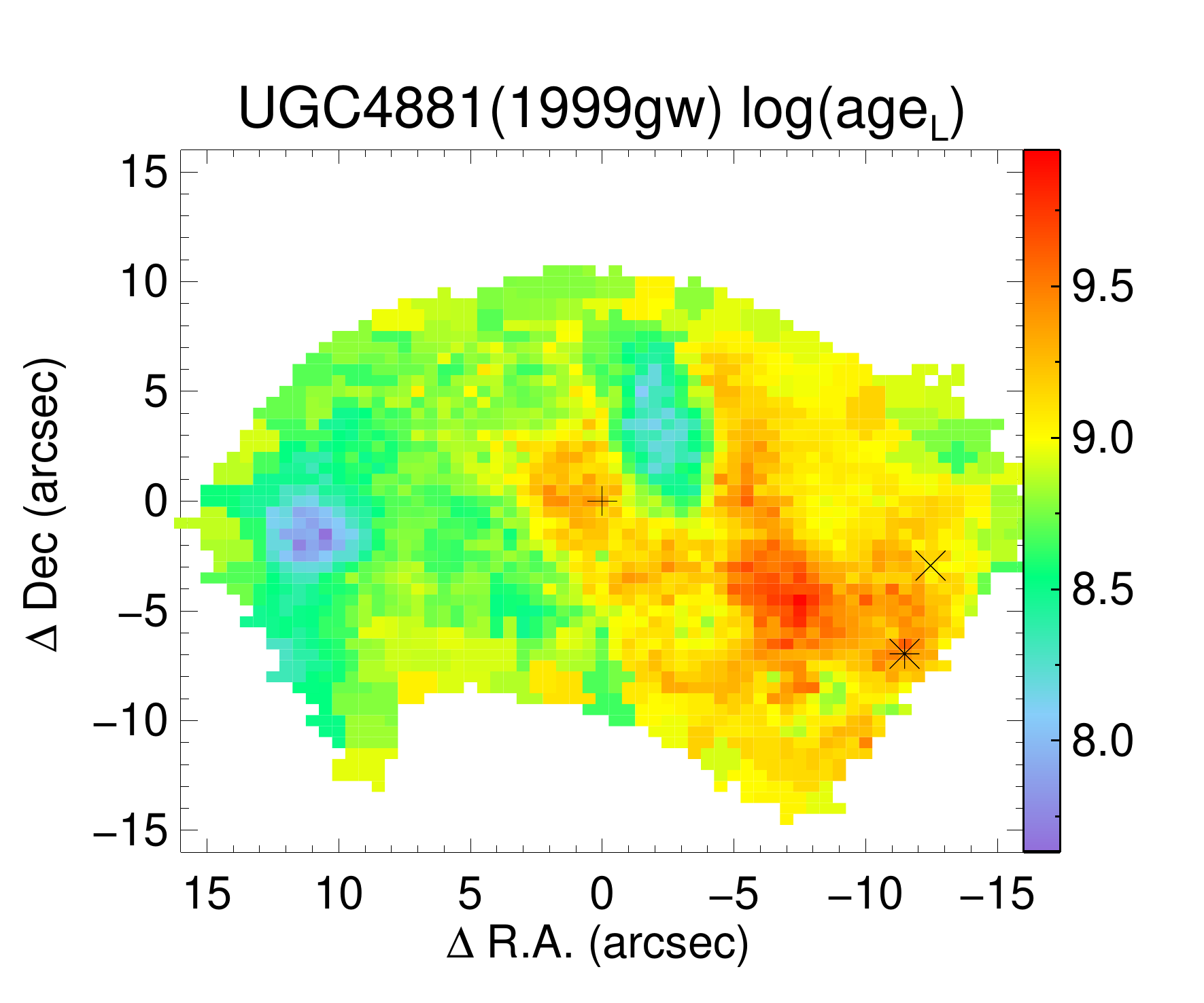}
\end{minipage}
\begin{minipage}{\textwidth}
\hspace{1.5cm}
\includegraphics[width=20mm]{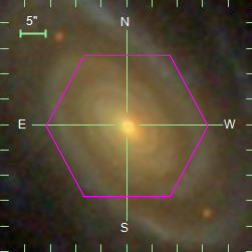}
\hspace{0.2cm}
\includegraphics[width=30mm]{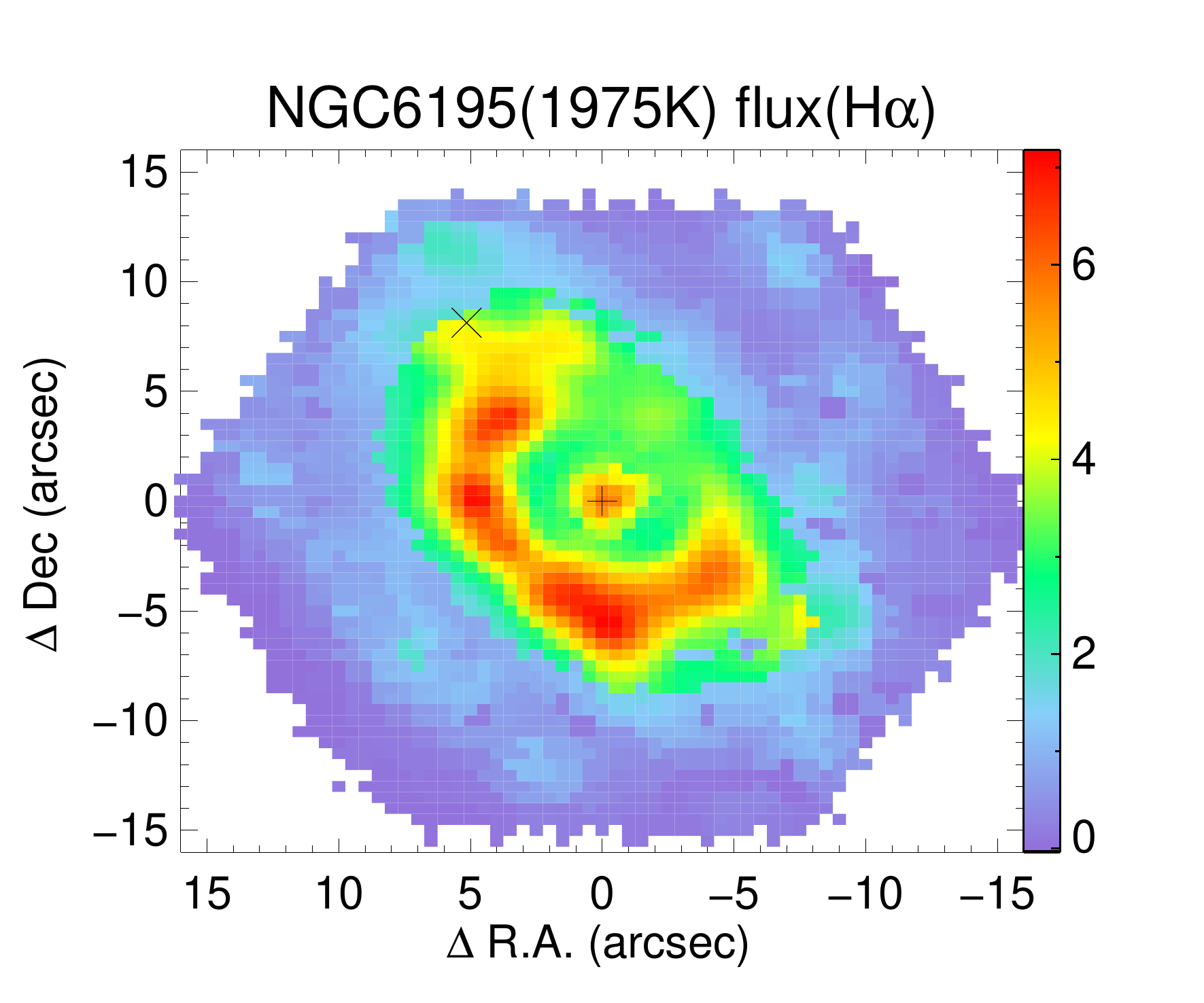}
\hspace{0.2cm}
\includegraphics[width=30mm]{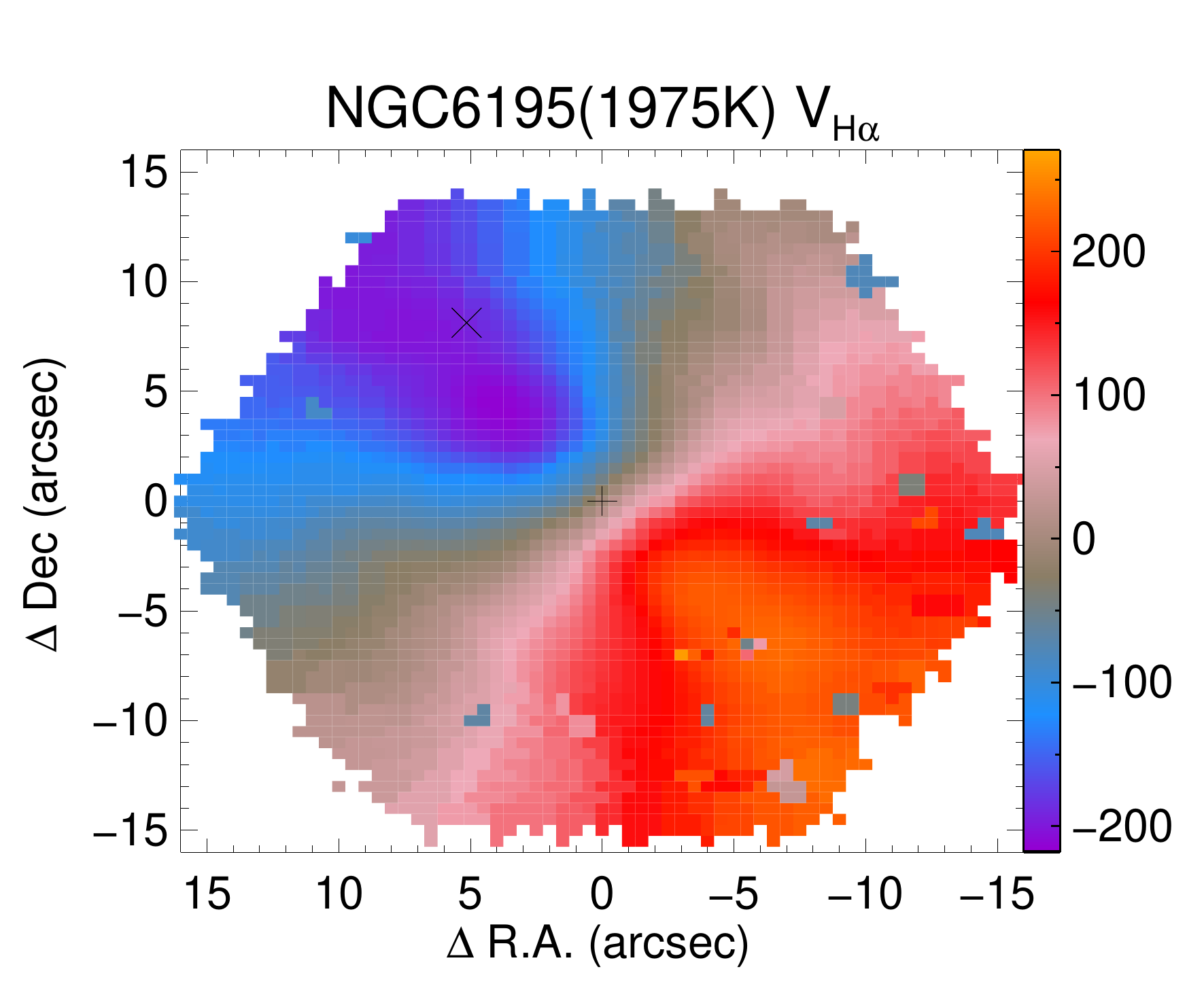}
\hspace{0.2cm}
\includegraphics[width=30mm]{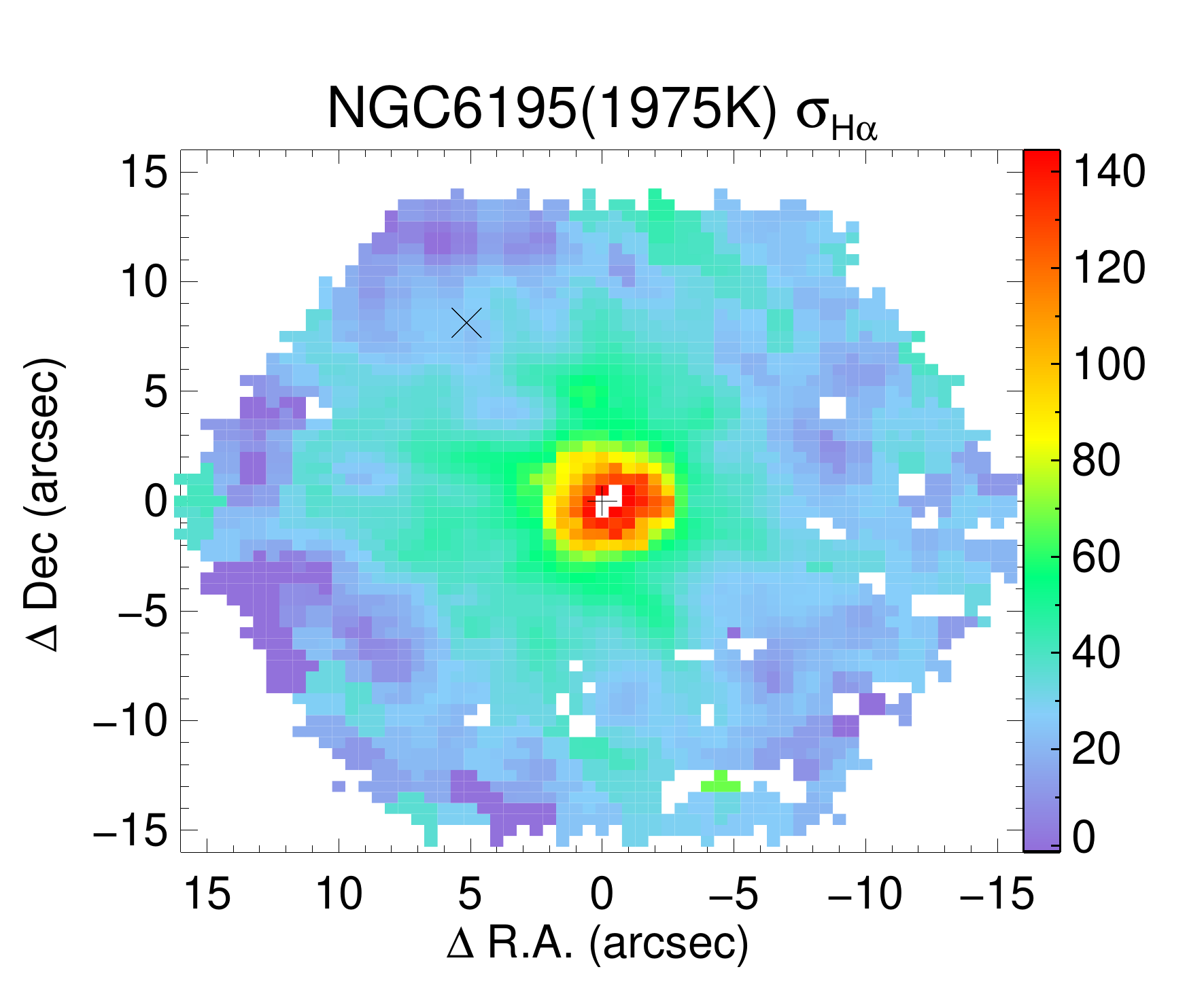}
\end{minipage}
\begin{minipage}{\textwidth}
\hspace{0.9cm}
\includegraphics[width=30mm]{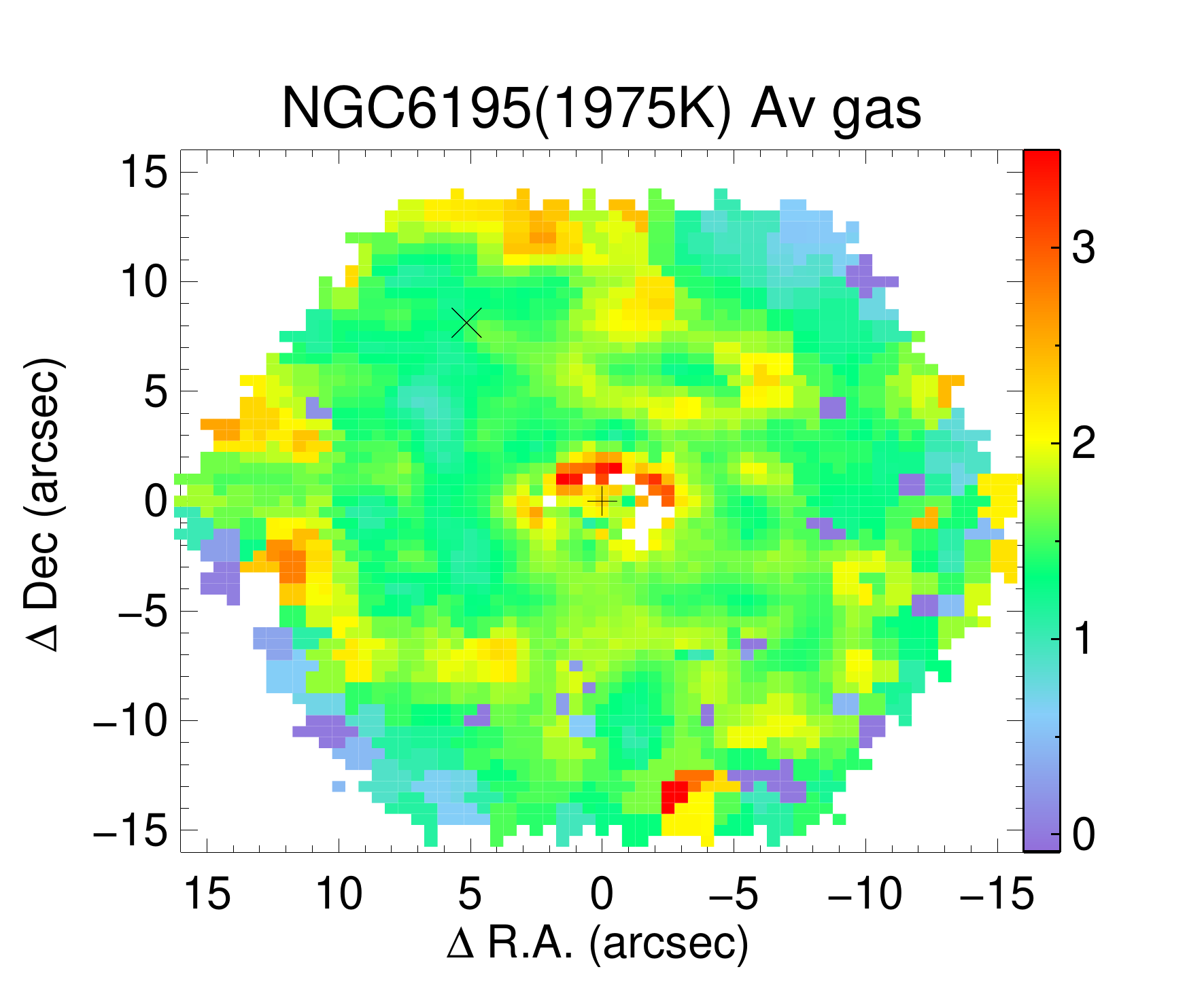}
\hspace{-0.2cm}
\includegraphics[width=30mm]{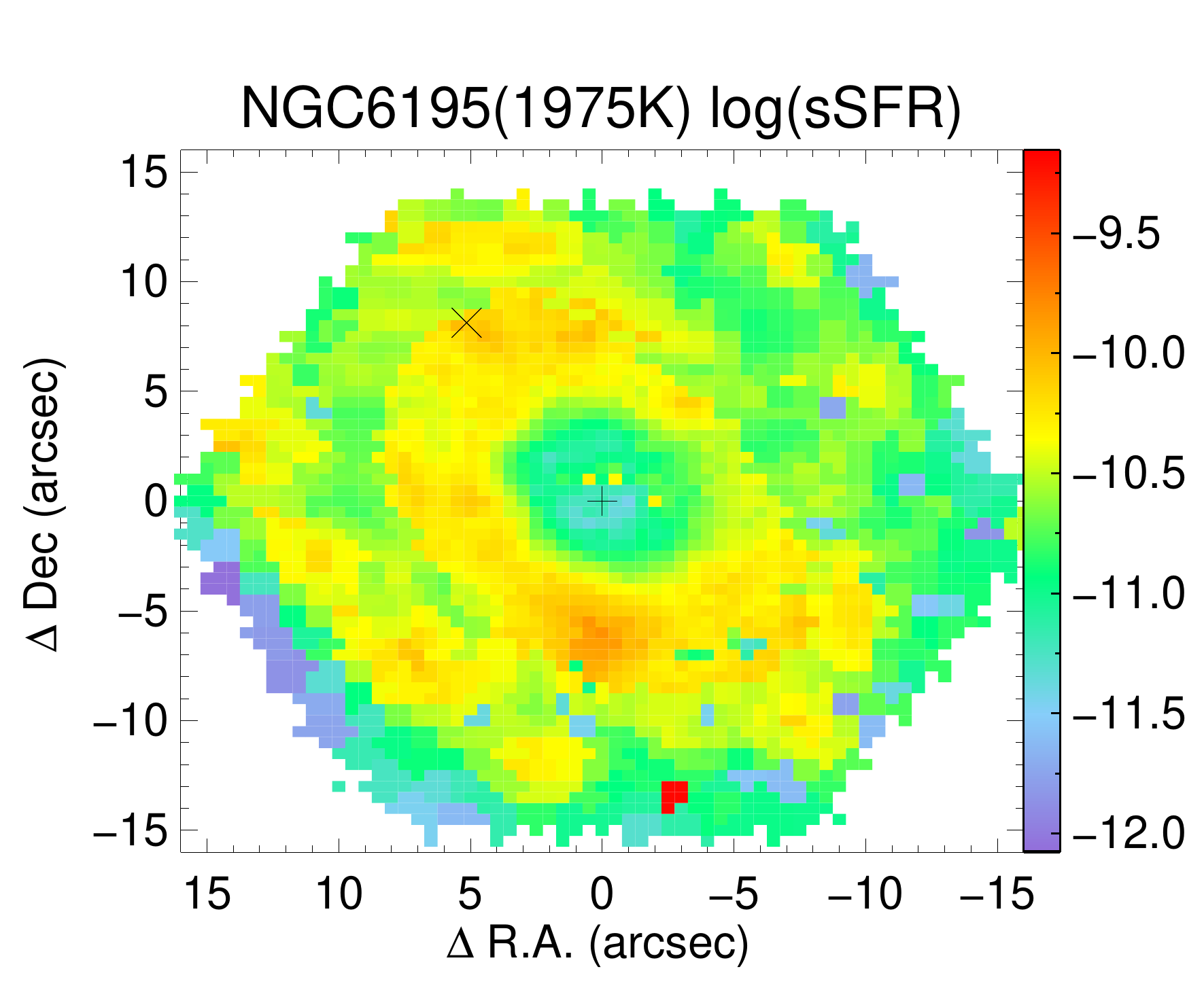}
\hspace{0.2cm}
\includegraphics[width=30mm]{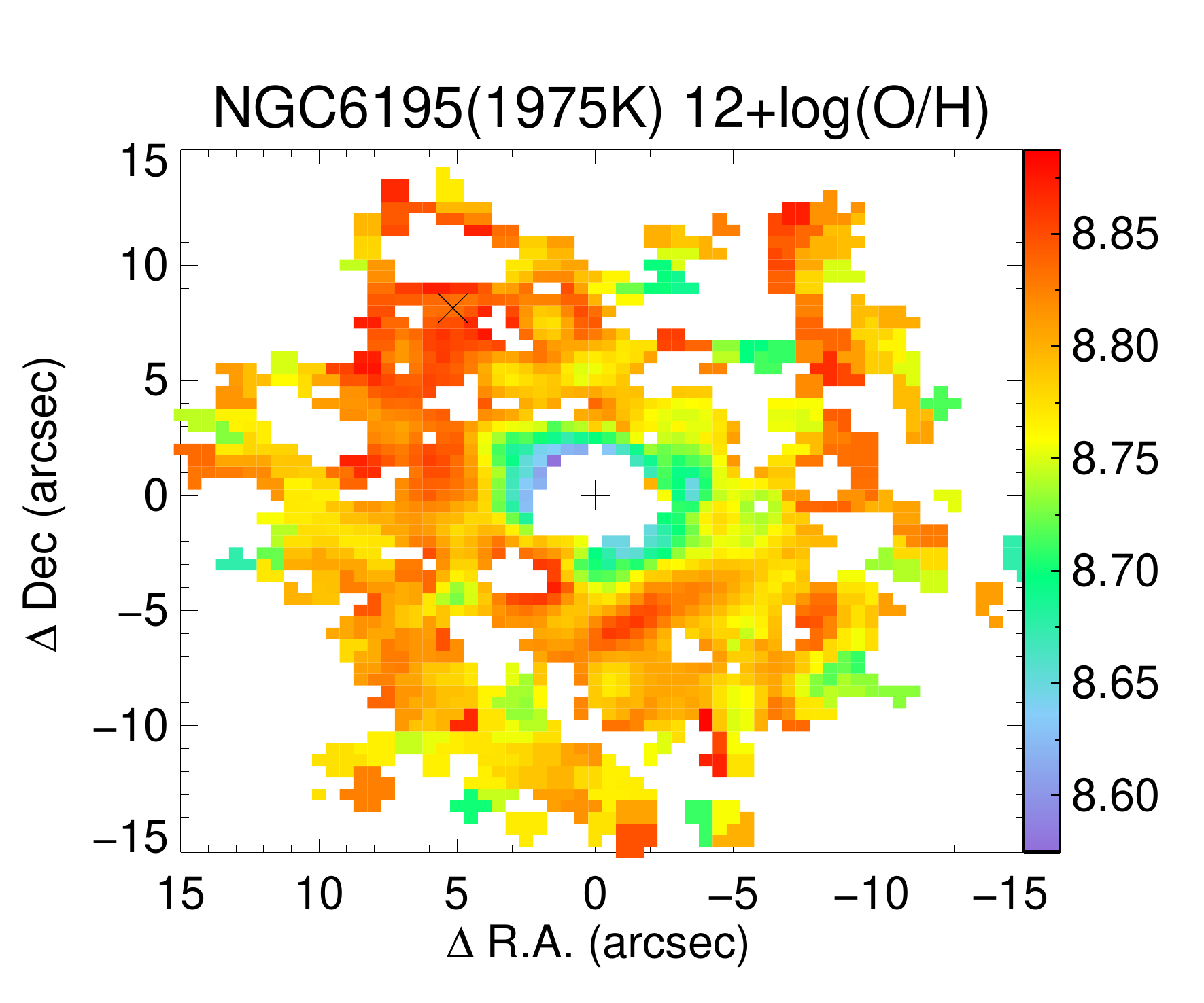}
\hspace{0.2cm}
\includegraphics[width=30mm]{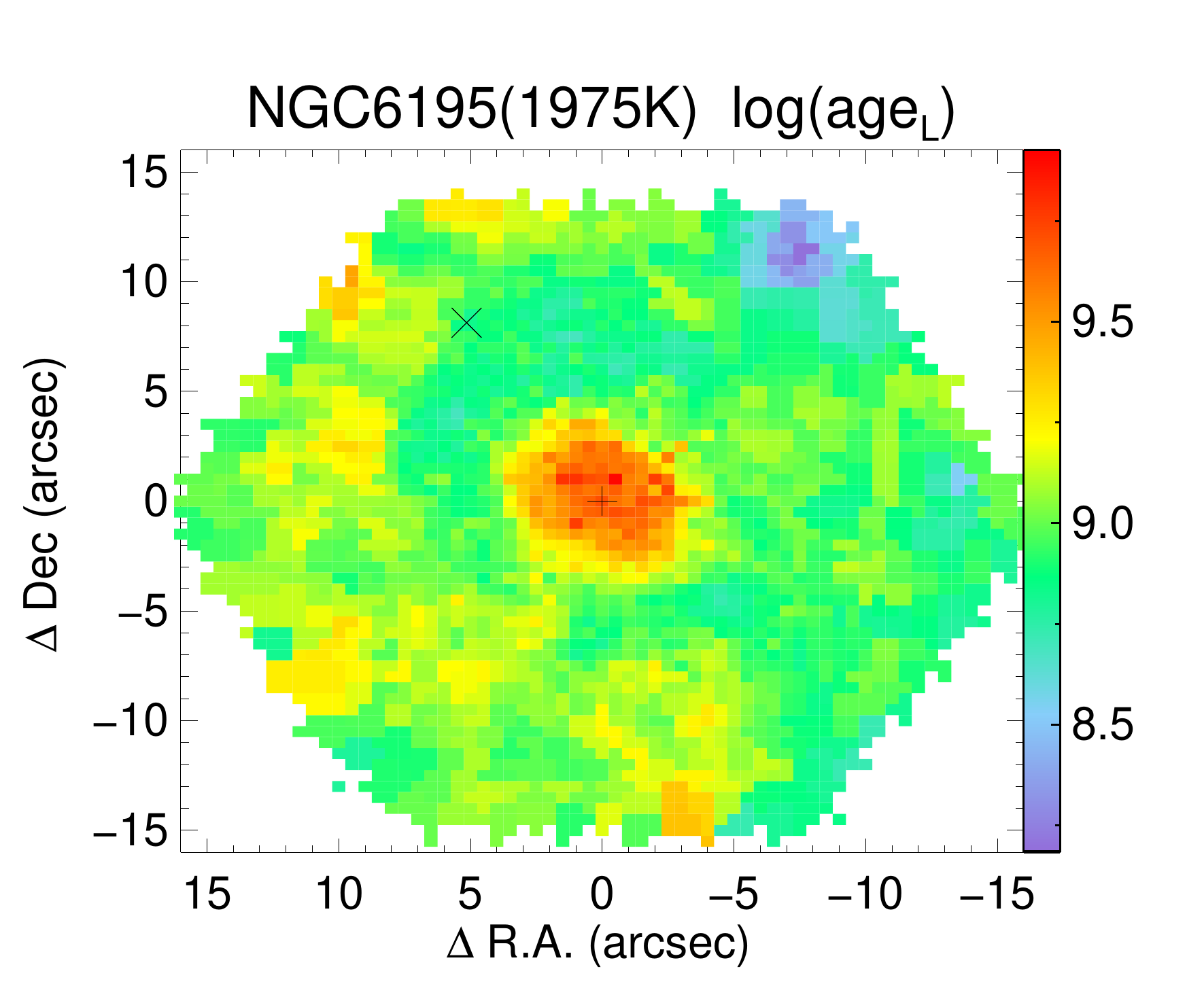}
\end{minipage}
\caption{Same as Fig.~\ref{fig.SNIa-havd} but for unclassified SN host galaxies. For the host galaxy of SN 1999gw, the plus marks the center of the field of view of the merger system, and the star marks the center of the other galaxy in the merger system, the real SN host of 1999gw (UGC 4881 NED02).}
\label{fig.unSN-havd}
\end{figure*}

The H$\alpha$ velocity distributions are displayed in the 2D maps. 
With the method of analysis of \citet{kr06}, we can distinguish whether the galaxy is pure disk rotation or not, which is shown in Table.~\ref{table.allsamples}

For distant galaxies, a simple kinematic classification was developed based on the 3D kinematics and the morphology by \citet{fl06}. \citet{ya08} and \citet{2017sdg..book.....H} summarized the differences between different classes. For rotating disks (RD), the velocity field presents an ordered gradient and the dynamical major axis is consistent with the morphological major axis. There is a single peak in the velocity dispersion ($\rm \sigma$) map, which locates close to the dynamical center. The features of the velocity field of perturbed rotations (PR) are similar with a RD, but there is no peak or the peak is apparently shifted away from the dynamical center in the $\rm \sigma$ map. For complex kinematics (CK), both the velocity field distribution and the $\rm \sigma$ map are irregular and not compatible with regular disk rotation. This classification is based on the large-scale structure of distant galaxies (\citealt{2017sdg..book.....H}). Therefore, we can roughly classify our host galaxies into RD, PR and CK as marked in Table.~\ref{table.allsamples}.

According to Fig.~\ref{fig.SNIa-havd}, the H$\alpha$ velocity gradient of the host galaxies of all the four Type Ia SNe 2007sw (UGC 7228), 2006iq (PGC 1380172), 2007R (PGC 21767) and 2005cc (NGC 5383) are smooth without irregularities. There are some spaxels that have a higher $\rm \sigma$ in the host galaxies and they distribute in the outer regions of the field of view. We infer that this may be caused by the low signal-to-noise (S/N) for the spaxels in the outer region. Roughly, the host galaxies of all the SNe Ia only have one peak of $\rm \sigma$ and the peaks locate at the center of the host galaxies. Therefore, all the SNe Ia host galaxies here are almost RD. That is to say, in our 4 sample galaxies, SNe Ia explode in normal galaxies without turbulence.

According to Fig.~\ref{fig.SNII-havd}, the H$\alpha$ velocity gradient of the host galaxies of all the five Type II SNe have almost regular H$\alpha$ velocity distribution. For SN 2004eb, there is more than one peak in the $\rm \sigma$ map of the host galaxy (NGC 6387), which indicates that the host galaxy of SN 2004eb (NGC 6387) is PR. It can also be shown from the images of MaNGA that NGC 6387 is in an interactive system. However, for the other four Type II SN host galaxies, there is only one peak in the $\rm \sigma$ map and the peak locates at the center of galaxies, which indicates that these host galaxies are RD. Since 4 out of 5 SNe II host galaxies in our sample are RD, and 1 as PR, which indicates that for our sample galaxies, SNe II tend to explode in normal galaxies or an interactive or merger system.

According to Fig.~\ref{fig.unSN-havd}, the H$\alpha$ velocity gradient of the host galaxy of unclassified type of SN 1975K (NGC 6195) is smooth without  irregularities. The $\rm \sigma$ map has only one peak in the galaxy center. Therefore, according to the velocity field and the $\rm \sigma$ map, NGC 6195 is almost RD without turbulence. While for the host galaxy of SN 1999gw (UGC 4881), the H$\alpha$ velocity distribution is irregular. From the maps of $\rm \sigma$, the host galaxy of SN 1999gw (UGC 4881) has more than one peak, so there is turbulence in this galaxy. Therefore, according to the classification criteria in \citet{ya08}, the kinematic type of UGC 4881 is CK, which can also be shown from the image of MaNGA that this galaxy is in merger system.

\subsection{H$\alpha$ flux, extinction, and star formation rate}
\label{flux, extinction, sfr}
We estimate star formation rate through H$\alpha$ flux, which has been corrected for dust extinction.
Here we classify our sample into three groups (galaxies that host SNe Ia, galaxies that host SNe II and galaxies that host unclassified type of supernovae) and analyze them below in three groups.

\subsubsection{SN Ia}
\label{sn ia}

UGC 7228, PGC 1380172, PGC 21767 and NGC 5383 are host galaxies of Type Ia SN 2007sw, 2006iq, 2007R and 2005cc, respectively.

\begin{table*}
\caption{The values of global SFR of galaxy ($\rm SFR_g$), the global sSFR of the host galaxies and the local sSFR at the SNe explosion sites, gas extinction at SN position ($\rm A_{Vl}$) and the global stellar mass of the host galaxies.}

\label{table.sfr}
\begin{threeparttable}
\resizebox{\textwidth}{!}{
\begin{tabular}{|l|l|l|l|r|r|r|r|r|r|}
\hline

  \multicolumn{1}{|c|}{SN} &
  \multicolumn{1}{c|}{SN type} &
  \multicolumn{1}{c|}{Host} &
  \multicolumn{1}{c|}{SFR$_g$} &
  \multicolumn{1}{c|}{log(sSFR$_g$)} &
  \multicolumn{1}{c|}{log(sSFR$_l$)} &
  \multicolumn{1}{c|}{$\rm A_{Vl}$} &
  \multicolumn{1}{c|}{log(M/$\rm M_\odot$)\tnote{a}} &
  \multicolumn{1}{c|}{log(M/$\rm M_\odot$)\tnote{b}} &
  \multicolumn{1}{c|}{log(M/$\rm M_\odot$)\tnote{c}} \\
  \multicolumn{1}{|c|}{}&
  \multicolumn{1}{c|}{}&
  \multicolumn{1}{c|}{}&
  \multicolumn{1}{c|}{($\rm M_\odot/yr$)}&
  \multicolumn{1}{c|}{($yr^{-1}$)}&
  \multicolumn{1}{c|}{($yr^{-1}$)}&
  \multicolumn{1}{c|}{($mag$)}&
  \multicolumn{1}{c|}{\tt STARLIGHT}&
  \multicolumn{1}{c|}{DRP}&
  \multicolumn{1}{c|}{MPA/JHU}\\

\hline
   2007sw &Ia & UGC 7228 &  3.42$\pm$0.003 & -9.96$\pm$0.001 & -10.00$\pm$0.001& 1.84$\pm$0.01 &10.49& 10.34 & ...\\
  2006iq &Ia &  PGC 1380172 &  3.41$\pm$ 0.029& -10.32$\pm$0.004 & -10.78$\pm$0.116&1.84$\pm$0.40 &10.85& 10.73 & 10.96\\
   2007R & Ia &PGC 21767 &  4.45$\pm$0.026 & -10.34$\pm$0.003 & -10.22$\pm$0.003 & 1.70$\pm$0.02 & 10.99&11.00 &11.16 \\
   2005cc & Ia pec & NGC 5383 &  0.53$\pm$0.005 & -9.98$\pm$0.004 & -9.74$\pm$0.008& 1.71$\pm$0.05 &9.7&10.40 &  7.50\\
 2000cs & II pec &  MCG +07-34-015 & 0.32$\pm$0.013& -11.15$\pm$0.018 &-11.37$\pm$0.171& 0.98$\pm$1.29 &10.6& 11.06 & 11.08\\
  2010ee &II &  UGC 8652 &1.18$\pm$0.011 & -10.42$\pm$0.004 & -10.39$\pm$0.004 & 1.11$\pm$0.09 & 10.49&10.28 & 10.61\\
   2012al &IIn & PGC 213664 & 0.04$\pm$0.002&-10.99$\pm$0.026 &-11.04$\pm$0.026 & 0.96$\pm$0.48 &9.55& 9.56 & 9.60\\
   2011cc &IIn & IC 4612 &  4.30$\pm$0.008 &-10.00$\pm$0.001 & -9.99$\pm$0.001& 0.97$\pm$0.06 & 10.63&10.56 & 10.74\\
   2004eb & II &NGC 6387 &5.62$\pm$0.038 &-8.97$\pm$0.003&-9.22$\pm$0.005& 0.40$\pm$0.08 & 9.72&9.61 & 9.61\\
   1999gw &U& UGC 4881 & 13.04$\pm$0.138& -9.93$\pm$0.005&-10.12$\pm$ 0.026& 1.98$\pm$0.15 &11.05&10.73 & 10.95 \\
 1975K & U&  NGC 6195 &  2.26$\pm$0.082 &-10.56$\pm$0.016&-10.26$\pm$0.015 & 1.38$\pm$0.18 & 10.91&11.05 & 11.17\\
\hline
\end{tabular}}
\begin{tablenotes}
 \footnotesize
 \item[a] The global stellar mass of the host galaxies estimated using {\tt STARLIGHT} code in this work. \item[b] The
 global stellar mass of the host galaxies taken from MaNGA Data Reduction Pipeline (DRP) catalogue  (\citealt{la16}). \item[c] The global stellar mass of the host galaxies taken from MPA/JHU  (\citealt{ka03b}, \citealt{sa07}).
\end{tablenotes}
\end{threeparttable}
\end{table*}

From Fig.~\ref{fig.SNIa-havd}, we can see that the $\rm H\alpha$ flux of these four galaxies have the highest value in the center except for the host galaxy of SN 2005cc (NGC 5383).
According to \cite{ja09}, there is an obvious $ \rm H\alpha$ flux deficit in the center of late-type spirals(Sc+),
but very different in Sa-type spirals that the $\rm H\alpha$ flux increases towards the center.
\cite{ja09} shows that the $\rm H\alpha$ flux radial distribution of unbarred Sb spirals is similar to Sa spirals,
but $\rm H\alpha$ flux of barred Sb spirals is strong in the center, then it decreases before $\rm H\alpha$ flux increases again.
According to ASC and \cite{2013OAP....26..187D}, the types of host galaxies of SN Ia 2007sw, 2006iq, 2007R and 2005cc belong to Sbc, SB, S0/a and SBb, respectively. The morphology of the host galaxies is consistent with the classification from the $\rm H\alpha$ flux distribution of the images provided by MaNGA based on the classification method in \citet{ja09}.

Table.~\ref{table.sfr} shows the global SFR of host galaxies in units of $\rm M_{\odot}/yr$, the global sSFR of the host galaxies and the local sSFR at the SNe explosion sites in units of $yr^{-1}$ and gas
extinction at SN position ($\rm A_{Vl}$ ) in units of $mag$.

As \cite{st12} mentioned that in their sample galaxies, the extinction would increase with the $\rm H\alpha$ flux, which was explained that extinction was expected to be observed in the star formation regions.
From the 2D maps of our Fig.~\ref{fig.SNIa-havd},
we can see that the gas dust extinction distribution is not always the same as the trend of $H\alpha$ flux in our sample galaxies, especially for the host galaxy of SN 2005cc (NGC 5383). The extinction $\rm A_{Vl}$ of the four Type Ia SNe positions are high up to about 1.84, which are given in Table.~\ref{table.sfr}.

Compared with the global sSFR of Type Ia SN host galaxies, the local sSFR at the SNe explosion sites is similar or higher for SNe 2007sw, 2007R and 2005cc, which explode close to the center of the host galaxies. While for SN 2006iq, which explodes in the outer region of the galaxy center, the local sSFR at the SN explosion site is lower than the global sSFR of the host galaxy. Fig.~\ref{fig.SNIa-havd} also gives the 2D maps of sSFR of SNe Ia host galaxies. 
From this figure, sSFR decreases towards outer regions for SNe 2007sw and 2006iq, while SNe 2007R and 2005cc have a lower sSFR in the center of the host galaxies than that in the outer region.
Also, the figures and Table.~\ref{table.sfr} indicate that the supernovae positions are close to the highest sSFR region except for supernova 2006iq,
which explodes at the outer region that has much lower sSFR than the global value.

\subsubsection{SN II}
\label{snii}

From Fig.~\ref{fig.SNII-havd}, the host galaxies of Type II SN 2000cs (MCG +07-34-015), 2010ee (UGC 8652), 2012al (PGC 213664) and 2004eb (NGC 6387) have the highest $H\alpha$ flux in the center in each galaxy,
which indicate that these galaxies are unbarred Sb-type spiral galaxies.
From the pseudo-color image of MaNGA, the host galaxy of 2004eb (NGC 6387) is in the process of merger, which will make effect on the SFR,
oxygen abundance etc. The supernova 2004eb is on the edge of the field of view of this galaxy.
The host galaxy of SN 2011cc (IC 4612) has a $H\alpha$ flux deficit in the center, and the supernova locates at the ring of the host galaxy. 

From Fig.~\ref{fig.SNII-havd}, we can see that the gas dust extinction distribution of Type II SN 2010ee host galaxy increases towards the center, which has the similar trend as the $H\alpha$ flux. However, according to the image and the inclination, the host of SN 2010ee is an edge-on galaxy, so it is difficult to estimate the extinction accurately.
For the other four Type II SN host galaxies, the trend of dust extinction is not always the same as $H\alpha$ flux. The local dust extinctions of the five Type II SNe explosion sites are lower than those of SNe Ia.

Fig.~\ref{fig.SNII-havd} also shows the 2D maps of sSFR of the five Type II SN host galaxies.
SNe 2000cs, 2004eb are on the edge of field of view of hosts and have lower sSFR at the SNe positions than those of global values of host galaxies, which are also shown in Table.~\ref{table.sfr}.
The sSFR at the SN position of SNe 2010ee, 2012al and 2011cc are almost the same as global value.

\subsubsection{Unclassified type of supernovae}
\label{unclassified sn}

The local dust extinctions of these two SNe explosion sites are higher than those of SNe II, which can be shown in Table.~\ref{table.sfr}.
From the pseudo-color image of MaNGA and H$\alpha$ flux 2D map, we can see that the host galaxy of SN 1999gw (UGC 4881) is in a merger system. The host galaxy of SN 1975K (NGC 6195) is a Sb-type spiral galaxy based on the morphology type in ASC, which also can be shown from MaNGA image.

Fig.~\ref{fig.unSN-havd} shows the 2D maps of the sSFR of these two unclassified type of SN host galaxies. According to Fig.~\ref{fig.unSN-havd},
SN 1999gw explodes close to the region that has the highest H$\alpha$ flux, but the H$\alpha$ flux at  the specific SN explosion site is much lower. There are double peaks of sSFR distribution in the host galaxy of SN 1999gw (UGC 4881) shown in Fig.~\ref{fig.unSN-havd}, which is consistent with the morphology of the merger system.
From Table.~\ref{table.sfr}, the sSFR at the SN position of SN 1999gw is lower than the global value.
The supernova 1975K locates in the region
where H$\alpha$ flux is lower than the center, but higher than the outer regions.
From Table.~\ref{table.sfr}, the sSFR at the SN position of SN 1975K is higher than the global value of the host galaxy NGC 6195. Fig.~\ref{fig.unSN-havd} shows that it has the highest value of sSFR in the spiral arm regions of galaxy and the sSFR of the central region of the galaxy is lower.

\subsection{Gas phase oxygen abundances}
\label{oh-result}
\begin{table*}
\caption{The gas-phase oxygen abundance estimated from global galaxy spectra, the central regions of SN host galaxies and SN positions using O3N2, N2O2 and $\rm R_{23}$ methods. }
\label{table.oh.pp04}
\centering
\begin{threeparttable}
\resizebox{\textwidth}{!}{
\begin{tabular}[b]{|l|l|l|r|r|r|r|r|r|r|r|r|}
\hline
  \multicolumn{1}{|c|}{SN} &
  \multicolumn{1}{c|}{SN type} &
  \multicolumn{1}{c|}{Host} &
  \multicolumn{3}{c|}{O3N2\_PP04}&
  \multicolumn{3}{c|}{N2O2} &
  \multicolumn{3}{c|}{$\rm R_{23}$} \\
  \multicolumn{1}{|c|}{} &
  \multicolumn{1}{c|}{} &
  \multicolumn{1}{c|}{} &
  \multicolumn{1}{c|}{12+log(O/H)$_{g}$\tnote{a}} &
  \multicolumn{1}{c|}{12+log(O/H)$_{c}$\tnote{b}} &
  \multicolumn{1}{c|}{12+log(O/H)$_{l}$\tnote{c}} &
  \multicolumn{1}{c|}{12+log(O/H)$_{g}$\tnote{a}} &
  \multicolumn{1}{c|}{12+log(O/H)$_{c}$\tnote{b}} &
  \multicolumn{1}{c|}{12+log(O/H)$_{l}$\tnote{c}} &
  \multicolumn{1}{c|}{12+log(O/H)$_{g}$\tnote{a}} &
  \multicolumn{1}{c|}{12+log(O/H)$_{c}$\tnote{b}} &
  \multicolumn{1}{c|}{12+log(O/H)$_{l}$\tnote{c}}  \\

\hline
 2007sw &Ia & UGC 7228 & 8.65 $\pm$ 0.13 &8.78 $\pm$ 0.03 &8.78 $\pm$ 0.04 & 8.88 $\pm$ 0.13 & 8.95 $\pm$ 0.03& 8.98 $\pm$ 0.05 & 8.90 $\pm$ 0.14 & 9.01 $\pm$ 0.04& 9.02 $\pm$ 0.06\\
  2006iq\tnote{d}& Ia & PGC 1380172 &  8.72 $\pm$ 0.18 &8.80 $\pm$ 0.03 & 8.66 $\pm$ 0.19  & 9.02 $\pm$ 0.16  &9.11 $\pm$ 0.04 & 9.01 $\pm$ 0.20 & 9.02 $\pm$ 0.16 &9.14 $\pm$ 0.03 &8.98 $\pm$ 0.20\\
  2007R &Ia &  PGC 21767 &  8.77 $\pm$ 0.12  &8.73 $\pm$ 0.07 & 8.78 $\pm$ 0.09  & 9.06 $\pm$ 0.13   & 9.01 $\pm$ 0.09 &9.06 $\pm$ 0.08 & 9.08 $\pm$ 0.13 &8.85 $\pm$ 0.13 &9.10 $\pm$ 0.06\\
   2005cc &Ia pec & NGC 5383 &  8.82 $\pm$ 0.04  &8.82 $\pm$ 0.02 & 8.81 $\pm$ 0.06  & 9.08 $\pm$ 0.05  &9.09 $\pm$ 0.03& 9.07 $\pm$ 0.05 & 9.12 $\pm$ 0.05 &9.13 $\pm$ 0.02 & 9.12 $\pm$ 0.04\\
   2000cs & II pec &MCG +07-34-015  & 8.68 $\pm$ 0.18  &... & 8.66 $\pm$ 0.20  & 9.04 $\pm$ 0.18  & ...&9.04 $\pm$ 0.20 & 9.04 $\pm$ 0.18 & ...&9.05 $\pm$ 0.19\\
   2010ee &II & UGC 8652 & 8.67 $\pm$ 0.15  &8.71 $\pm$ 0.04 & 8.70 $\pm$ 0.11  & 8.93 $\pm$ 0.15  &8.97 $\pm$ 0.07 & 8.94 $\pm$ 0.12 & 8.92 $\pm$ 0.17 &8.85 $\pm$ 0.12 &8.95 $\pm$ 0.13\\
   2012al & IIn &PGC 213664 & 8.58 $\pm$ 0.18 &8.67 $\pm$ 0.17 &  8.58 $\pm$ 0.18  & 8.87 $\pm$ 0.18    &8.90 $\pm$ 0.17 & 8.86 $\pm$ 0.16 & 8.91 $\pm$ 0.18 &8.95 $\pm$ 0.18 & 8.90 $\pm$ 0.17\\
   2011cc &IIn & IC 4612 &  8.73 $\pm$ 0.12 &8.75 $\pm$ 0.02 & 8.76 $\pm$ 0.04  & 8.99 $\pm$ 0.13    &  9.03 $\pm$ 0.03 &9.04 $\pm$ 0.05 & 8.93 $\pm$ 0.15 &8.93 $\pm$ 0.07 & 8.99 $\pm$ 0.09\\
  2004eb & II & NGC 6387 &  8.43 $\pm$ 0.07 &8.52 $\pm$ 0.02  & 8.48 $\pm$ 0.15  & 8.73 $\pm$ 0.07   &8.81 $\pm$ 0.02 & 8.75 $\pm$ 0.12 & 8.68 $\pm$ 0.09 &8.78 $\pm$ 0.03 &8.73 $\pm$ 0.15\\
  1999gw & U& UGC 4881 &  8.75 $\pm$ 0.13 &8.79 $\pm$ 0.08 & 8.77 $\pm$ 0.07  & 8.97 $\pm$ 0.14    & 8.97 $\pm$ 0.08 & 8.89 $\pm$ 0.08  & 8.90 $\pm$ 0.17 &8.91 $\pm$ 0.11 & 8.71 $\pm$ 0.11\\
  1975K & U& NGC 6195 &  8.72 $\pm$ 0.16  &... &8.85 $\pm$ 0.06  & 8.99 $\pm$ 0.17  &... &9.09 $\pm$ 0.06 & 9.00 $\pm$ 0.17 &...& 9.10 $\pm$ 0.05\\
\hline\end{tabular}
}
\begin{tablenotes}
 \footnotesize
 \item[a] Global gas-phase oxygen abundance of the host galaxies.
 \item[b] The gas-phase oxygen abundance of the central regions of the host galaxies. No measures because the galaxies harbor AGNs \\
 in the center.
 \item[c] Local gas-phase oxygen abundance of SNe explosion sites.
 \item[d] As explained in the caption of Fig.~\ref{fig.SNIa-havd}, we do not cut the spaxels with S/N $\leq$ 3 when estimating the metallicity for this \\
 SN host galaxy.
\end{tablenotes}
\end{threeparttable}

\end{table*}

Gas-phase oxygen abundance distributions estimated by O3N2 method of \cite{pp04} in 2D maps are displayed from
Fig.~\ref{fig.SNIa-havd} to Fig.~\ref{fig.unSN-havd} for our sample galaxies.
The global gas-phase oxygen abundance of the host galaxies and local value at SNe explosion sites  estimated by O3N2, N2O2 and $\rm R_{23}$ methods are displayed in Table.~\ref{table.oh.pp04}.
According to the BPT diagram, the emission lines ratio of the centers of SNe 2000cs and 1975K host galaxies locate in the regions of AGNs.
Even though the effect of AGNs is small, the central regions of these two host galaxies, MCG +07-34-015 and NGC 6195, which harbor AGNs in the center,
are masked in the process of calculating oxygen abundance.
The existence of AGNs will make bias on the global spectra and
make deviation when estimating the gas-phase oxygen abundance.
Also, emission lines with S/N less than 3 are masked to reduce computational errors.

\subsubsection{SN Ia}
\label{snia-oh}
Fig.~\ref{fig.SNIa-havd} shows 2D maps of gas-phase oxygen abundance of 4 SNe Ia host galaxies.
The oxygen abundance increases towards the center of the host galaxy of 2007sw (UGC 7228), which indicate an inside-outside formation of the galaxy.
In the center, the metalliciy reaches the highest value and the supernova 2007sw locates close to the highest metallicity.
The oxygen abundances decrease towards the central regions for the host galaxies of SN Ia 2006iq (PGC 1380172) and 2007R (PGC 21767), which
means an outside-inside formation of the galaxies.
The host galaxy of 2007R (PGC 21767) has been analyzed in \cite{st12}.
Our result for this galaxy agree well with \cite{st12}.
The host galaxy of 2005cc (NGC 5383) shows an irregular distribution of metallicity.
There is a spindly region where presents apparently lower metallicity,
which is believed to be where the spiral arm locates.

Table.~\ref{table.oh.pp04} shows the metallicity estimated from the global galaxies spectra, central regions and the supernovae locations using O3N2, N2O2 and $\rm R_{23}$ methods.
From Table.~\ref{table.oh.pp04}, we can see that the local gas-phase oxygen abundance at the SN 2007sw explosion site is higher than the global value. For SN 2006iq, the local oxygen abundance is lower than the global value. The local oxygen abundance at the SN explosion sites of SNe 2007R and 2005cc are nearly the same as that of the global of their host galaxies.

\subsubsection{SN II}
\label{snii-oh}

For the 5 Type II SNe, Fig.~\ref{fig.SNII-havd} presents the 2D maps of gas-phase oxygen abundance. We can see from the 2D map of the host galaxy of 2000cs (MCG +07-34-015) that the center is masked due to the presence of AGNs, and the metallicity increase in outer regions.
There are two peaks of gas-phase oxygen abundance in the host galaxy of 2010ee (UGC 8652),
which may be in the spiral arms, but the center has lower metallicity.
The gas phase oxygen abundances of host galaxies of SN 2012al (PGC 213664) and 2004eb (NGC 6387) increase towards the centers.
SN 2011cc locates close to the region that has the highest metallicity.
The metallicity of the host galaxy of SN 2011cc (IC 4612) increases from the center, and decreases after running up to the peaks.
We suppose that the peaks are on the ring of the host galaxy.

Shown from Table.~\ref{table.oh.pp04}, the local oxygen abundance is almost the same as the global value estimated by O3N2, N2O2 and $\rm R_{23}$ methods for SNe 2000cs, 2010ee, 2012al, 2011cc and 2004eb.

\subsubsection{Unclassified type of supernovae}
\label{unsn-oh}

The host galaxy of SN 1999gw (UGC 4881)
is in a merger system. The 2D map of the gas-phase oxygen abundance in this merger system is fanshaped.
Fig.~\ref{fig.unSN-havd} shows that there are double peaks of gas-phase oxygen abundance, one is in the central regions of the field of view of MaNGA (UGC 4881 NED01, marked by plus), and the other locates at the right bottom of the field (UGC 4881 NED02, marked by star).
The supernova 1999gw locates close to the peak.
From Table.~\ref{table.oh.pp04}, the difference between local gas-phase oxygen abundance estimated by O3N2 method of SN 1999gw and the global value of host galaxy UGC 4881 is small.
According to Fig.~\ref{fig.unSN-havd}, the gas phase oxygen abundance of SN 1975K host galaxy decreases from outside towards central regions. From Table.~\ref{table.oh.pp04}, we can see that the local gas phase oxygen abundance of SN 1975K is a little higher than that of global value.

\subsection{Stellar mass}
\label{mass}

In this work, we estimate the current stellar mass using {\tt STARLIGHT} fits (\citealt{ci05}).
The global stellar mass of SN galaxies are shown in Table.~\ref{table.sfr}.
For comparison, we have also given the stellar mass taken from MaNGA Data Reduction Pipeline (DRP) catalogue (\citealt{la16}),
and the total stellar mass of SN galaxies taken from MPA/JHU (\citealt{ka03b,sa07}) in Table.~\ref{table.sfr}.
There is no stellar mass data for SN 2007sw host galaxy (UGC 7228) in MPA/JHU. The difference between these three stellar mass calculated using three different methods is very small except for the host galaxy of SN 2005cc (NGC 5383). The stellar mass of this galaxy is ${10^{9.7}}M{_\odot}$ calculated by {\tt STARLIGHT}, ${10^{10.4}}M{_\odot}$ taken from DRP (\citealt{la16}) and ${10^{7.5}}M{_\odot}$ taken from MPA/JHU. The Field of View of MaNGA bundles only cover the inside 1.5 Re or 2.5 Re of the galaxy, so we can not derive the whole galaxy mass based on the IFU data. The galaxy stellar mass given by DRP is based on the photometry data, which is calculated by using the photometry image of the whole galaxy and can hence provide us the mass of the whole galaxy. 
The masses from MPA/JHU are measured from a single fiber spectrum of the nucleus, scaled to the photometry of the whole galaxy, which may bias the final result because it assumes the same stellar populations and L/M along the galaxy. The lower mass of NGC 5383 from MPA/JHU may be mainly resulted from the wrong photometry. 

Here we should note that the calculated stellar mass for SN 1999gw host galaxy is for the merger system, including the host galaxy UGC 4881 NED02 and UGC 4881 NED01.

\subsection{Stellar age}
\label{age}
We estimate stellar age using {\tt STARLIGHT} fits. The 2D maps of light-weighted stellar age of the SN galaxies are presented from Fig.~\ref{fig.SNIa-havd} to Fig.~\ref{fig.unSN-havd}.

As \cite{ci05} pointed out that it is very uncertain for the individual components of the stellar population vectors estimated by {\tt STARLIGHT}. The scatter of the measurements estimated from every single pixel is too large to derive the authentic rules.
The scatter increases with the galactocentric distance increases, which is caused by the low S/N in the outer region of the galaxy. Following \cite{ci05}, to provide a more robust description of the current stellar population of the galaxies, \cite{st12} presented a roughly binned version of the stellar population vectors, which was three bins of stellar population vectors: young stellar population vector (age$_L$ $\leq$ 300 Myr), intermediate stellar population vector (300 Myr $<$ age$_L$ $<$ 2.4 Gyr), and old stellar population vector (age$_L$ $\geq$ 2.4 Gyr).

We provide the mass-weighted and light-weighted stellar population ages of the global of host galaxies and the local SNe explosion sites, and the fractions of spaxels in the bins of young, intermediate and old stellar populations in Table~\ref{table.age}.
From Table~\ref{table.age}, the light-weighted stellar population ages are lower than the mass-weighted ones. The light-weighted age has a higher weight for the younger stellar population.
In our sample galaxies, the mean mass-weighted stellar population age is more than $10^{9.5}$ yr except for the host galaxy of Type II SN 2004eb (NGC 6387), whose mass-weighted stellar population age is $10^{9.11}$ ($\pm$ 0.39) yr, which has a high (95.9\%) fraction of young stellar populations.
The scatter for the stellar population is too large to give an authentic conclusion and the sample size is insufficient, so we will not give a deep discuss for stellar age.

\begin{table*}
\centering
\caption{The mass-weighted and light-weighted stellar population age estimated from global galaxy spectra and SN positions and the fractions of spaxels in the range of stellar population age (young: age$_L$ $\leq$ 300 Myr, intermediate: 300 Myr $<$ age$_L$ $<$ 2.4 Gyr, old: age$_L$ $\geq$ 2.4 Gyr). It also presents the global and central values of Dn(4000) and $H\delta_A$ of the sample galaxies.}
\resizebox{\textwidth}{!}{
\begin{tabular}[b]{|cccccccccccccc|}
\hline
  \multicolumn{1}{|c}{SN} &
  \multicolumn{1}{c}{SN} &
  \multicolumn{1}{c}{Host} &
  \multicolumn{1}{c}{age$_{M,g}$} &
  \multicolumn{1}{c}{age$_{M,l}$} &
  \multicolumn{1}{c}{age$_{L,g}$} &
  \multicolumn{1}{c}{age$_{L,l}$} &
  \multicolumn{1}{c}{age$_{L,young}$} &
  \multicolumn{1}{c}{age$_{L,int}$} &
  \multicolumn{1}{c}{age$_{L,old}$} &
  \multicolumn{1}{c}{$Dn(4000)_g$} &
  \multicolumn{1}{c}{$H\delta_{A,g}$} &
  \multicolumn{1}{c}{$Dn(4000)_c$} &
  \multicolumn{1}{c|}{$H\delta_{A,c}$} \\
  \multicolumn{1}{|c}{} &
  \multicolumn{1}{c}{type} &
  \multicolumn{1}{c}{} &
  \multicolumn{1}{c}{log[yr]} &
  \multicolumn{1}{c}{log[yr]} &
  \multicolumn{1}{c}{log[yr]} &
  \multicolumn{1}{c}{log[yr]} &
  \multicolumn{1}{c}{fraction} &
  \multicolumn{1}{c}{fraction} &
  \multicolumn{1}{c}{fraction} &
  \multicolumn{1}{c}{} &
  \multicolumn{1}{c}{\AA} &
  \multicolumn{1}{c}{} &
  \multicolumn{1}{c|}{\AA} \\
\hline
2007sw &  Ia &  UGC 7228 &  9.79 $\pm$ 0.13   & 9.92 $\pm$ 0.08  & 9.01 $\pm$ 0.28   & 9.26 $\pm$ 0.12   & 1.79 & 88.03 & 10.18 & 1.39 $\pm$ 0.05 & 5.03 $\pm$ 0.82 &1.47 $\pm$ 0.01 &4.21 $\pm$ 0.20\\
 2006iq &  Ia & PGC 1380172 &  9.79 $\pm$ 0.15   & 9.73 $\pm$ 0.20  & 9.22 $\pm$ 0.17   & 9.35 $\pm$ 0.10   & 0.00 & 80.50 & 19.50 & 1.53 $\pm$ 0.26 & 3.20 $\pm$ 1.38 & 1.54 $\pm$ 0.03 &2.81 $\pm$ 0.28 \\
 2007R &  Ia & PGC 21767 &  9.85 $\pm$ 0.11    & 9.87 $\pm$ 0.05  & 8.95 $\pm$ 0.21   & 8.98 $\pm$ 0.08   & 0.25 & 96.36 & 3.39 & 1.51 $\pm$ 0.10 & 3.22 $\pm$  0.90 &1.84 $\pm$ 0.05 &-0.32 $\pm$ 0.42\\
 2005cc &   Ia pec &NGC 5383 & 9.89 $\pm$ 0.12    & 9.82 $\pm$ 0.13  & 8.64 $\pm$ 0.34   & 8.44 $\pm$ 0.14   & 27.25 & 72.75 & 0.00 & 1.37 $\pm$ 0.08 & 4.61 $\pm$ 1.01 &1.53$\pm$0.03 & 2.68 $\pm$ 0.39\\
 2000cs &  II pec & MCG +07-34-015 &9.75 $\pm$ 0.17    & 9.72 $\pm$ 0.13  & 9.23 $\pm$ 0.20   & 9.18 $\pm$ 0.06   & 0.00 & 80.07 & 19.93 & 1.57 $\pm$ 0.14 & 2.59 $\pm$ 1.57 & 2.02 $\pm$ 0.03 & -1.27 $\pm$ 0.17\\
 2010ee &   II &UGC 8652 &  9.75 $\pm$ 0.17    & 9.60 $\pm$ 0.18  & 8.93 $\pm$ 0.34   & 8.52 $\pm$ 0.20   & 13.14 & 79.97 & 6.89 & 1.41 $\pm$ 0.11 & 4.79 $\pm$ 1.33 & 1.68 $\pm$ 0.04 & 1.37 $\pm$ 0.30\\
 2012al &   IIn &PGC 213664 &  9.92 $\pm$ 0.14  & 9.98 $\pm$ 0.11  & 9.36 $\pm$ 0.23   & 9.39 $\pm$ 0.16   & 0.00 & 52.10 & 47.90 & 1.34 $\pm$ 0.06 & 5.24 $\pm$ 0.97 &1.45 $\pm$ 0.02 & 4.04 $\pm$ 0.23\\
 2011cc &   IIn &IC 4612 &  9.66 $\pm$ 0.13   & 9.82 $\pm$ 0.07  & 8.82 $\pm$ 0.40  & 8.63 $\pm$ 0.16  &  20.94 & 73.99 & 5.07 & 1.43 $\pm$ 0.11  & 3.78 $\pm$ 1.33 & 1.56 $\pm$ 0.05 & 1.57 $\pm$ 0.32\\
 2004eb & II &  NGC 6387 &  9.11 $\pm$ 0.39  & 8.97 $\pm$ 0.18  & 7.82 $\pm$ 0.37    & 8.05 $\pm$ 0.17  &  95.90 & 4.10 & 0.00 & 1.20 $\pm$ 0.05 & 5.43 $\pm$ 0.82 & 1.19 $\pm$ 0.01 & 5.15 $\pm$ 0.14\\
   1999gw &U& UGC 4881 &  9.64 $\pm$ 0.30   & 9.81 $\pm$ 0.15  & 8.92 $\pm$ 0.30   & 9.17 $\pm$ 0.14  &  6.54 & 88.03 & 5.43 & 1.40 $\pm$ 0.08 & 5.66 $\pm$ 0.99 & 1.45 $\pm$ 0.03 & 3.57 $\pm$ 0.32 \\
   1975K &U& NGC 6195 &  9.79 $\pm$ 0.11    & 9.78 $\pm$ 0.09  & 9.01 $\pm$ 0.19   & 8.97 $\pm$ 0.10  &  0.98 & 95.07 & 3.95 & 1.50 $\pm$ 0.09 & 3.70 $\pm$ 1.22 & 1.93 $\pm$ 0.03 & -0.79 $\pm$ 0.29\\
\hline\end{tabular}
}
\label{table.age}
\end{table*}

\subsection{Dn(4000) \& $\rm H\delta_A$}
\label{d4000}

\cite{br83} defined Dn(4000) as the average flux density ratio of the two bands of 4050-4250 and 3750-3950 $\AA$.
Later the average flux density ratio of another two narrower bands, 4000-4100 and 3850-3950 $\AA$, was defined as Dn(4000) by \cite{bal99}.
The narrower definition has a significant point that this ratio is less sensitive to reddening effects.
In galaxies which had experienced a star formation burst that ended about 0.1-1 Gyr ago, a strong $\rm H\delta$ absorption line arose.
The peak occurs once hot O and B stars have terminated their evolution (\citealt{ka03a}).
A $\rm H\delta$ absorption line was defined using a central bandpass bracketed through two pseudo-continuum bandpasses by \cite{wo97}.

\begin{figure}[]
\centering
\includegraphics[angle=0,width=7.8cm]{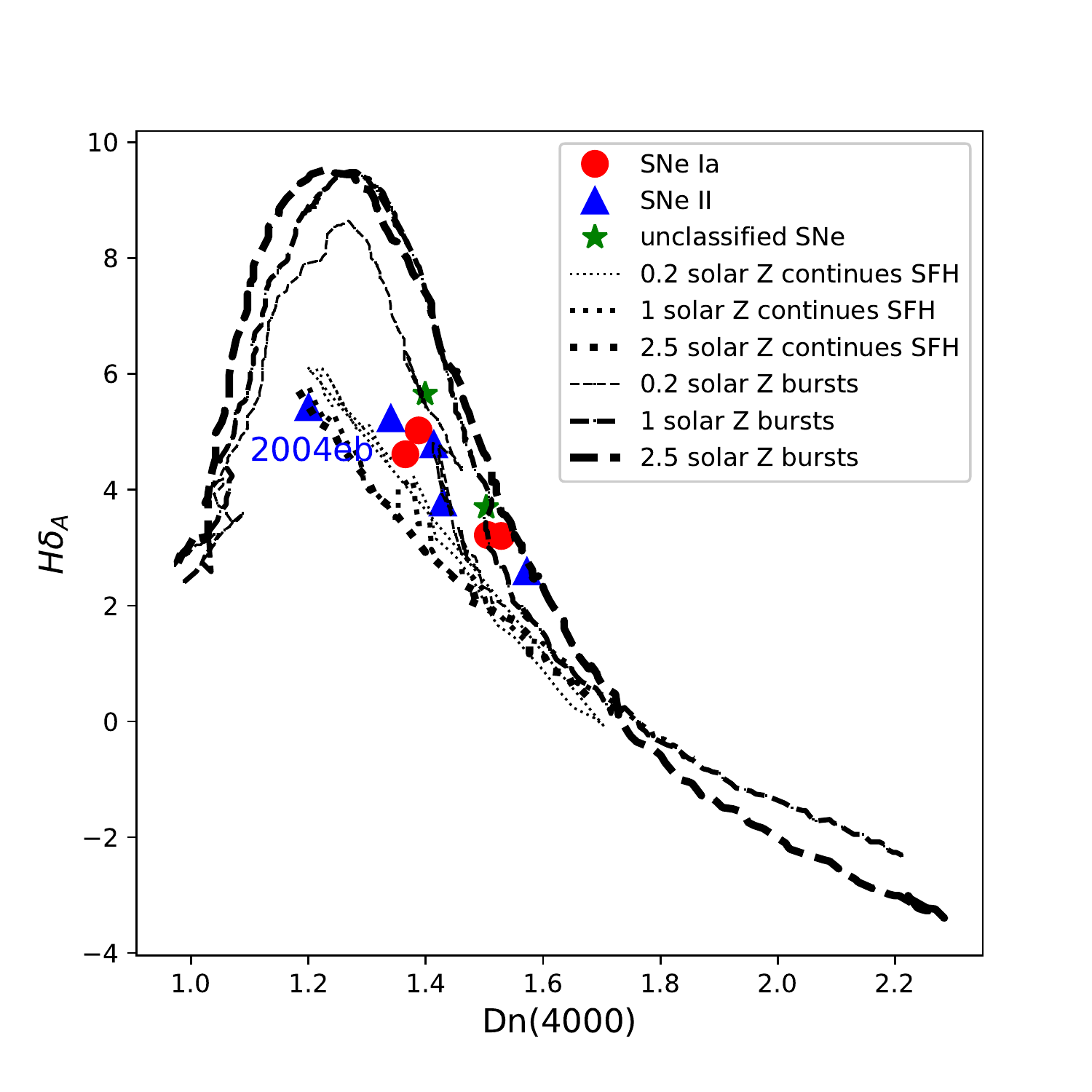}
\caption {The distribution of the global Dn(4000) and $\rm H\delta_A$ for our SN galaxies. The dashed  lines are the relations between Dn(4000) and $\rm H\delta_A$ for 20 per cent solar, solar and 2.5 times solar metallicity bursts, respectively. The dotted lines are the relations between Dn(4000) and $\rm H\delta_A$ for 20 per cent solar, solar and 2.5 solar metallicity continuous star formation histories. These lines are from \cite{ka03b}. }
\label{fig.d4000_hdelta}
\end{figure}
As \cite{ka03a} pointed out that the depth of the 4000 $\AA$ break, Dn(4000), and the equivalent width of the $\rm H\delta$ line, $\rm H\delta_A$,
are significant sensitive indicators of stellar populations that have different ages. 
Fig.~\ref{fig.d4000_hdelta} presents the distribution of the global Dn(4000) and $\rm H\delta_A$ for our sample galaxies. The lines are taken from \cite{ka03b}, which present the relations between Dn(4000) and $\rm H\delta_A$ for pure burst star formation histories and for continuous star formation histories with different metallicity. Our sample galaxies distribute  inside the region of Dn(4000) and $\rm H\delta_A$ plane for pure burst and continues star formation histories with different metallicity. From this figure, there is no significant difference of the global Dn(4000) and $\rm H\delta_A$ for different types of SN galaxies. Table~\ref{table.age} shows that SN 2004eb host galaxy has the lowest value of global Dn(4000) of 1.2 among our sample host galaxies, which is consistent with the result of stellar population age estimated using {\tt STARLIGHT} that this galaxy has a larger fraction of young stellar population.

Here the Dn(4000) and $\rm H\delta_A$ of the SN host galaxies in this study are from \cite{li15}. 
According to \cite{li15}, the standard to classify galaxies into centrally quiescent and centrally star-forming is
whether the Dn(4000) of the center of the galaxies is larger than 1.6 or not.
Among all the 11 SN host galaxies, we can see from Table~\ref{table.age} that the central Dn(4000) of the host galaxies of SN 2007R (PGC 21767), 2000cs (MCG +07-34-015), 2010ee (UGC 8652) and 1975K (NGC 6195)
are larger than 1.6, so these galaxies should belong to centrally quiescent.

\section{Discussion}
\label{discussion}

\subsection{Compare our sample with other works}
\label{comparision}

Compared with SN host galaxies in \cite{ga16b} from CALIFA, our sample size of SN galaxies from MaNGA is much smaller (132/939 vs. 14/1390).
To explore the reason for this, we compare the redshift distributions of these data base galaxies observed with CALIFA, MaNGA in SDSS DR13, and SN host galaxies obtained from the supernovae catalogue in Fig.~\ref{fig.hist_z}. ASC provides redshift information only for 1460 distant galaxies with redshift greater than 0.1. SAI supernovae catalogue provides redshift information for 5790 galaxies, including 1442 galaxies which are also provided by ASC. Therefore, we apply the redshift from SAI supernovae catalogue for comparison here.

From Fig.~\ref{fig.hist_z}, we can see that there is a peak of the redshift of SN galaxies locating at about 0.02 (SAI catalog).
Most of the CALIFA DR3 galaxies redshifts range between 0 and 0.03 with a peak about 0.01.
However, the redshifts of galaxies from MaNGA mainly range from 0 to 0.08 with peak about 0.03.
Therefore, the main reason that our sample size of SN hosts is much smaller than the CALIFA data, is due to the different redshift ranges of the sample. MaNGA selects more distant galaxies, among which less ones could be observed SN explosion inside than the much local ones, such as the CALIFA galaxies.

\begin{figure}[H]
\centering
\includegraphics[angle=0,width=7.8cm]{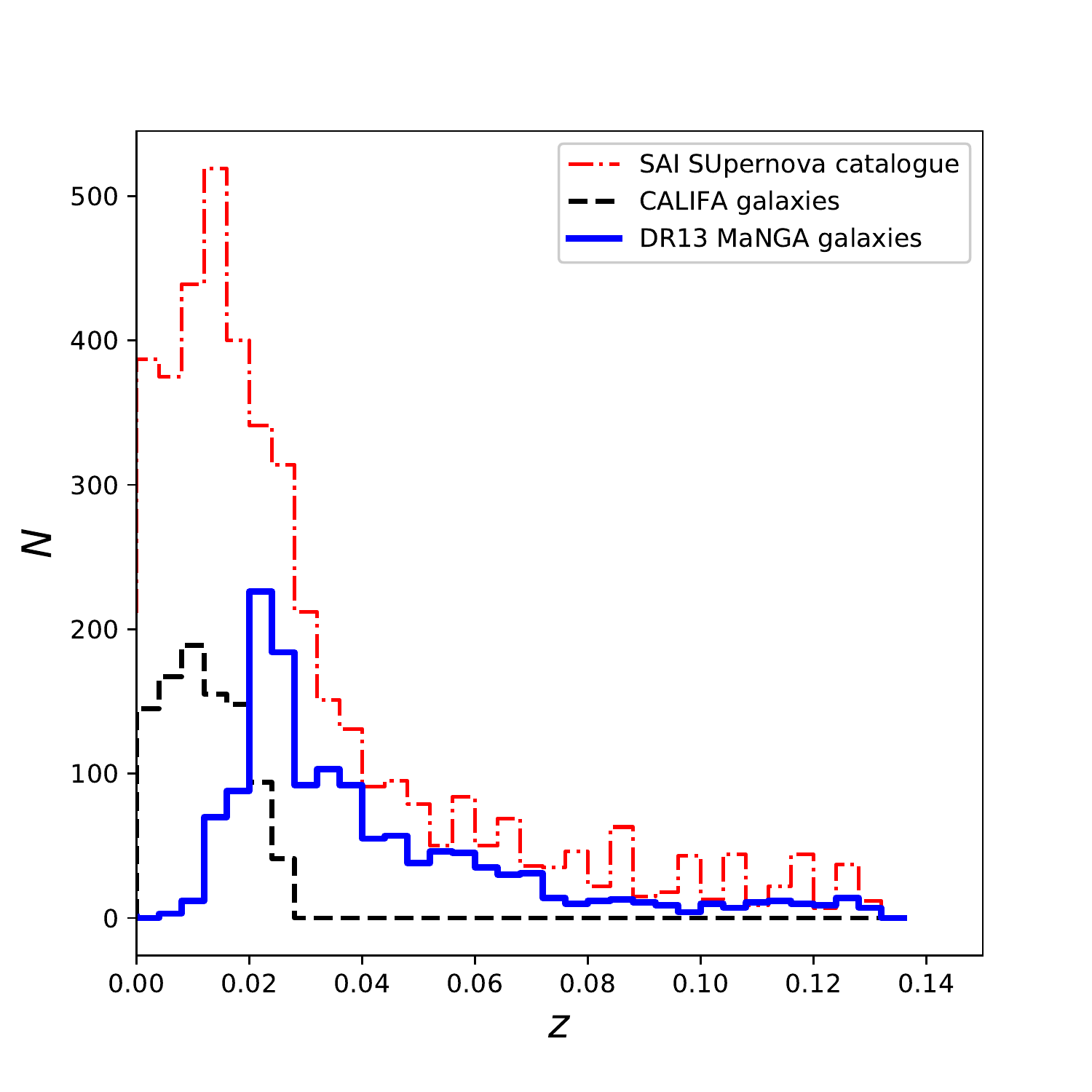}
\caption {The histogram distribution of redshift of galaxies from SAI supernova catalogue, CALIFA, MaNGA galaxies in SDSS DR13, respectively.}
\label{fig.hist_z}
\end{figure}

Also, we check the effect of the diameter of the FoV on different numbers of SNe explored by CALIFA and MaNGA. The largest diameter of the IFU size of MaNGA is 32 $arcsec$ and the FoV of CALIFA is 1.3 $arcmin^2$. We obtain 138 SNe by cross-correlating SNe in ASC with 939 galaxies in CALIFA and 23 SNe  among 1390 galaxies in MaNGA with the same matching radius of 50 $arcsec$. Therefore, the smaller sample size of SN galaxies from MaNGA than that from CALIFA is not mainly caused by the smaller FoV of MaNGA.

\subsection{Star formation rate}
\label{sfr-discussion}
\begin{figure}[]
\centering
\includegraphics[angle=0,width=7.8cm]{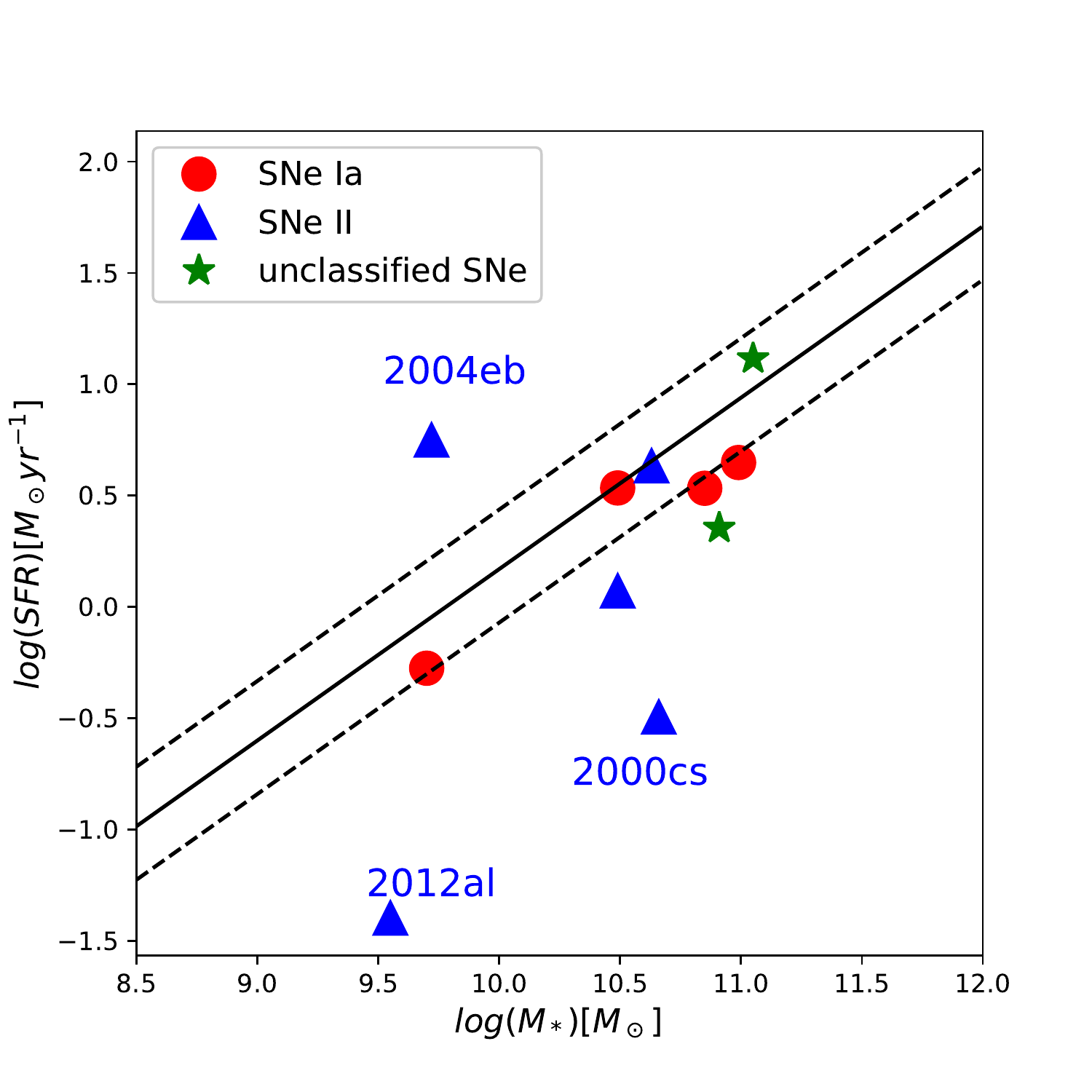}
\caption {The relation between global SFR and stellar mass estimated using {\tt STARLIGHT} code.
The blue SDSS galaxies at $ 0.015 \leq z \leq 0.1$ follow the relation (solid line): $SFR[M_\odot yr^{-1}]= 8.7[-3.7,+7.4] \times (M_\star / 10^{11} M_\odot)^{0.77}$ at the 68\% confidence level (dashed lines, \citealt{el07}).
The red circles, blue triangles and green stars represent SNe Ia,
SNe II and unclassified type of SNe, respectively.}
\label{fig.sfr_m}
\end{figure}

Fig.~\ref{fig.sfr_m} shows the distribution of the global SFR and the stellar mass estimated using {\tt STARLIGHT} code. From this figure, most of our sample galaxies locate within or close to the 68\% confidence level of the locus of blue SDSS galaxies at $ 0.015 \leq z \leq 0.1$ in \cite{el07} except for host galaxies of SN 2000cs, 2012al and 2004eb.
The central region of host galaxy of SN 2000cs locates in the AGN region, which can be shown from Fig.~\ref{fig.BPT}.
According to Table.~\ref{table.age}, the light-weighted stellar age of SNe 2000cs and 2012al host galaxies are older than other galaxies in our sample. The host galaxy of SN 2012al is faint,  which can be shown from the MaNGA image, and the $\rm H\alpha$ flux is very low. Thus, for the host galaxies of SN 2000cs and 2012al, the SFR estimated from $\rm H\alpha$ flux are much lower than normal SF galaxies.
The host galaxy of SN 2004eb is in an interactive system, which could excite star forming activity and result in a higher SFR than normal galaxies.

\subsection{Gas-phase oxygen abundance}
\label{oh-discussion}

Fig.~\ref{fig.oh_delta} presents the local gas-phase oxygen abundance at the SNe explosion sites and the global gas-phase oxygen abundance of the host galaxies estimated using O3N2 method. From the figure, we can see that in our sample, SNe Ia tend to explode in galaxies which have higher gas-phase oxygen abundance than SNe II. 
The gas-phase oxygen abundance of unclassified SN host galaxies is similar with that of SN Ia host galaxies in our sample.
The average local gas-phase oxygen of SNe Ia (12 + log(O/H) $\sim$ 8.76) in the sample host galaxies is also higher than those of SNe II (12 + log(O/H) $\sim$ 8.64). 
Most of our SN host galaxies are metallicity-richer ones with 12+log(O/H)$>$8.5, except one, the host galaxy of 2004eb (NGC 6387), has lower oxygen abundance as 8.43.

The dashed lines in Fig.~\ref{fig.oh_delta} represent the ratio between global and local gas-phase oxygen abundance of 1:1. This figure shows that most of the SN host galaxies locate near the diagonal line. There is a little difference between the local metallicity at the SNe explosion site and the global metallicity of the host galaxy.

\begin{figure}[]
\centering
\includegraphics[angle=0,width=7.8cm]{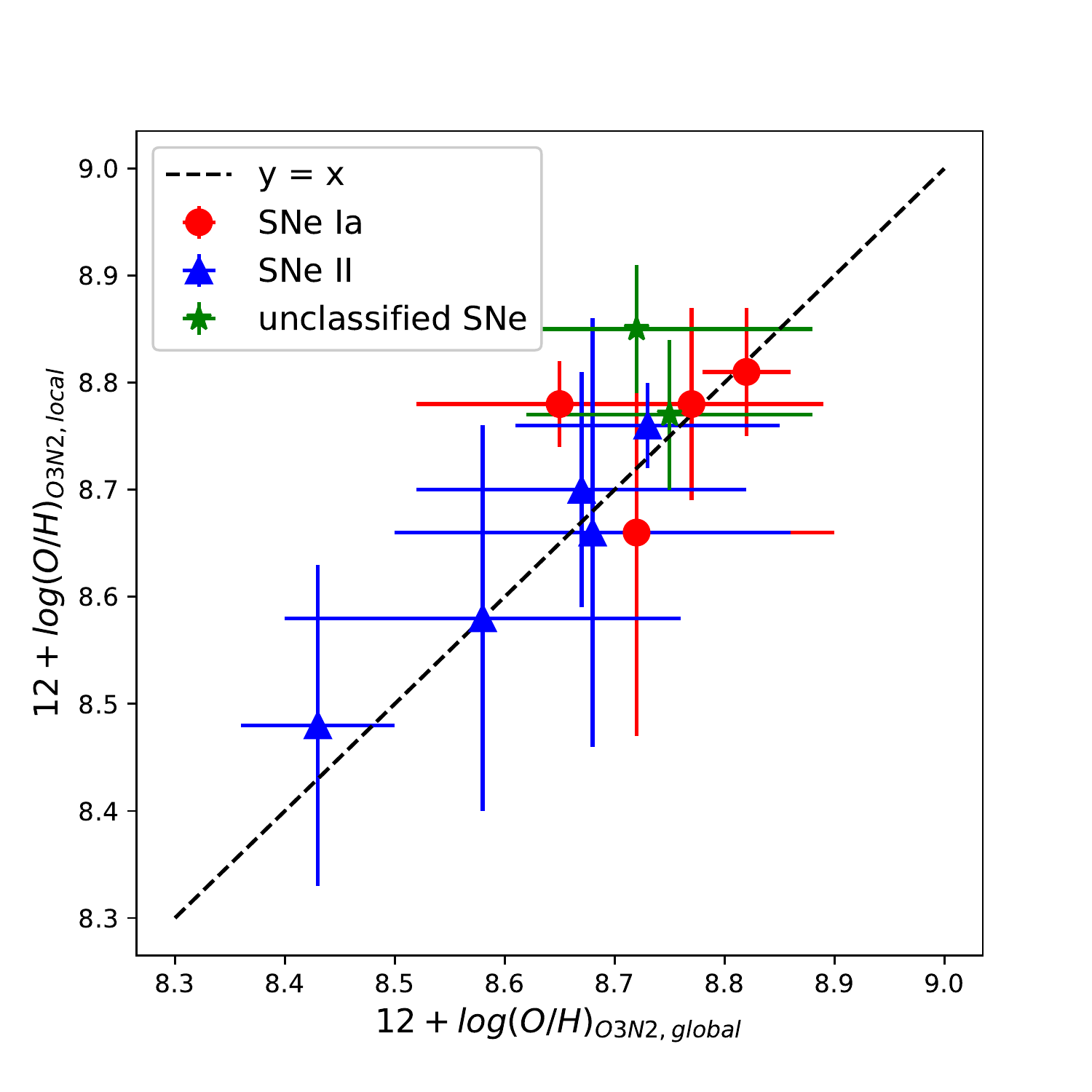}
\caption {The diagram of the gas-phase oxygen abundance estimated by O3N2 method. 
It reveals the difference between the local gas-phase oxygen abundance of
supernovae explosion sites and that of the global host galaxies of all the sample.
The red circles, blue triangles and green stars represent SN Ia,
SN II and unclassified type of SNe, respectively. }
\label{fig.oh_delta}
\end{figure}

\subsection{The relation between stellar mass and gas-phase oxygen abundance}
\label{m-oh}

To compare our SN host galaxies with the SDSS main sample galaxies from \cite{tr04}, we present the distributions of stellar mass
described with nsa-mstar in the DRP catalogue, which could provide the stellar mass of the whole galaxy (see details in Sect.~\ref{mass}), and the global gas-phase oxygen abundance of the host galaxies estimated using $\rm R_{23}$ method in Fig.~\ref{fig.oh_m}. 

In this figure, red dots, blue triangles and green stars represent SNe Ia, SNe II and unclassified SNe, respectively.
The dashed line represents the polynomial fitting to the median value in bins of 0.1 dex in stellar mass from \cite{tr04}. The dotted lines show the contours that enclose 68\% and 95\% of the data from \cite{tr04}.
According to Fig.~\ref{fig.oh_m}, there is a positive relation between stellar mass and metallicity, which is consistent with \cite{tr04}. Most of our sample galaxies locate in the region of 95\% of the data of \cite{tr04} except for the host galaxy of unclassified type of SN 1999gw, which is in a merger system.
The gas-phase oxygen abundance of peculiar Type Ia SN 2005cc host galaxy estimated using $R_{23}$ method is a little higher than those of other sample galaxies. This host galaxy also shows the highest gas-phase oxygen abundance estimated using O3N2 and N2O2 methods.

All the Type Ia SNe in our sample explode in the galaxies with the stellar mass higher than ${10^{10}}M{_\odot}$, which is consistent with \cite{ga14}. According to \cite{su06}, SNe Ia also explode in galaxies that have lower mass (than ${10^{10}}M{_\odot}$). The fact that our sample galaxies lack of low-mass galaxies may be
caused by the small sample size.

\begin{figure}[]
\centering
\includegraphics[angle=0,width=7.8cm]{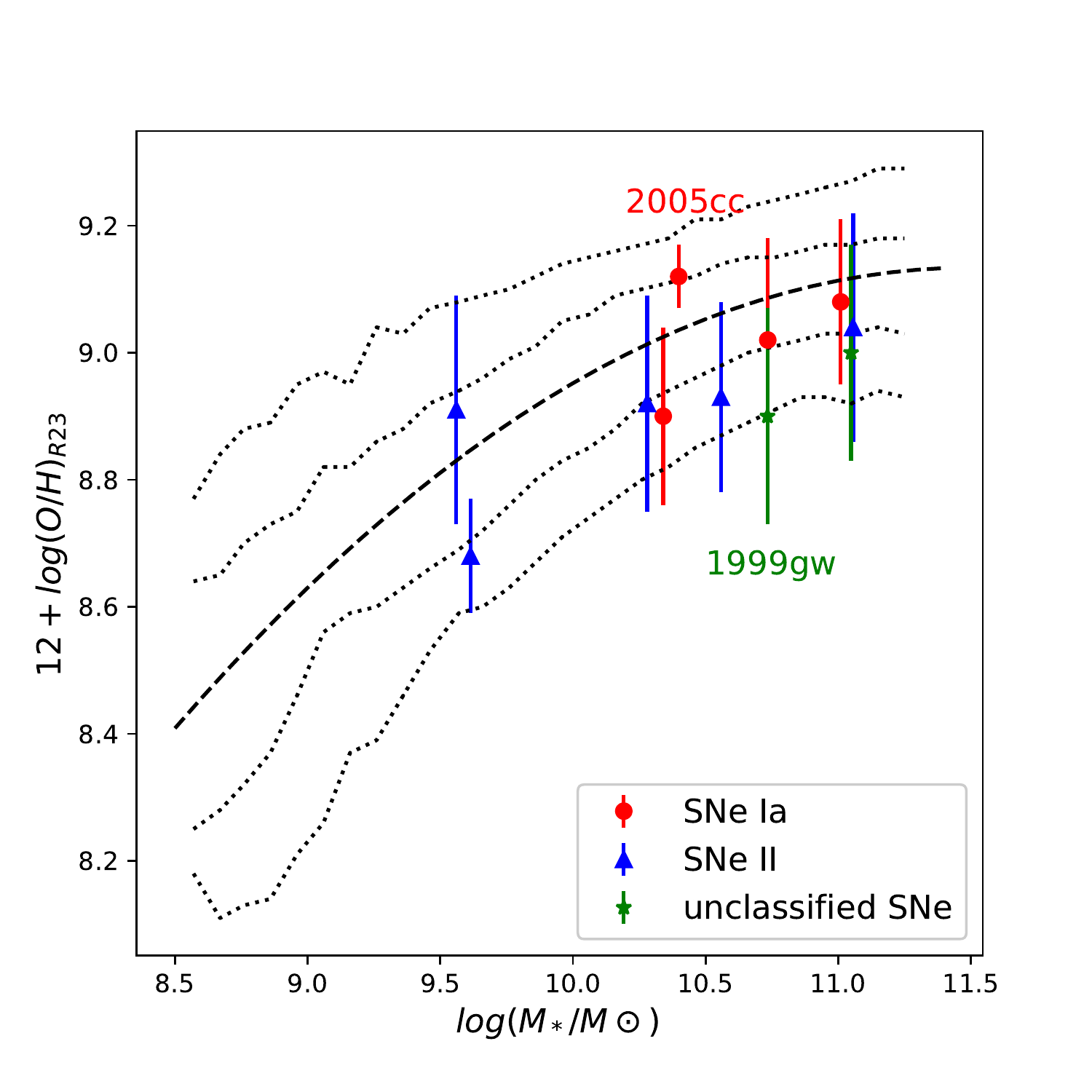}
\caption {The distributions of the stellar mass and the global gas-phase oxygen abundance of the host galaxies in our sample.
The stellar mass is taken from MaNGA DRP catalogue (\citealt{la16}). The gas-phase oxygen abundance is estimated using $\rm R_{23}$ method. The dashed line is the polynomial fitting to the median value in bins of 0.1 dex in mass from \cite{tr04}. The dotted lines are the contours that enclose 68\% and 95\% of the data from \cite{tr04}.}
\label{fig.oh_m}
\end{figure}

\section{Conclusion}
\label{conclusion}

In this paper,
we analyze the local properties of explosion sites of 11 supernovae and global properties of
their host galaxies using IFS of MaNGA.

There are some significant advantages in our work.
Compared with multicolor broad-band imaging or integrated spectroscopy, 
we could derive the 2D maps of parameters for SN galaxies using the spatially resolved spectroscopy of MaNGA to compare the local properties at the SNe explosion sites and the global properties of the SN galaxies.
What's more, the higher redshift distribution up to median as 0.03 allows us to obtain sample and information about more distant galaxies which host different types of SNe. Thanks to the 2D maps of MaNGA observations, we can analyze these SN host galaxies one by one in much details including their $H\alpha$ velocity, sSFR, gas-phase oxygen abundance and stellar population age etc. for a sample of 11 sample galaxies. The results are concluded as follows.

With the little differences between local metallicity at the SNe explosion sites and that of global galaxy,
the metallicity estimated from integrated spectra can represent the local metallicity at SNe explosion sites with small bias, which is in consistent with \cite{ga16b}. Here global refers to the whole area inside the field of view of MaNGA, which cover more than 1.5 $R_e$ or 2.5 $R_e$ of the galaxies. 
For our sample, SNe tend to explode in rich-metallicity galaxies. 

From the velocity field and velocity dispersion map,
we can conclude that both SNe Ia and SNe II in our sample could explode in normal galaxies.
SNe II also could explode in an interactive or merger system, which have recent star formation.
For our sample galaxies, the global and local gas-phase oxygen abundance of SN Ia host galaxies are a little higher than those of SN II host galaxies.
On average, the stellar mass of SN Ia host galaxies in our sample is a little higher than that of SN II host galaxies.
SN Ia in our sample could explode in more massive galaxies, while SNe II can explode in both high mass and low mass galaxies. More sample galaxies with lower masses are needed for more information.

The MaNGA survey will provide a larger sample of SN
galaxies, which could provide us statistical conclusion of the differences of the explosion environment between different types of SNe
will be obtained
in the following work.

\begin{acknowledgements}
We appreciate the referee who provided very constructive and
helpful comments and suggestions, which helped to improve very
well our work. We thank Weibin Shi, Wei Du, Hao Tian and Bo Zhang for helpful and significant discussions on this work. We thank Cheng Li, Enci Wang, Ting Xiao and Lin Lin for providing the data of Dn(4000) and $\rm H\delta_A$ and we thank Yanbin Yang, Francois Hammer, Michel Dennefeld for the helpful discussion on the kinematics of the galaxies. We thank Zheng Zheng for the helpful discussions and comments.
This work was supported by the National Natural
Science Foundation of China (Grant Nos.11733006, 11273026, 11233004
, U1631105).

Funding for the Sloan Digital Sky Survey IV has been provided by the Alfred P. Sloan Foundation, the U.S. Department of Energy Office of Science, and the Participating Institutions. SDSS acknowledges support and resources from the Center for High-Performance Computing at the University of Utah. The SDSS web site is www.sdss.org.

SDSS is managed by the Astrophysical Research Consortium for the Participating Institutions of the SDSS Collaboration including the Brazilian Participation Group, the Carnegie Institution for Science, Carnegie Mellon University, the Chilean Participation Group, the French Participation Group, Harvard-Smithsonian Center for Astrophysics, Instituto de Astrofísica de Canarias, The Johns Hopkins University, Kavli Institute for the Physics and Mathematics of the Universe (IPMU) / University of Tokyo, the Korean Participation Group, Lawrence Berkeley National Laboratory, Leibniz Institut für Astrophysik Potsdam (AIP), Max-Planck-Institut für Astronomie (MPIA Heidelberg), Max-Planck-Institut für Astrophysik (MPA Garching), Max-Planck-Institut für Extraterrestrische Physik (MPE), National Astronomical Observatories of China, New Mexico State University, New York University, University of Notre Dame, Observatório Nacional / MCTI, The Ohio State University, Pennsylvania State University, Shanghai Astronomical Observatory, United Kingdom Participation Group, Universidad Nacional Autónoma de México, University of Arizona, University of Colorado Boulder, University of Oxford, University of Portsmouth, University of Utah, University of Virginia, University of Washington, University of Wisconsin, Vanderbilt University, and Yale University.
\end{acknowledgements}

\label{lastpage}
\end{document}